\documentclass[longauth]{aa}

\usepackage{graphicx}
\usepackage{natbib}
\usepackage{scalerel}
\usepackage{float}
\usepackage[table]{xcolor}
\usepackage{booktabs}
\usepackage{multirow}
\usepackage{cuted}
\usepackage{txfonts}
\usepackage[pdfencoding=auto,psdextra]{hyperref}
\hypersetup{
    colorlinks=true,
    linkcolor=blue,
    filecolor=magenta,      
    urlcolor=blue,
    citecolor=blue
}
\makeatletter
\renewcommand*\aa@pageof{, page \thepage{} of \pageref*{LastPage}}

\makeatother

\usepackage[utf8]{inputenc}

\usepackage[switch, modulo]{lineno}

\usepackage{euclid}
\usepackage{orcidlink}
\usepackage{amsmath} 
\usepackage{amssymb}
\usepackage{mathtools}
\usepackage{array}
\usepackage{makecell}
\usepackage{siunitx}
\usepackage{color}
\usepackage[caption=false]{subfig}
\usepackage{tabularx}
\usepackage{booktabs} 
\usepackage{enumitem}

\begin{document}
%
%

\title{Euclid Data Release 1 (DR1)}\subtitle{Enhancing deep-learning models: the impact of image processing and training data selection for strong-lens detection}

   
\newcommand{\orcid}[1]{\,\orcidlink{#1}} 
\author{Euclid Collaboration: M.~Fogliardi\orcid{0009-0006-4964-5311}\thanks{\email{fglmhl@unife.it}}\inst{\ref{aff1},\ref{aff2}}
\and P.~Holloway\orcid{0009-0002-8896-6100}\inst{\ref{aff3}}
\and S.~Schuldt\orcid{0000-0003-2497-6334}\inst{\ref{aff4},\ref{aff5},\ref{aff6}}
\and P.~Rosati\orcid{0000-0002-6813-0632}\inst{\ref{aff1},\ref{aff2}}
\and S.~H.~Vincken\orcid{0009-0005-7305-2359}\inst{\ref{aff3},\ref{aff7},\ref{aff8}}
\and T.~Li\orcid{0009-0005-5008-0381}\inst{\ref{aff3}}
\and N.~E.~P.~Lines\orcid{0009-0004-7751-1914}\inst{\ref{aff3}}
\and M.~Meneghetti\orcid{0000-0003-1225-7084}\inst{\ref{aff2},\ref{aff9}}
\and J.~W.~Nightingale\orcid{0000-0002-8987-7401}\inst{\ref{aff10}}
\and A.~Nersesian\orcid{0000-0001-6843-409X}\inst{\ref{aff11},\ref{aff12}}
\and T.~E.~Collett\orcid{0000-0001-5564-3140}\inst{\ref{aff3}}
\and A.~Verma\orcid{0000-0002-0730-0781}\inst{\ref{aff13}}
\and L.~Bazzanini\orcid{0000-0003-0727-0137}\inst{\ref{aff1},\ref{aff2}}
\and G.~Angora\orcid{0000-0002-0316-6562}\inst{\ref{aff14},\ref{aff1}}
\and P.~Bergamini\orcid{0000-0003-1383-9414}\inst{\ref{aff2}}
\and H.~Degaudenzi\orcid{0000-0002-5887-6799}\inst{\ref{aff15}}
\and G.~Despali\orcid{0000-0001-6150-4112}\inst{\ref{aff16},\ref{aff2},\ref{aff9}}
\and R.~B.~Metcalf\orcid{0000-0003-3167-2574}\inst{\ref{aff16},\ref{aff2}}
\and K.~Rojas\orcid{0000-0003-1391-6854}\inst{\ref{aff17}}
\and P.~Casenove\orcid{0009-0006-6736-1670}\inst{\ref{aff18}}
\and L.~R.~Ecker\orcid{0009-0005-3508-2469}\inst{\ref{aff19},\ref{aff20},\ref{aff21}}
\and M.~Grespan\orcid{0000-0002-8213-0217}\inst{\ref{aff13},\ref{aff22}}
\and F.~Courbin\orcid{0000-0003-0758-6510}\inst{\ref{aff23},\ref{aff24},\ref{aff25}}
\and R.~Gavazzi\orcid{0000-0002-5540-6935}\inst{\ref{aff26},\ref{aff27}}
\and A.~Acebron\orcid{0000-0003-3108-9039}\inst{\ref{aff28},\ref{aff6}}
\and G.~Di~Rosa\orcid{0009-0001-9416-0923}\inst{\ref{aff1}}
\and B.~Altieri\orcid{0000-0003-3936-0284}\inst{\ref{aff29}}
\and S.~Andreon\orcid{0000-0002-2041-8784}\inst{\ref{aff30}}
\and C.~Baccigalupi\orcid{0000-0002-8211-1630}\inst{\ref{aff31},\ref{aff32},\ref{aff33},\ref{aff34}}
\and M.~Baldi\orcid{0000-0003-4145-1943}\inst{\ref{aff35},\ref{aff2},\ref{aff9}}
\and A.~Balestra\orcid{0000-0002-6967-261X}\inst{\ref{aff36}}
\and S.~Bardelli\orcid{0000-0002-8900-0298}\inst{\ref{aff2}}
\and P.~Battaglia\orcid{0000-0002-7337-5909}\inst{\ref{aff2}}
\and A.~Biviano\orcid{0000-0002-0857-0732}\inst{\ref{aff32},\ref{aff31}}
\and E.~Branchini\orcid{0000-0002-0808-6908}\inst{\ref{aff37},\ref{aff38},\ref{aff30}}
\and M.~Brescia\orcid{0000-0001-9506-5680}\inst{\ref{aff39},\ref{aff14}}
\and S.~Camera\orcid{0000-0003-3399-3574}\inst{\ref{aff40},\ref{aff41},\ref{aff42}}
\and V.~Capobianco\orcid{0000-0002-3309-7692}\inst{\ref{aff42}}
\and C.~Carbone\orcid{0000-0003-0125-3563}\inst{\ref{aff6}}
\and J.~Carretero\orcid{0000-0002-3130-0204}\inst{\ref{aff43},\ref{aff44}}
\and M.~Castellano\orcid{0000-0001-9875-8263}\inst{\ref{aff45}}
\and G.~Castignani\orcid{0000-0001-6831-0687}\inst{\ref{aff2}}
\and S.~Cavuoti\orcid{0000-0002-3787-4196}\inst{\ref{aff14},\ref{aff46}}
\and A.~Cimatti\inst{\ref{aff47}}
\and C.~Colodro-Conde\inst{\ref{aff48}}
\and G.~Congedo\orcid{0000-0003-2508-0046}\inst{\ref{aff49}}
\and L.~Conversi\orcid{0000-0002-6710-8476}\inst{\ref{aff50},\ref{aff29}}
\and Y.~Copin\orcid{0000-0002-5317-7518}\inst{\ref{aff51}}
\and H.~M.~Courtois\orcid{0000-0003-0509-1776}\inst{\ref{aff52}}
\and M.~Cropper\orcid{0000-0003-4571-9468}\inst{\ref{aff53}}
\and A.~Da~Silva\orcid{0000-0002-6385-1609}\inst{\ref{aff54},\ref{aff55}}
\and G.~De~Lucia\orcid{0000-0002-6220-9104}\inst{\ref{aff32}}
\and H.~Dole\orcid{0000-0002-9767-3839}\inst{\ref{aff56},\ref{aff57}}
\and F.~Dubath\orcid{0000-0002-6533-2810}\inst{\ref{aff15}}
\and X.~Dupac\inst{\ref{aff29}}
\and S.~Escoffier\orcid{0000-0002-2847-7498}\inst{\ref{aff58}}
\and M.~Fabricius\orcid{0000-0002-7025-6058}\inst{\ref{aff20},\ref{aff19}}
\and M.~Farina\orcid{0000-0002-3089-7846}\inst{\ref{aff59}}
\and R.~Farinelli\inst{\ref{aff2}}
\and F.~Faustini\orcid{0000-0001-6274-5145}\inst{\ref{aff45}}
\and S.~Ferriol\inst{\ref{aff51}}
\and F.~Finelli\orcid{0000-0002-6694-3269}\inst{\ref{aff2},\ref{aff60}}
\and M.~Frailis\orcid{0000-0002-7400-2135}\inst{\ref{aff32}}
\and E.~Franceschi\orcid{0000-0002-0585-6591}\inst{\ref{aff2}}
\and M.~Fumana\orcid{0000-0001-6787-5950}\inst{\ref{aff6}}
\and S.~Galeotta\orcid{0000-0002-3748-5115}\inst{\ref{aff32}}
\and K.~George\orcid{0000-0002-1734-8455}\inst{\ref{aff61}}
\and C.~Giocoli\orcid{0000-0002-9590-7961}\inst{\ref{aff2},\ref{aff9}}
\and P.~G\'omez-Alvarez\orcid{0000-0002-8594-5358}\inst{\ref{aff62},\ref{aff29}}
\and J.~Gracia-Carpio\orcid{0000-0003-4689-3134}\inst{\ref{aff20}}
\and A.~Grazian\orcid{0000-0002-5688-0663}\inst{\ref{aff36}}
\and F.~Grupp\inst{\ref{aff20},\ref{aff19}}
\and L.~Guzzo\orcid{0000-0001-8264-5192}\inst{\ref{aff63},\ref{aff30},\ref{aff64}}
\and W.~G.~Hartley\orcid{0000-0002-4563-5101}\inst{\ref{aff15}}
\and S.~V.~H.~Haugan\orcid{0000-0001-9648-7260}\inst{\ref{aff65}}
\and S.~Hemmati\orcid{0000-0003-2226-5395}\inst{\ref{aff66}}
\and W.~Holmes\orcid{0009-0007-8554-4646}\inst{\ref{aff67}}
\and F.~Hormuth\inst{\ref{aff68}}
\and A.~Hornstrup\orcid{0000-0002-3363-0936}\inst{\ref{aff69},\ref{aff70}}
\and K.~Jahnke\orcid{0000-0003-3804-2137}\inst{\ref{aff71}}
\and M.~Jhabvala\inst{\ref{aff72}}
\and A.~Kiessling\orcid{0000-0002-2590-1273}\inst{\ref{aff67}}
\and B.~Kubik\orcid{0009-0006-5823-4880}\inst{\ref{aff51}}
\and M.~K\"ummel\orcid{0000-0003-2791-2117}\inst{\ref{aff19}}
\and M.~Kunz\orcid{0000-0002-3052-7394}\inst{\ref{aff73}}
\and H.~Kurki-Suonio\orcid{0000-0002-4618-3063}\inst{\ref{aff5},\ref{aff74}}
\and A.~M.~C.~Le~Brun\orcid{0000-0002-0936-4594}\inst{\ref{aff75}}
\and S.~Ligori\orcid{0000-0003-4172-4606}\inst{\ref{aff42}}
\and P.~B.~Lilje\orcid{0000-0003-4324-7794}\inst{\ref{aff65}}
\and V.~Lindholm\orcid{0000-0003-2317-5471}\inst{\ref{aff5},\ref{aff74}}
\and I.~Lloro\orcid{0000-0001-5966-1434}\inst{\ref{aff76}}
\and M.~Magliocchetti\orcid{0000-0001-9158-4838}\inst{\ref{aff59}}
\and G.~Mainetti\orcid{0000-0003-2384-2377}\inst{\ref{aff77}}
\and O.~Mansutti\orcid{0000-0001-5758-4658}\inst{\ref{aff32}}
\and O.~Marggraf\orcid{0000-0001-7242-3852}\inst{\ref{aff78}}
\and M.~Martinelli\orcid{0000-0002-6943-7732}\inst{\ref{aff45},\ref{aff79}}
\and N.~Martinet\orcid{0000-0003-2786-7790}\inst{\ref{aff26}}
\and F.~Marulli\orcid{0000-0002-8850-0303}\inst{\ref{aff16},\ref{aff2},\ref{aff9}}
\and R.~J.~Massey\orcid{0000-0002-6085-3780}\inst{\ref{aff80}}
\and N.~Mauri\orcid{0000-0001-8196-1548}\inst{\ref{aff47},\ref{aff9}}
\and E.~Medinaceli\orcid{0000-0002-4040-7783}\inst{\ref{aff2}}
\and S.~Mei\orcid{0000-0002-2849-559X}\inst{\ref{aff81},\ref{aff82}}
\and E.~Merlin\orcid{0000-0001-6870-8900}\inst{\ref{aff36}}
\and G.~Meylan\orcid{0000-0001-6503-0209}\inst{\ref{aff83}}
\and P.~Monaco\orcid{0000-0003-2083-7564}\inst{\ref{aff84},\ref{aff32},\ref{aff33},\ref{aff31}}
\and M.~Moresco\orcid{0000-0002-7616-7136}\inst{\ref{aff16},\ref{aff2}}
\and C.~Moretti\orcid{0000-0003-3314-8936}\inst{\ref{aff32},\ref{aff31},\ref{aff33}}
\and L.~Moscardini\orcid{0000-0002-3473-6716}\inst{\ref{aff16},\ref{aff2},\ref{aff9}}
\and C.~Neissner\orcid{0000-0001-8524-4968}\inst{\ref{aff85},\ref{aff44}}
\and S.-M.~Niemi\orcid{0009-0005-0247-0086}\inst{\ref{aff86}}
\and C.~Padilla\orcid{0000-0001-7951-0166}\inst{\ref{aff85}}
\and S.~Paltani\orcid{0000-0002-8108-9179}\inst{\ref{aff15}}
\and F.~Pasian\orcid{0000-0002-4869-3227}\inst{\ref{aff32}}
\and W.~J.~Percival\orcid{0000-0002-0644-5727}\inst{\ref{aff87},\ref{aff88},\ref{aff89}}
\and V.~Pettorino\orcid{0000-0002-4203-9320}\inst{\ref{aff86}}
\and S.~Pires\orcid{0000-0002-0249-2104}\inst{\ref{aff90}}
\and G.~Polenta\orcid{0000-0003-4067-9196}\inst{\ref{aff91}}
\and M.~Poncet\inst{\ref{aff18}}
\and L.~A.~Popa\inst{\ref{aff92}}
\and F.~Raison\orcid{0000-0002-7819-6918}\inst{\ref{aff20}}
\and A.~Renzi\orcid{0000-0001-9856-1970}\inst{\ref{aff93},\ref{aff94},\ref{aff2}}
\and J.~Rhodes\orcid{0000-0002-4485-8549}\inst{\ref{aff67}}
\and G.~Riccio\inst{\ref{aff14}}
\and I.~Risso\orcid{0000-0003-2525-7761}\inst{\ref{aff37},\ref{aff38},\ref{aff30}}
\and F.~Rizzo\orcid{0000-0002-9407-585X}\inst{\ref{aff32}}
\and E.~Romelli\orcid{0000-0003-3069-9222}\inst{\ref{aff32}}
\and M.~Roncarelli\orcid{0000-0001-9587-7822}\inst{\ref{aff2}}
\and B.~Rusholme\orcid{0000-0001-7648-4142}\inst{\ref{aff66}}
\and R.~Saglia\orcid{0000-0003-0378-7032}\inst{\ref{aff19},\ref{aff20}}
\and Z.~Sakr\orcid{0000-0002-4823-3757}\inst{\ref{aff95},\ref{aff96},\ref{aff97}}
\and D.~Sapone\orcid{0000-0001-7089-4503}\inst{\ref{aff98}}
\and B.~Sartoris\orcid{0000-0003-1337-5269}\inst{\ref{aff99}}
\and M.~Schirmer\orcid{0000-0003-2568-9994}\inst{\ref{aff71}}
\and P.~Schneider\orcid{0000-0001-8561-2679}\inst{\ref{aff78}}
\and A.~Secroun\orcid{0000-0003-0505-3710}\inst{\ref{aff58}}
\and E.~Sihvola\orcid{0000-0003-1804-7715}\inst{\ref{aff5},\ref{aff74}}
\and C.~Sirignano\orcid{0000-0002-0995-7146}\inst{\ref{aff93},\ref{aff94}}
\and G.~Sirri\orcid{0000-0003-2626-2853}\inst{\ref{aff9}}
\and L.~Stanco\orcid{0000-0002-9706-5104}\inst{\ref{aff94}}
\and P.~Tallada-Cresp\'{i}\orcid{0000-0002-1336-8328}\inst{\ref{aff43},\ref{aff44}}
\and A.~N.~Taylor\inst{\ref{aff49}}
\and I.~Tereno\orcid{0000-0002-4537-6218}\inst{\ref{aff54},\ref{aff55}}
\and S.~Toft\orcid{0000-0003-3631-7176}\inst{\ref{aff100},\ref{aff101}}
\and R.~Toledo-Moreo\orcid{0000-0002-2997-4859}\inst{\ref{aff102},\ref{aff103}}
\and F.~Torradeflot\orcid{0000-0003-1160-1517}\inst{\ref{aff44},\ref{aff43}}
\and I.~Tutusaus\orcid{0000-0002-3199-0399}\inst{\ref{aff104},\ref{aff105},\ref{aff96}}
\and J.~Valiviita\orcid{0000-0001-6225-3693}\inst{\ref{aff5},\ref{aff74}}
\and T.~Vassallo\orcid{0000-0001-6512-6358}\inst{\ref{aff32},\ref{aff61}}
\and G.~Verdoes~Kleijn\orcid{0000-0001-5803-2580}\inst{\ref{aff106}}
\and A.~Veropalumbo\orcid{0000-0003-2387-1194}\inst{\ref{aff30},\ref{aff38},\ref{aff37}}
\and Y.~Wang\orcid{0000-0002-4749-2984}\inst{\ref{aff66}}
\and J.~Weller\orcid{0000-0002-8282-2010}\inst{\ref{aff19},\ref{aff20}}
\and A.~Zacchei\orcid{0000-0003-0396-1192}\inst{\ref{aff32},\ref{aff31}}
\and G.~Zamorani\orcid{0000-0002-2318-301X}\inst{\ref{aff2}}
\and F.~M.~Zerbi\orcid{0000-0002-9996-973X}\inst{\ref{aff30}}
\and E.~Zucca\orcid{0000-0002-5845-8132}\inst{\ref{aff2}}
\and V.~Allevato\orcid{0000-0001-7232-5152}\inst{\ref{aff14}}
\and I.~T.~Andika\orcid{0000-0001-6102-9526}\inst{\ref{aff19}}
\and M.~Ballardini\orcid{0000-0003-4481-3559}\inst{\ref{aff1},\ref{aff107},\ref{aff2}}
\and S.~Borgani\orcid{0000-0001-6151-6439}\inst{\ref{aff84},\ref{aff31},\ref{aff32},\ref{aff33},\ref{aff108}}
\and E.~Bozzo\orcid{0000-0002-8201-1525}\inst{\ref{aff15}}
\and C.~Burigana\orcid{0000-0002-3005-5796}\inst{\ref{aff109},\ref{aff60}}
\and M.~Calabrese\orcid{0000-0002-2637-2422}\inst{\ref{aff110},\ref{aff6}}
\and A.~Cappi\inst{\ref{aff99},\ref{aff2}}
\and F.~Caro\orcid{0009-0003-1053-0507}\inst{\ref{aff45}}
\and T.~Castro\orcid{0000-0002-6292-3228}\inst{\ref{aff111},\ref{aff32},\ref{aff31}}
\and D.~Di~Ferdinando\inst{\ref{aff9}}
\and J.~A.~Escartin~Vigo\inst{\ref{aff20}}
\and J.~Garc\'ia-Bellido\orcid{0000-0002-9370-8360}\inst{\ref{aff95}}
\and C.~M.~Gutierrez\orcid{0000-0001-7854-783X}\inst{\ref{aff48},\ref{aff112}}
\and A.~Hall\orcid{0000-0002-3139-8651}\inst{\ref{aff49}}
\and Y.~Kang\orcid{0009-0000-8588-7250}\inst{\ref{aff113}}
\and G.~Leroy\orcid{0009-0004-2523-4425}\inst{\ref{aff114},\ref{aff80}}
\and A.~Loureiro\orcid{0000-0002-4371-0876}\inst{\ref{aff115},\ref{aff116}}
\and J.~F.~Macias-Perez\orcid{0000-0002-5385-2763}\inst{\ref{aff117}}
\and R.~Maoli\orcid{0000-0002-6065-3025}\inst{\ref{aff118},\ref{aff45}}
\and M.~P\"ontinen\orcid{0000-0001-5442-2530}\inst{\ref{aff5}}
\and E.~Sarpa\orcid{0000-0002-1256-655X}\inst{\ref{aff32}}
\and V.~Scottez\orcid{0009-0008-3864-940X}\inst{\ref{aff119},\ref{aff120}}
\and M.~Sereno\orcid{0000-0003-0302-0325}\inst{\ref{aff2},\ref{aff9}}
\and K.~Tanidis\orcid{0000-0001-9843-5130}\inst{\ref{aff121}}
\and M.~Tenti\orcid{0000-0002-4254-5901}\inst{\ref{aff9}}
\and A.~Troja\orcid{0000-0003-0239-4595}\inst{\ref{aff32}}
\and M.~Tucci\inst{\ref{aff15}}
\and M.~Viel\orcid{0000-0002-2642-5707}\inst{\ref{aff31},\ref{aff32},\ref{aff34},\ref{aff33},\ref{aff108}}
\and M.~Wiesmann\orcid{0009-0000-8199-5860}\inst{\ref{aff65}}
\and Abdurro'uf\orcid{0000-0002-5258-8761}\inst{\ref{aff122}}
\and J.~A.~Acevedo~Barroso\orcid{0000-0002-9654-1711}\inst{\ref{aff67}}
\and Y.~Akrami\orcid{0000-0002-2407-7956}\inst{\ref{aff95},\ref{aff123}}
\and M.~Archidiacono\orcid{0000-0003-4952-9012}\inst{\ref{aff63},\ref{aff64}}
\and G.~Aric\`o\orcid{0000-0002-2802-2928}\inst{\ref{aff9}}
\and F.~Atrio-Barandela\orcid{0000-0002-2130-2513}\inst{\ref{aff124}}
\and M.~Aubert\orcid{0009-0002-7667-8814}\inst{\ref{aff51}}
\and M.~Baes\orcid{0000-0002-3930-2757}\inst{\ref{aff12}}
\and N.~Bartolo\inst{\ref{aff93},\ref{aff94},\ref{aff36}}
\and J.~R.~Bermejo-Climent\inst{\ref{aff125},\ref{aff126}}
\and D.~Bertacca\orcid{0000-0002-2490-7139}\inst{\ref{aff93},\ref{aff36},\ref{aff94}}
\and M.~Bethermin\orcid{0000-0002-3915-2015}\inst{\ref{aff127}}
\and F.~Beutler\orcid{0000-0003-0467-5438}\inst{\ref{aff49}}
\and A.~Blanchard\orcid{0000-0001-8555-9003}\inst{\ref{aff96}}
\and H.~B\"ohringer\orcid{0000-0001-8241-4204}\inst{\ref{aff20},\ref{aff61},\ref{aff128}}
\and F.~Bouch\`e\orcid{0000-0002-4663-1786}\inst{\ref{aff129},\ref{aff46}}
\and M.~L.~Brown\orcid{0000-0002-0370-8077}\inst{\ref{aff130}}
\and S.~Bruton\orcid{0000-0002-6503-5218}\inst{\ref{aff131}}
\and A.~Calabro\orcid{0000-0003-2536-1614}\inst{\ref{aff45}}
\and C.~S.~Carvalho\inst{\ref{aff132}}
\and N.~Chartab\orcid{0000-0003-3691-937X}\inst{\ref{aff66}}
\and F.~Cogato\orcid{0000-0003-4632-6113}\inst{\ref{aff2}}
\and P.~Corcho-Caballero\orcid{0000-0001-6327-7080}\inst{\ref{aff106}}
\and M.~Correnti\orcid{0000-0001-6464-3257}\inst{\ref{aff45},\ref{aff91}}
\and B.~Csizi\orcid{0000-0003-3227-6581}\inst{\ref{aff113}}
\and O.~Cucciati\orcid{0000-0002-9336-7551}\inst{\ref{aff2}}
\and T.~de~Boer\orcid{0000-0001-5486-2747}\inst{\ref{aff133}}
\and F.~De~Paolis\orcid{0000-0001-6460-7563}\inst{\ref{aff134},\ref{aff135},\ref{aff136}}
\and G.~Desprez\orcid{0000-0001-8325-1742}\inst{\ref{aff106}}
\and A.~D\'iaz-S\'anchez\orcid{0000-0003-0748-4768}\inst{\ref{aff137}}
\and S.~Di~Domizio\orcid{0000-0003-2863-5895}\inst{\ref{aff37},\ref{aff38}}
\and J.~M.~Diego\orcid{0000-0001-9065-3926}\inst{\ref{aff28}}
\and P.-A.~Duc\orcid{0000-0003-3343-6284}\inst{\ref{aff127}}
\and V.~Duret\orcid{0009-0009-0383-4960}\inst{\ref{aff138}}
\and A.~Enia\orcid{0000-0002-0200-2857}\inst{\ref{aff139},\ref{aff2}}
\and Y.~Fang\orcid{0000-0002-0334-6950}\inst{\ref{aff19}}
\and A.~Farina\orcid{0009-0000-3420-929X}\inst{\ref{aff30},\ref{aff38}}
\and A.~Finoguenov\orcid{0000-0002-4606-5403}\inst{\ref{aff5}}
\and A.~Franco\orcid{0000-0002-4761-366X}\inst{\ref{aff134},\ref{aff135},\ref{aff136}}
\and N.~Frusciante\orcid{0000-0002-7375-1230}\inst{\ref{aff140},\ref{aff141}}
\and A.~Fumagalli\orcid{0009-0004-0300-2535}\inst{\ref{aff32}}
\and Y.~Fu\orcid{0000-0002-0759-0504}\inst{\ref{aff142},\ref{aff106}}
\and K.~Ganga\orcid{0000-0001-8159-8208}\inst{\ref{aff81}}
\and E.~Gaztanaga\orcid{0000-0001-9632-0815}\inst{\ref{aff104},\ref{aff105},\ref{aff3}}
\and Z.~Ghaffari\orcid{0000-0002-6467-8078}\inst{\ref{aff32},\ref{aff31}}
\and F.~Giacomini\orcid{0000-0002-3129-2814}\inst{\ref{aff9}}
\and F.~Gianotti\orcid{0000-0003-4666-119X}\inst{\ref{aff2}}
\and G.~Gozaliasl\orcid{0000-0002-0236-919X}\inst{\ref{aff143},\ref{aff5}}
\and A.~Gruppuso\orcid{0000-0001-9272-5292}\inst{\ref{aff2},\ref{aff9}}
\and M.~Guidi\orcid{0000-0001-9408-1101}\inst{\ref{aff35},\ref{aff2}}
\and C.~Hern\'andez-Monteagudo\orcid{0000-0001-5471-9166}\inst{\ref{aff112},\ref{aff48}}
\and J.~Hjorth\orcid{0000-0002-4571-2306}\inst{\ref{aff144}}
\and M.~Jauzac\orcid{0000-0003-1974-8732}\inst{\ref{aff114},\ref{aff80},\ref{aff145},\ref{aff146},\ref{aff96}}
\and J.~J.~E.~Kajava\orcid{0000-0002-3010-8333}\inst{\ref{aff147},\ref{aff4},\ref{aff148}}
\and D.~Karagiannis\orcid{0000-0002-4927-0816}\inst{\ref{aff149},\ref{aff1},\ref{aff150}}
\and J.~Kim\orcid{0000-0003-2776-2761}\inst{\ref{aff13}}
\and C.~C.~Kirkpatrick\inst{\ref{aff151}}
\and A.~Kov\'acs\orcid{0000-0002-5825-579X}\inst{\ref{aff126},\ref{aff125}}
\and I.~Kova{\v{c}}i{\'{c}}\orcid{0000-0001-6751-3263}\inst{\ref{aff12}}
\and K.~Koyama\orcid{0000-0001-6727-6915}\inst{\ref{aff3}}
\and S.~Kruk\orcid{0000-0001-8010-8879}\inst{\ref{aff29}}
\and M.~C.~Lam\orcid{0000-0002-9347-2298}\inst{\ref{aff49}}
\and S.~Lee\orcid{0000-0002-8289-740X}\inst{\ref{aff152},\ref{aff67}}
\and M.~D.~Lepinzan\orcid{0000-0003-1287-9801}\inst{\ref{aff32}}
\and G.~F.~Lesci\orcid{0000-0002-4607-2830}\inst{\ref{aff109}}
\and J.~Lesgourgues\orcid{0000-0001-7627-353X}\inst{\ref{aff153}}
\and T.~I.~Liaudat\orcid{0000-0002-9104-314X}\inst{\ref{aff154}}
\and L.~Linke\orcid{0000-0002-2622-8113}\inst{\ref{aff113}}
\and E.~A.~Magnier\orcid{0000-0002-7965-2815}\inst{\ref{aff133}}
\and A.~Manj\'on-Garc\'ia\orcid{0000-0002-7413-8825}\inst{\ref{aff137}}
\and F.~Mannucci\orcid{0000-0002-4803-2381}\inst{\ref{aff155}}
\and C.~J.~A.~P.~Martins\orcid{0000-0002-4886-9261}\inst{\ref{aff156},\ref{aff157}}
\and M.~Miluzio\inst{\ref{aff29},\ref{aff158}}
\and G.~Morgante\inst{\ref{aff2}}
\and K.~Naidoo\orcid{0000-0002-9182-1802}\inst{\ref{aff3},\ref{aff71}}
\and P.~Natoli\orcid{0000-0003-0126-9100}\inst{\ref{aff1},\ref{aff107}}
\and A.~Navarro-Alsina\orcid{0000-0002-3173-2592}\inst{\ref{aff78}}
\and S.~Nesseris\orcid{0000-0002-0567-0324}\inst{\ref{aff95}}
\and L.~Nicastro\orcid{0000-0001-8534-6788}\inst{\ref{aff2}}
\and F.~Oppizzi\orcid{0000-0003-3904-8370}\inst{\ref{aff38},\ref{aff94},\ref{aff93}}
\and F.~Pace\orcid{0000-0001-8039-0480}\inst{\ref{aff40},\ref{aff41},\ref{aff42},\ref{aff55}}
\and D.~Paoletti\orcid{0000-0003-4761-6147}\inst{\ref{aff2},\ref{aff60}}
\and K.~Paterson\orcid{0000-0001-8340-3486}\inst{\ref{aff71}}
\and C.~Pattison\orcid{0000-0003-3272-2617}\inst{\ref{aff3}}
\and A.~Pisani\orcid{0000-0002-6146-4437}\inst{\ref{aff58}}
\and L.~Pizzuti\orcid{0000-0001-5654-7580}\inst{\ref{aff159}}
\and J.~J.~Popp\orcid{0000-0002-3724-1727}\inst{\ref{aff160}}
\and D.~Potter\orcid{0000-0002-0757-5195}\inst{\ref{aff161}}
\and G.~W.~Pratt\inst{\ref{aff90}}
\and S.~Quai\orcid{0000-0002-0449-8163}\inst{\ref{aff16},\ref{aff2}}
\and M.~Radovich\orcid{0000-0002-3585-866X}\inst{\ref{aff36}}
\and G.~Rodighiero\orcid{0000-0002-9415-2296}\inst{\ref{aff93},\ref{aff36}}
\and W.~Roster\orcid{0000-0002-9149-6528}\inst{\ref{aff20}}
\and M.~Sahl\'en\orcid{0000-0003-0973-4804}\inst{\ref{aff162}}
\and D.~B.~Sanders\orcid{0000-0002-1233-9998}\inst{\ref{aff133}}
\and A.~Schneider\orcid{0000-0001-7055-8104}\inst{\ref{aff161}}
\and E.~Sellentin\orcid{0009-0002-2655-3458}\inst{\ref{aff163},\ref{aff142}}
\and S.~Serjeant\orcid{0000-0002-0517-7943}\inst{\ref{aff160}}
\and L.~C.~Smith\orcid{0000-0002-3259-2771}\inst{\ref{aff164}}
\and I.~Szapudi\orcid{0000-0003-2274-0301}\inst{\ref{aff133}}
\and M.~Talia\orcid{0000-0003-4352-2063}\inst{\ref{aff16},\ref{aff2}}
\and F.~Tarsitano\orcid{0000-0002-5919-0238}\inst{\ref{aff165},\ref{aff166}}
\and G.~Testera\orcid{0000-0003-2970-766X}\inst{\ref{aff38}}
\and R.~Teyssier\orcid{0000-0001-7689-0933}\inst{\ref{aff167}}
\and S.~Tosi\orcid{0000-0002-7275-9193}\inst{\ref{aff37},\ref{aff30},\ref{aff38}}
\and C.~Uhlemann\orcid{0000-0001-7831-1579}\inst{\ref{aff168}}
\and C.~Valieri\inst{\ref{aff9}}
\and D.~Vergani\orcid{0000-0003-0898-2216}\inst{\ref{aff2}}
\and G.~Vernardos\orcid{0000-0001-8554-7248}\inst{\ref{aff169},\ref{aff170}}
\and S.~Vinciguerra\orcid{0009-0005-4018-3184}\inst{\ref{aff26}}
\and M.~von~Wietersheim-Kramsta\orcid{0000-0003-4986-5091}\inst{\ref{aff80},\ref{aff114}}
\and N.~A.~Walton\orcid{0000-0003-3983-8778}\inst{\ref{aff164}}
\and A.~H.~Wright\orcid{0000-0001-7363-7932}\inst{\ref{aff171}}}
										   
\institute{Dipartimento di Fisica e Scienze della Terra, Universit\`a degli Studi di Ferrara, Via Giuseppe Saragat 1, 44122 Ferrara, Italy\label{aff1}
\and
INAF-Osservatorio di Astrofisica e Scienza dello Spazio di Bologna, Via Piero Gobetti 93/3, 40129 Bologna, Italy\label{aff2}
\and
Institute of Cosmology and Gravitation, University of Portsmouth, Portsmouth PO1 3FX, UK\label{aff3}
\and
Finnish Centre for Astronomy with ESO (FINCA), Quantum, Vesilinnantie 5, University of Turku, 20014 Turku, Finland\label{aff4}
\and
Department of Physics, P.O. Box 64, University of Helsinki, 00014 Helsinki, Finland\label{aff5}
\and
INAF-IASF Milano, Via Alfonso Corti 12, 20133 Milano, Italy\label{aff6}
\and
Fontys University of Applied Sciences, School of ICT, Rachelsmolen 1, 5612 MA Eindhoven, The Netherlands\label{aff7}
\and
University of Applied Sciences and Arts of Northwestern Switzerland, School of Engineering, 5210 Windisch, Switzerland\label{aff8}
\and
INFN-Sezione di Bologna, Viale Berti Pichat 6/2, 40127 Bologna, Italy\label{aff9}
\and
School of Mathematics, Statistics and Physics, Newcastle University, Herschel Building, Newcastle-upon-Tyne, NE1 7RU, UK\label{aff10}
\and
STAR Institute, University of Li{\`e}ge, Quartier Agora, All\'ee du six Ao\^ut 19c, 4000 Li\`ege, Belgium\label{aff11}
\and
Universiteit Gent, Department of Physics and Astronomy, Proeftuinstraat 86 N3, 9000 Ghent, Belgium
\label{aff12}
\and
Department of Physics, University of Oxford, Keble Road, Oxford OX1 3RH, UK\label{aff13}
\and
INAF-Osservatorio Astronomico di Capodimonte, Via Moiariello 16, 80131 Napoli, Italy\label{aff14}
\and
Department of Astronomy, University of Geneva, ch. d'Ecogia 16, 1290 Versoix, Switzerland\label{aff15}
\and
Dipartimento di Fisica e Astronomia "Augusto Righi" - Alma Mater Studiorum Universit\`a di Bologna, via Piero Gobetti 93/2, 40129 Bologna, Italy\label{aff16}
\and
University of Applied Sciences and Arts of Northwestern Switzerland, School of Computer Science, 5210 Windisch, Switzerland\label{aff17}
\and
Centre National d'Etudes Spatiales -- Centre spatial de Toulouse, 18 avenue Edouard Belin, 31401 Toulouse Cedex 9, France\label{aff18}
\and
Universit\"ats-Sternwarte M\"unchen, Fakult\"at f\"ur Physik, Ludwig-Maximilians-Universit\"at M\"unchen, Scheinerstr.~1, 81679 M\"unchen, Germany\label{aff19}
\and
Max Planck Institute for Extraterrestrial Physics, Giessenbachstr. 1, 85748 Garching, Germany\label{aff20}
\and
Max-Planck-Institut f\"ur Astrophysik, Karl-Schwarzschild-Str.~1, 85748 Garching, Germany\label{aff21}
\and
National Centre for Nuclear Research, ul. Pasteura 7, 02-093, Warsaw, Poland\label{aff22}
\and
Institut de Ci\`{e}ncies del Cosmos (ICCUB), Universitat de Barcelona (IEEC-UB), Mart\'{i} i Franqu\`{e}s 1, 08028 Barcelona, Spain\label{aff23}
\and
Instituci\'o Catalana de Recerca i Estudis Avan\c{c}ats (ICREA), Passeig de Llu\'{\i}s Companys 23, 08010 Barcelona, Spain\label{aff24}
\and
Institut de Ciencies de l'Espai (IEEC-CSIC), Campus UAB, Carrer de Can Magrans, s/n Cerdanyola del Vall\'es, 08193 Barcelona, Spain\label{aff25}
\and
Aix-Marseille Universit\'e, CNRS, CNES, LAM, Marseille, France\label{aff26}
\and
Institut d'Astrophysique de Paris, UMR 7095, CNRS, and Sorbonne Universit\'e, 98 bis boulevard Arago, 75014 Paris, France\label{aff27}
\and
Instituto de F\'isica de Cantabria, Edificio Juan Jord\'a, Avenida de los Castros, 39005 Santander, Spain\label{aff28}
\and
ESAC/ESA, Camino Bajo del Castillo, s/n., Urb. Villafranca del Castillo, 28692 Villanueva de la Ca\~nada, Madrid, Spain\label{aff29}
\and
INAF-Osservatorio Astronomico di Brera, Via Brera 28, 20122 Milano, Italy\label{aff30}
\and
IFPU, Institute for Fundamental Physics of the Universe, via Beirut 2, 34151 Trieste, Italy\label{aff31}
\and
INAF-Osservatorio Astronomico di Trieste, Via G. B. Tiepolo 11, 34143 Trieste, Italy\label{aff32}
\and
INFN, Sezione di Trieste, Via Valerio 2, 34127 Trieste TS, Italy\label{aff33}
\and
SISSA, International School for Advanced Studies, Via Bonomea 265, 34136 Trieste TS, Italy\label{aff34}
\and
Dipartimento di Fisica e Astronomia, Universit\`a di Bologna, Via Gobetti 93/2, 40129 Bologna, Italy\label{aff35}
\and
INAF-Osservatorio Astronomico di Padova, Via dell'Osservatorio 5, 35122 Padova, Italy\label{aff36}
\and
Dipartimento di Fisica, Universit\`a di Genova, Via Dodecaneso 33, 16146, Genova, Italy\label{aff37}
\and
INFN-Sezione di Genova, Via Dodecaneso 33, 16146, Genova, Italy\label{aff38}
\and
Department of Physics "E. Pancini", University Federico II, Via Cinthia 6, 80126, Napoli, Italy\label{aff39}
\and
Dipartimento di Fisica, Universit\`a degli Studi di Torino, Via P. Giuria 1, 10125 Torino, Italy\label{aff40}
\and
INFN-Sezione di Torino, Via P. Giuria 1, 10125 Torino, Italy\label{aff41}
\and
INAF-Osservatorio Astrofisico di Torino, Via Osservatorio 20, 10025 Pino Torinese (TO), Italy\label{aff42}
\and
Centro de Investigaciones Energ\'eticas, Medioambientales y Tecnol\'ogicas (CIEMAT), Avenida Complutense 40, 28040 Madrid, Spain\label{aff43}
\and
Port d'Informaci\'{o} Cient\'{i}fica, Campus UAB, C. Albareda s/n, 08193 Bellaterra (Barcelona), Spain\label{aff44}
\and
INAF-Osservatorio Astronomico di Roma, Via Frascati 33, 00078 Monteporzio Catone, Italy\label{aff45}
\and
INFN -- Sezione di Napoli, Via Cinthia 6, 80126, Napoli, Italy\label{aff46}
\and
Dipartimento di Fisica e Astronomia "Augusto Righi" - Alma Mater Studiorum Universit\`a di Bologna, Viale Berti Pichat 6/2, 40127 Bologna, Italy\label{aff47}
\and
Instituto de Astrof\'{\i}sica de Canarias, E-38205 La Laguna, Tenerife, Spain\label{aff48}
\and
Institute for Astronomy, University of Edinburgh, Royal Observatory, Blackford Hill, Edinburgh EH9 3HJ, UK\label{aff49}
\and
European Space Agency/ESRIN, Largo Galileo Galilei 1, 00044 Frascati, Roma, Italy\label{aff50}
\and
Universit\'e Lyon 1, CNRS/IN2P3, IP2I Lyon, Villeurbanne, France\label{aff51}
\and
Universit\'e Lyon 1, CNRS/IN2P3, IUF, IP2I Lyon, Villeurbanne, France \label{aff52}
\and
Mullard Space Science Laboratory, University College London, Holmbury St Mary, Dorking, Surrey RH5 6NT, UK\label{aff53}
\and
Departamento de F\'isica, Faculdade de Ci\^encias, Universidade de Lisboa, Edif\'icio C8, Campo Grande, PT1749-016 Lisboa, Portugal\label{aff54}
\and
Instituto de Astrof\'isica e Ci\^encias do Espa\c{c}o, Faculdade de Ci\^encias, Universidade de Lisboa, Campo Grande, 1749-016 Lisboa, Portugal\label{aff55}
\and
Universit\'e Paris-Saclay, CNRS, Institut d'astrophysique spatiale, 91405, Orsay, France\label{aff56}
\and
Institut universitaire de France (IUF), 1 rue Descartes, 75231 PARIS CEDEX 05, France\label{aff57}
\and
Aix-Marseille Universit\'e, CNRS/IN2P3, CPPM, Marseille, France\label{aff58}
\and
INAF-Istituto di Astrofisica e Planetologia Spaziali, via del Fosso del Cavaliere, 100, 00100 Roma, Italy\label{aff59}
\and
INFN-Bologna, Via Irnerio 46, 40126 Bologna, Italy\label{aff60}
\and
University Observatory, LMU Faculty of Physics, Scheinerstr.~1, 81679 Munich, Germany\label{aff61}
\and
FRACTAL S.L.N.E., calle Tulip\'an 2, Portal 13 1A, 28231, Las Rozas de Madrid, Spain\label{aff62}
\and
Dipartimento di Fisica "Aldo Pontremoli", Universit\`a degli Studi di Milano, Via Celoria 16, 20133 Milano, Italy\label{aff63}
\and
INFN-Sezione di Milano, Via Celoria 16, 20133 Milano, Italy\label{aff64}
\and
Institute of Theoretical Astrophysics, University of Oslo, P.O. Box 1029 Blindern, 0315 Oslo, Norway\label{aff65}
\and
Caltech/IPAC, 1200 E. California Blvd., Pasadena, CA 91125, USA\label{aff66}
\and
Jet Propulsion Laboratory, California Institute of Technology, 4800 Oak Grove Drive, Pasadena, CA, 91109, USA\label{aff67}
\and
Felix Hormuth Engineering, Goethestr. 17, 69181 Leimen, Germany\label{aff68}
\and
Technical University of Denmark, Elektrovej 327, 2800 Kgs. Lyngby, Denmark\label{aff69}
\and
Cosmic Dawn Center (DAWN), Denmark\label{aff70}
\and
Max-Planck-Institut f\"ur Astronomie, K\"onigstuhl 17, 69117 Heidelberg, Germany\label{aff71}
\and
NASA Goddard Space Flight Center, Greenbelt, MD 20771, USA\label{aff72}
\and
Universit\'e de Gen\`eve, D\'epartement de Physique Th\'eorique and Centre for Astroparticle Physics, 24 quai Ernest-Ansermet, CH-1211 Gen\`eve 4, Switzerland\label{aff73}
\and
Helsinki Institute of Physics, Gustaf H{\"a}llstr{\"o}min katu 2, University of Helsinki, 00014 Helsinki, Finland\label{aff74}
\and
Laboratoire d'etude de l'Univers et des phenomenes eXtremes, Observatoire de Paris, Universit\'e PSL, Sorbonne Universit\'e, CNRS, 92190 Meudon, France\label{aff75}
\and
SKAO, Jodrell Bank, Lower Withington, Macclesfield SK11 9FT, UK\label{aff76}
\and
Centre de Calcul de l'IN2P3/CNRS, 21 avenue Pierre de Coubertin 69627 Villeurbanne Cedex, France\label{aff77}
\and
Universit\"at Bonn, Argelander-Institut f\"ur Astronomie, Auf dem H\"ugel 71, 53121 Bonn, Germany\label{aff78}
\and
INFN-Sezione di Roma, Piazzale Aldo Moro, 2 - c/o Dipartimento di Fisica, Edificio G. Marconi, 00185 Roma, Italy\label{aff79}
\and
Department of Physics, Institute for Computational Cosmology, Durham University, South Road, Durham, DH1 3LE, UK\label{aff80}
\and
Universit\'e Paris Cit\'e, CNRS, Astroparticule et Cosmologie, 75013 Paris, France\label{aff81}
\and
CNRS-UCB International Research Laboratory, Centre Pierre Bin\'etruy, IRL2007, CPB-IN2P3, Berkeley, USA\label{aff82}
\and
Institute of Physics, Laboratory of Astrophysics, Ecole Polytechnique F\'ed\'erale de Lausanne (EPFL), Observatoire de Sauverny, 1290 Versoix, Switzerland\label{aff83}
\and
Dipartimento di Fisica - Sezione di Astronomia, Universit\`a di Trieste, Via Tiepolo 11, 34131 Trieste, Italy\label{aff84}
\and
Institut de F\'{i}sica d'Altes Energies (IFAE), The Barcelona Institute of Science and Technology, Campus UAB, 08193 Bellaterra (Barcelona), Spain\label{aff85}
\and
European Space Agency/ESTEC, Keplerlaan 1, 2201 AZ Noordwijk, The Netherlands\label{aff86}
\and
Waterloo Centre for Astrophysics, University of Waterloo, Waterloo, Ontario N2L 3G1, Canada\label{aff87}
\and
Department of Physics and Astronomy, University of Waterloo, Waterloo, Ontario N2L 3G1, Canada\label{aff88}
\and
Perimeter Institute for Theoretical Physics, Waterloo, Ontario N2L 2Y5, Canada\label{aff89}
\and
Universit\'e Paris-Saclay, Universit\'e Paris Cit\'e, CEA, CNRS, AIM, 91191, Gif-sur-Yvette, France\label{aff90}
\and
Space Science Data Center, Italian Space Agency, via del Politecnico snc, 00133 Roma, Italy\label{aff91}
\and
Institute of Space Science, Str. Atomistilor, nr. 409 M\u{a}gurele, Ilfov, 077125, Romania\label{aff92}
\and
Dipartimento di Fisica e Astronomia "G. Galilei", Universit\`a di Padova, Via Marzolo 8, 35131 Padova, Italy\label{aff93}
\and
INFN-Padova, Via Marzolo 8, 35131 Padova, Italy\label{aff94}
\and
Instituto de F\'isica Te\'orica UAM-CSIC, Campus de Cantoblanco, 28049 Madrid, Spain\label{aff95}
\and
Institut de Recherche en Astrophysique et Plan\'etologie (IRAP), Universit\'e de Toulouse, CNRS, UPS, CNES, 14 Av. Edouard Belin, 31400 Toulouse, France\label{aff96}
\and
Universit\'e St Joseph; Faculty of Sciences, Beirut, Lebanon\label{aff97}
\and
Departamento de F\'isica, FCFM, Universidad de Chile, Blanco Encalada 2008, Santiago, Chile\label{aff98}
\and
Universit\'e C\^{o}te d'Azur, Observatoire de la C\^{o}te d'Azur, CNRS, Laboratoire Lagrange, Bd de l'Observatoire, CS 34229, 06304 Nice cedex 4, France\label{aff99}
\and
Cosmic Dawn Center (DAWN)\label{aff100}
\and
Niels Bohr Institute, University of Copenhagen, Jagtvej 128, 2200 Copenhagen, Denmark\label{aff101}
\and
Universidad Polit\'ecnica de Cartagena, Departamento de Electr\'onica y Tecnolog\'ia de Computadoras,  Plaza del Hospital 1, 30202 Cartagena, Spain\label{aff102}
\and
European University of Technology EUt+, European Union\label{aff103}
\and
Institute of Space Sciences (ICE, CSIC), Campus UAB, Carrer de Can Magrans, s/n, 08193 Barcelona, Spain\label{aff104}
\and
Institut d'Estudis Espacials de Catalunya (IEEC),  Edifici RDIT, Campus UPC, 08860 Castelldefels, Barcelona, Spain\label{aff105}
\and
Kapteyn Astronomical Institute, University of Groningen, PO Box 800, 9700 AV Groningen, The Netherlands\label{aff106}
\and
Istituto Nazionale di Fisica Nucleare, Sezione di Ferrara, Via Giuseppe Saragat 1, 44122 Ferrara, Italy\label{aff107}
\and
ICSC - Centro Nazionale di Ricerca in High Performance Computing, Big Data e Quantum Computing, Via Magnanelli 2, Bologna, Italy\label{aff108}
\and
INAF, Istituto di Radioastronomia, Via Piero Gobetti 101, 40129 Bologna, Italy\label{aff109}
\and
Astronomical Observatory of the Autonomous Region of the Aosta Valley (OAVdA), Loc. Lignan 39, I-11020, Nus (Aosta Valley), Italy\label{aff110}
\and
Department of Mathematical Physics, Institute of Physics, University of S\~ao Paulo, R. do Mat\~ao 1371, 05508-090, S\~ao Paulo, SP, Brazil\label{aff111}
\and
Universidad de La Laguna, Dpto. Astrof\'\i sica, E-38206 La Laguna, Tenerife, Spain\label{aff112}
\and
Universit\"at Innsbruck, Institut f\"ur Astro- und Teilchenphysik, Technikerstr. 25/8, 6020 Innsbruck, Austria\label{aff113}
\and
Department of Physics, Centre for Extragalactic Astronomy, Durham University, South Road, Durham, DH1 3LE, UK\label{aff114}
\and
Oskar Klein Centre for Cosmoparticle Physics, Department of Physics, Stockholm University, Stockholm, SE-106 91, Sweden\label{aff115}
\and
Astrophysics Group, Blackett Laboratory, Imperial College London, London SW7 2AZ, UK\label{aff116}
\and
Univ. Grenoble Alpes, CNRS, Grenoble INP, LPSC-IN2P3, 53, Avenue des Martyrs, 38000 Grenoble, France\label{aff117}
\and
Dipartimento di Fisica, Sapienza Universit\`a di Roma, Piazzale Aldo Moro 2, 00185 Roma, Italy\label{aff118}
\and
Institut d'Astrophysique de Paris, 98bis Boulevard Arago, 75014, Paris, France\label{aff119}
\and
ICL, Junia, Universit\'e Catholique de Lille, LITL, 59000 Lille, France\label{aff120}
\and
Center for Astrophysics and Cosmology, University of Nova Gorica, Nova Gorica, Slovenia\label{aff121}
\and
Department of Astronomy, Indiana University, 727 East Third Street, Bloomington, IN 47405, USA\label{aff122}
\and
CERCA/ISO, Department of Physics, Case Western Reserve University, 10900 Euclid Avenue, Cleveland, OH 44106, USA\label{aff123}
\and
Departamento de F{\'\i}sica Fundamental. Universidad de Salamanca. Plaza de la Merced s/n. 37008 Salamanca, Spain\label{aff124}
\and
Konkoly Observatory, HUN-REN CSFK, MTA Centre of Excellence, Budapest, Konkoly Thege Mikl\'os {\'u}t 15-17. H-1121, Hungary\label{aff125}
\and
MTA-CSFK Lend\"ulet Large-Scale Structure Research Group, Konkoly-Thege Mikl\'os \'ut 15-17, H-1121 Budapest, Hungary\label{aff126}
\and
Universit\'e de Strasbourg, CNRS, Observatoire astronomique de Strasbourg, UMR 7550, 67000 Strasbourg, France\label{aff127}
\and
Max-Planck-Institut f\"ur Physik, Boltzmannstr. 8, 85748 Garching, Germany\label{aff128}
\and
Scuola Superiore Meridionale, Via Mezzocannone 4, 80138, Napoli, Italy\label{aff129}
\and
Jodrell Bank Centre for Astrophysics, Department of Physics and Astronomy, University of Manchester, Oxford Road, Manchester M13 9PL, UK\label{aff130}
\and
California Institute of Technology, 1200 E California Blvd, Pasadena, CA 91125, USA\label{aff131}
\and
Instituto de Astrof\'isica e Ci\^encias do Espa\c{c}o, Faculdade de Ci\^encias, Universidade de Lisboa, Tapada da Ajuda, 1349-018 Lisboa, Portugal\label{aff132}
\and
Institute for Astronomy, University of Hawaii, 2680 Woodlawn Drive, Honolulu, HI 96822, USA\label{aff133}
\and
Department of Mathematics and Physics E. De Giorgi, University of Salento, Via per Arnesano, CP-I93, 73100, Lecce, Italy\label{aff134}
\and
INFN, Sezione di Lecce, Via per Arnesano, CP-193, 73100, Lecce, Italy\label{aff135}
\and
INAF-Sezione di Lecce, c/o Dipartimento Matematica e Fisica, Via per Arnesano, 73100, Lecce, Italy\label{aff136}
\and
Departamento F\'isica Aplicada, Universidad Polit\'ecnica de Cartagena, Campus Muralla del Mar, 30202 Cartagena, Murcia, Spain\label{aff137}
\and
Univ. Grenoble Alpes, CNRS, IPAG, 38000 Grenoble, France\label{aff138}
\and
CEA Saclay, DFR/IRFU, Service d'Astrophysique, Bat. 709, 91191 Gif-sur-Yvette, France\label{aff139}
\and
Dipartimento di Fisica "E. Pancini", Universita degli Studi di Napoli Federico II, Via Cinthia 6, 80126, Napoli, Italy\label{aff140}
\and
INFN Sezione di Napoli, Universita degli Studi di Napoli "Federico II", Compl. Univ. di Monte S. Angelo, Edificio G, Via Cinthia, I-80126, Napoli, Italy\label{aff141}
\and
Leiden Observatory, Leiden University, Einsteinweg 55, 2333 CC Leiden, The Netherlands\label{aff142}
\and
Department of Computer Science, Aalto University, PO Box 15400, Espoo, FI-00 076, Finland\label{aff143}
\and
DARK, Niels Bohr Institute, University of Copenhagen, Jagtvej 155, 2200 Copenhagen, Denmark\label{aff144}
\and
Astrophysics Research Centre, University of KwaZulu-Natal, Westville Campus, Durban 4041, South Africa\label{aff145}
\and
School of Mathematics, Statistics \& Computer Science, University of KwaZulu-Natal, Westville Campus, Durban 4041, South Africa\label{aff146}
\and
Department of Physics and Astronomy, Vesilinnantie 5, University of Turku, 20014 Turku, Finland\label{aff147}
\and
Serco for European Space Agency (ESA), Camino bajo del Castillo, s/n, Urbanizacion Villafranca del Castillo, Villanueva de la Ca\~nada, 28692 Madrid, Spain\label{aff148}
\and
Van Swinderen Institute, University of Groningen, Nijenborgh 3, 9747 AG Groningen, The Netherlands\label{aff149}
\and
Department of Physics and Astronomy, University of the Western Cape, Bellville, Cape Town, 7535, South Africa\label{aff150}
\and
Department of Physics and Helsinki Institute of Physics, Gustaf H\"allstr\"omin katu 2, University of Helsinki, 00014 Helsinki, Finland\label{aff151}
\and
Ohio University, Physics \& Astronomy Department,1 Ohio University, Athens, OH 45701, USA\label{aff152}
\and
Institute for Theoretical Particle Physics and Cosmology (TTK), RWTH Aachen University, 52056 Aachen, Germany\label{aff153}
\and
IRFU, CEA, Universit\'e Paris-Saclay 91191 Gif-sur-Yvette Cedex, France\label{aff154}
\and
INAF-Osservatorio Astrofisico di Arcetri, Largo E. Fermi 5, 50125, Firenze, Italy\label{aff155}
\and
Centro de Astrof\'{\i}sica da Universidade do Porto, Rua das Estrelas, 4150-762 Porto, Portugal\label{aff156}
\and
Instituto de Astrof\'isica e Ci\^encias do Espa\c{c}o, Universidade do Porto, CAUP, Rua das Estrelas, PT4150-762 Porto, Portugal\label{aff157}
\and
HE Space for European Space Agency (ESA), Camino bajo del Castillo, s/n, Urbanizacion Villafranca del Castillo, Villanueva de la Ca\~nada, 28692 Madrid, Spain\label{aff158}
\and
Dipartimento di Fisica ``G. Occhialini", Universit\`a degli Studi di Milano Bicocca, Piazza della Scienza 3, 20126 Milano, Italy\label{aff159}
\and
School of Physical Sciences, The Open University, Milton Keynes, MK7 6AA, UK\label{aff160}
\and
Department of Astrophysics, University of Zurich, Winterthurerstrasse 190, 8057 Zurich, Switzerland\label{aff161}
\and
Theoretical astrophysics, Department of Physics and Astronomy, Uppsala University, Box 524, 751 20 Uppsala, Sweden\label{aff162}
\and
Mathematical Institute, University of Leiden, Einsteinweg 55, 2333 CA Leiden, The Netherlands\label{aff163}
\and
Institute of Astronomy, University of Cambridge, Madingley Road, Cambridge CB3 0HA, UK\label{aff164}
\and
Kobayashi-Maskawa Institute for the Origin of Particles and the Universe, Nagoya University, Chikusa-ku, Nagoya, 464-8602, Japan\label{aff165}
\and
Institute for Particle Physics and Astrophysics, Dept. of Physics, ETH Zurich, Wolfgang-Pauli-Strasse 27, 8093 Zurich, Switzerland\label{aff166}
\and
Department of Astrophysical Sciences, Peyton Hall, Princeton University, Princeton, NJ 08544, USA\label{aff167}
\and
Fakult\"at f\"ur Physik, Universit\"at Bielefeld, Postfach 100131, 33501 Bielefeld, Germany\label{aff168}
\and
Department of Physics and Astronomy, Lehman College of the CUNY, Bronx, NY 10468, USA\label{aff169}
\and
American Museum of Natural History, Department of Astrophysics, New York, NY 10024, USA\label{aff170}
\and
Ruhr University Bochum, Faculty of Physics and Astronomy, Astronomical Institute (AIRUB), German Centre for Cosmological Lensing (GCCL), 44780 Bochum, Germany\label{aff171}}    

\date{Received xxx; accepted xxx}

%
%
\abstract{
The \Euclid mission is expected to discover ${\sim}\,10^5$ galaxy--galaxy strong gravitational lenses across its ${\sim}\,14\,000$\,deg$^2$ survey, requiring efficient automated searches with high completeness and low false-positive rates. Although deep-learning lens finders have achieved remarkable success in recent years, the impact of training-set composition, image representation, and preprocessing on survey-scale lens detection remains poorly understood.
We investigate these methodological choices through a systematic ablation study of a convolutional neural network developed for the ECLIPSE strong-lens detection pipeline, systematically varying its training configuration while evaluating the resulting models on a fixed test set. Using simulated and real \Euclid data, we compare VIS-only and pseudo-RGB representations, FITS and JPEG formats, training-set composition, hard-negative selection, and multi-stage training strategies. We further assess model ensembling and evaluate the final classifiers using approximately $67$ million DR1 cutouts.
We find that training-set composition and training strategy strongly affect high-Recall performance. Omitting simulation based pre-training reduces high-Recall weighted average Precision by $83.7\%$ for VIS-only and $74.5\%$ for pseudo-RGB models, while negative-sample selection also substantially influences performance. FITS-trained models achieve modest improvements over JPEG-trained counterparts at high Recall (${\sim}\,2$--$5\%$). On the full DR1 survey, the logit-mean ensemble improves average Precision by $28.0\%$ and $39.9\%$ relative to the individual VIS-only and pseudo-RGB models, respectively. Among the ${\sim}\,70\,000$ highest-ranked candidates selected for visual inspection, it recovers $11\,256$ unique grade A (secure) and B (probable) lenses, compared with $9650$ and $10\,395$ for the individual models, corresponding to $1606$ ($+\,11.3\%$) and $861$ ($+\,6.0\%$) additional systems. The two branches exhibit complementary ranking preferences, particularly for lenses benefiting from near-infrared information.
We conclude that careful training-set design, realistic negative examples, and multi-band ensembling improve survey-scale strong-lens searches. Our final ensemble achieves ${\sim}\,80\%$ Recall at the operational selection threshold, outperforming ablation models with higher global classification metrics and highlighting the importance of optimising performance for realistic survey conditions.

}
%
%
%
%
   \titlerunning{Enhancing deep-learning models for DR1}
   \authorrunning{Euclid Collaboration: M. Fogliardi et al.}
   
   \maketitle
   \nolinenumbers
%
%
%
%

\section{\label{sc:intro}Introduction}

Strong gravitational lensing provides a unique probe of the distribution of matter in the Universe across a wide range of scales. Galaxy-scale strong lenses enable detailed measurements of the total mass profiles of galaxies and constraints on the nature and distribution of dark matter \citep{treu2010, auger2010}, studies of highly magnified high-redshift sources \citep{pettini2000,quider2009}, and independent cosmological measurements through lensing statistics and time-delay analyses \citep{oguri2010,suyu2017,sonnenfeld2021}. The scientific return of strong-lens samples therefore scales directly with the number, diversity, and completeness of the systems that can be identified.

The advent of large-area imaging surveys has transformed the search for strong gravitational lenses from a targeted observational effort into a large-scale data mining problem. Surveys such as CFHTLS, DES, HSC, and KiDS have motivated the development of increasingly automated lens-finding techniques capable of processing millions of galaxies \citep{jacobs2019, petrillo2019a, huang2021}. Among these approaches, deep-learning (DL) methods have emerged as one of the most effective solutions, consistently achieving state-of-the-art performance in lens-identification tasks \citep{lanusse2018,petrillo2019b,canameras2021}.

\Euclid \citep{Laureijs11, EuclidSkyOverview} represents the next major step in this progression. The survey is expected to image approximately $14\,000$\,deg$^2$ of the extragalactic sky with unprecedented combinations of area, depth, and angular resolution. Simulations predict that \Euclid will discover of order $10^5$ galaxy--galaxy strong lenses, increasing the currently known population by an order of magnitude \citep{collett2015}. However, realising this scientific potential requires highly efficient automated search strategies. Even a classifier with excellent overall accuracy must operate at extremely low false-positive rates when applied to hundreds of millions of galaxies, while simultaneously maintaining high completeness to avoid losing rare lens systems.

The importance of understanding the best methodological choices extends beyond \Euclid itself. In the coming years, \Euclid observations will be complemented by data from the Vera C. Rubin Observatory Legacy Survey of Space and Time (LSST; \citealp{LSST2019}) and the \textit{Nancy Grace Roman} Space Telescope (\citealp{roman2015}). Together, these facilities will enable strong-lens searches over unprecedented combinations of area, depth, time, wavelength coverage, and angular resolution. Their complementary optical and near-infrard (NIR) observations will then provide further opportunities for multi-wavelength lens searches.

Recent \Euclid lens-finding efforts have therefore focused on developing and validating machine-learning (ML) pipelines capable of operating at survey scale. The Euclid Quick Data Release (Q1; \citealt{Q1-TP001}) strong-lens search \citep{Q1-SP048} combined multiple complementary approaches, including convolutional neural networks (CNNs, \citealt{Q1-SP053,Q1-Xu}), vision transformers (ViTs, \citealt{Q1-Vincken}), and visual-inspection campaigns featuring also citizen science-assisted classification workflows \citep{Q1-SP052,Q1-SP059}, to produce the first \Euclid galaxy--galaxy strong-lens (GGSL) catalogue \citep{Q1-SP048,Q1-Ecker}. 

After Q1, the \Euclid strong-lens search has now been extended to the first major public data release (DR1; \citealt{DR1-TP001}), covering approximately $2000$\,deg$^2$ of extragalactic sky. At this scale, exhaustive visual inspection is no longer feasible, and candidate selection relies on automated ranking pipelines capable of processing tens of millions of galaxies. To construct the ECLIPSE strong-lens catalogue \citep{DR1-SL01}, multiple independently developed ML models were applied to the full survey footprint, followed by citizen science and expert visual-inspection stages \citep{DR1-SL02}. This effort yielded a catalogue containing more than $15\,000$ strong-lens candidates, complete with lensing-model fits \citep{DR1-SL03} and photometric redshifts \citep{DR1-SL04}, increasing the number of known candidates by more than an order of magnitude relative to the Q1 release.

The network presented in this work formed one of the ML components used in the DR1 ECLIPSE pipeline (described in Euclid Collaboration: Vincken et al. in prep.), achieving the highest single performance overall. While the operational performance of the DR1 pipeline demonstrates the effectiveness of DL approaches for large-scale lens discovery, it also raises a number of important methodological questions. For this reason, we investigated which aspects of the training strategy contribute most strongly to performance, how much information is lost when astronomical images are compressed or reformatted for ML applications, to what extent does the selection of positive and negative training examples influence the final candidate rankings, and how can complementary representations of the same astronomical scene be combined to maximise completeness while maintaining manageable candidate volumes. Addressing these aspects is particularly relevant for \Euclid observations. The survey simultaneously provides high-resolution VIS imaging \citep{EuclidSkyVIS} and multi-band NIR photometry \citep{EuclidSkyNISP}, enabling multiple strategies for constructing ML inputs. Furthermore, the rarity of galaxy-scale strong lenses (approximately 1 lens every $10\,000$ ML-scored images) means that the dominant challenge is not distinguishing lenses from random galaxies, but from visually similar non-lenses that populate the high-score tail of candidate rankings. As a result, the selection of realistic hard negatives and the preservation of subtle lensing signatures may be as important as the choice of network architecture itself.

In this work, we investigate how training-set composition, image representation, preprocessing, and multi-stage training affect the performance of a Zoobot-based CNN \citep{walmsley2020} developed for Euclid Wide Survey observations (EWS; \citealt{Scaramella-EP1}). Using Q1 and DR1 observations together with simulated lenses, we evaluate these choices through a controlled ablation study, in which we systematically modify selected components of the training dataset while keeping the remaining configuration fixed. By comparing the resulting models, we quantify the impact of each modification on classification performance, focusing on high-Recall ranking performance and the complementarity between VIS-only and pseudo-RGB representations.

This paper is organised as follows. Section~\ref{sc:training-dataset} describes the data sets, while Sect.~\ref{sc:network-training} presents the network architecture, preprocessing, and training procedures. Section~\ref{sc:ablation-analysis} reports the ablation results, Sect.~\ref{sc:model-ensemble} evaluates the final models and their ensembles on DR1, and Sect.~\ref{sc:discussion-and-conclusion} summarises the main findings.

\section{\label{sc:training-dataset} Data}

The \Euclid telescope provides imaging in one wide, visible-wavelength band, \IE, captured by the Visible Camera (VIS), alongside three NIR bands, \YE, \JE, and \HE, captured by the Near-Infrared Spectrometer and Photometer (NISP). 

To train our networks, we extracted cutouts from the background-subtracted merged mosaic DR1 tiles, wherein the images from the four \Euclid bands are matched to the resolution of VIS \IE. For reference, the point-spread function (PSF) full width at half maximum (FWHM) is \ang{;;0.16} for \IE, and \ang{;;0.35}, \ang{;;0.34}, and \ang{;;0.35} for \YE, \JE, and \HE, respectively, with details regarding the interpolation provided in \citet{Q1-TP004}.

The performance and reliability of CNNs depend critically on the quality, volume, and composition of the data used during their optimisation. In the context of strong gravitational lens discovery, compiling a sufficient number of high-confidence examples presents a formidable challenge. To overcome this, we have constructed a comprehensive, hybrid training data set that leverages both highly realistic simulations and carefully vetted empirical observations. The training procedure is designed in two complementary stages: an initial pre-training phase driven by extensive simulated data, followed by a rigorous fine-tuning phase that relies exclusively on real \Euclid observations. In this section, we detail the composition of these data sets. For consistency across all training, validation, and test samples, every image cutout (simulated lenses, real lenses, and non-lens systems) is represented as a $100\times100$ pixel image, corresponding to a $10\arcsec \times 10\arcsec$ field of view.

\subsection{\label{sc:simulated-ggsl-events}Simulated GGSL events}
The current number of real spectroscopically or visually confirmed strong lenses in \Euclid is insufficient to properly train a complex network from scratch. To prevent this limitation, we extensively rely on simulated data for the first training stage. These simulated GGSLs required for the initial pre-training phase are then broadly categorised by their host environment, which we detail in the subsequent sections: cluster environments (Sect.~\ref{sc:simulated-ggsl-events-in-clusters}) and the general field (Sect.~\ref{sc:simulated-ggsl-events-in-field}). Representative examples of both simulation populations are shown in Appendix~\ref{app:simulated_lenses}.

\subsubsection{\label{sc:simulated-ggsl-events-in-clusters}Simulated GGSL events in clusters}
To create a robust sample of simulated GGSL events within cluster environments, we adopted the pipeline developed by \citet{angora2023}. We refer the reader to their work for a comprehensive description of the methodology, while we briefly summarise the key steps here. 

The simulation framework relies on nine existing cluster lens models, optimised using {\textsc{lenstool}}~\citep{kneib1996, jullo2007, jullo2009}, to provide accurate cluster deflection angle maps. The process of generating mock lenses is carried out using the \texttt{Python}\footnote{\url{https://www.python.org/}} library \texttt{PyLensLib} \citep{meneghetti2021} and can be outlined as follows: 
\begin{enumerate}[label=(\roman*)]
    \item Critical curve identification: convergence and shear maps are derived from the deflection angle maps to identify the critical curves where the lensing magnification diverges.
    \item Lens selection: secondary critical lines belonging to cluster members, with equivalent Einstein radii in the range $ \text{\ang{;;0.2}} < \theta_\sfont{E} < \text{\ang{;;3.5}}$, are selected. To build a mass-limited sample and avoid photometric biases, the extraction probability is set proportional to the size of $\theta_\sfont{E}$.
    \item Caustic mapping: the selected secondary critical lines are mapped onto the source plane via the lens equation to trace the corresponding caustics.
    \item Source injection: background sources, simulated with Sérsic surface brightness profiles, are injected within the caustics. Since gravitational lensing conserves surface brightness, the lensed images are computationally generated, convolved with the appropriate instrumental PSF, and finally co-added into the real archival \HST (HST) images.
\end{enumerate}

To ensure the physical realism of the simulated lenses, the injected background sources are assigned magnitudes and redshifts drawn from probability density functions based on the COSMOS photometric catalogue \citep{scoville2007,laigle2016}, enforcing a minimum source redshift of $z_{\rm src} \ge z_{\rm cls} + 0.4$ \citep{meneghetti2022}. Furthermore, their sizes are assigned following observationally motivated size-redshift relations \citep{shibuya2015, tortorelli2018}

This procedure allowed us to construct a large knowledge base of $5011$ mock GGSLs encompassing a wide variety of lensing configurations, including adversarial examples (such as faint sources or small-scale lenses) to prevent network overfitting. 

The resulting $5011$ simulated GGSLs were converted into \Euclid-like observations using the \texttt{HST2EUCLID} \texttt{Python} package \citep{EP-Bergamini}. This pipeline reproduces the instrumental response of \Euclid through PSF matching, HST-filter combinations approximating the VIS and NISP bands, resampling to the corresponding pixel scales ($100$ and $300$ mas px$^{-1}$, respectively), and noise adjustments consistent with the EWS depth. The original images were drawn from CLASH \citep{postman2012} and the \textit{Hubble} Frontier Fields \citep{koekemoer2014}. Further implementation details are provided in \citet{EP-Bergamini}.

\subsubsection{\label{sc:simulated-ggsl-events-in-field}Simulated GGSL events in the field}
To train and validate our lens-finding algorithms, we used also two distinct sets of simulated field GGSL systems, hereafter referred to as S1 and S2, as described in \citet{Q1-SP053}. These simulations were designed to capture a broad range of lensing configurations and morphologies likely to be encountered in the EWS.

The first simulation set (S1) was generated following the methodology described by \citet{Q1-SP052} using the \texttt{Lenstronomy} package \citep{2018BIRRER,birrer2021}. Specifically, observed luminous red galaxies were used as deflectors and paired with high-resolution background sources drawn from HST imaging and HSC photometry. The systems were lensed using singular isothermal ellipsoid mass models and rendered to match the \Euclid VIS and NISP imaging characteristics. After augmentation and combination with an earlier iteration of the data set, S1 contains $11\,057$ simulated lenses. We refer the reader to \citet{Q1-SP053} for a more detailed description of the simulation procedure.

The second simulation set (S2) was produced using the \texttt{GLAMER} code \citep{metcalfpetkova2014,petkova2014}. In this case, observed galaxies were matched to counterparts in the \Euclid Flagship simulation \citep{EuclidSkyFlagship} to construct physically motivated baryonic and dark matter mass models, while background sources were generated using Sérsic-profile realisations based on the COSMOS 2020 catalogue \citep{weaver2022}. The resulting lensed systems were rendered using realistic \Euclid observing conditions and subjected to a final visual quality control. The final S2 sample contains $5363$ simulated lenses. Further details can be found in \citet{Q1-SP053}.

For the purpose of training our networks, we aimed to construct a balanced data set composed of approximately $50$\% cluster GGSLs and $50$\% field GGSLs. To match the available number of cluster mock lenses, we extracted a specific subset of these field simulations rather than using the full S1 and S2 catalogues. Specifically, we used $2585$ simulations from the S1 data set (strictly omitting the augmented, rotated versions) and $2500$ simulations from the S2 set, yielding a representative field sample that evenly complements the cluster data set.

\subsection{\label{sc:non-lenses}Non-lens systems}

\subsubsection{\label{sc:hard-negatives}Hard negatives} Our hard-negative training set consists of 13\,000 objects accurately extracted from the SLDE classification catalogues of the Q1 fields \citep{Q1-SP048}. This sample is divided into two distinct categories based on their SLDE classification grade: $8800$ grade ‘XX' objects (hard negatives) and $4200$ grade ‘X' objects (very hard negatives). The grade XX systems represent objects that were initially flagged by earlier ML pipelines but were subsequently rejected by citizen-scientists during the Space Warps project (SW; \citealt{Q1-SP059}) after failing to reach a consensus for lensing features. SW is the \Euclid citizen science project on Zooniverse\footnote{\url{https://www.zooniverse.org/projects/aprajita/space-warps-esa-euclid}} for finding strong gravitational lenses.

In contrast, grade X objects are highly deceptive systems that managed to pass through both the initial automated ML selection and the SW citizen science vetting phase, only to be discarded upon rigorous visual inspection and grading by our expert scientific team during the internal Galaxy Judges (GJ) campaign, the final visual inspection reserved to consortium members.

The strategic rationale for using this specific sample rather than random galaxies is rooted in the ‘hard-negative mining' strategy \citep{srivastava2016}. Because every object in this non-lens sample was originally flagged as a lens candidate by previous ML models, they possess morphological visual features (such as ring-like artefacts, blue spiral arms, and serendipitous chance alignments) that are known to actively confuse ML algorithms. Training on this challenging data set forces the network to capture more specific differences between true strong lenses and false positives that are morphologically similar. This approach results in a considerably more robust and accurate model compared to those trained on a random sample.

\subsubsection{\label{sc:artefacts}Artefacts} 
In addition to the non-lenses from the SLDE catalogue, we also manually selected $2496$ artefacts and non-lens cutouts from five tiles located near the centers of the Euclid Deep Field (EDF) patches: one in EDF-South, three in EDF-Fornax, and one in EDF-North. The sample primarily contains instrumental artefacts and objects near stellar diffraction spikes, together with some mergers and spiral galaxies whose morphologies can mimic lensing features \citep{Q1-SP067}. Although some of these contaminants may already be represented among the grade XX objects, the manually selected sample deliberately increases their representation, whose frequency in the XX pool was not explicitly controlled.

\subsection{\label{sc:euclid-ggsls} Euclid GGSLs}
While simulated lenses play a crucial role in the initial pre-training of DL models, transitioning these models to correctly identify strong lensing in actual survey data requires a fine-tuning stage using a sufficiently large sample of observational lenses. To construct a high-purity, completely empirical data set of real \Euclid GGSLs, we aggregated high-confidence candidate systems from two distinct observational releases: the initial Q1 fields and a targeted preliminary subset of the DR1 data.  

\subsubsection{\label{sc:q1-lenses}Q1 lenses}

The discovery of our high-confidence lens sample was facilitated by the multi-stage SLDE pipeline, which was applied to a parent sample of approximately $1.09 \times 10^6$ galaxies identified in the Q1 release. This search combined eight DL classifiers, followed by citizen-science classification through SW and expert visual inspection through GJ. Crucially for the present work, these models were trained exclusively on simulated field lenses: they did not include explicit training on simulated cluster-scale environments or their specific contaminants. The resulting catalogue contains 250 grade A (high-confidence) and 247 grade B (probable) lenses. Lower-confidence grade C candidates were excluded from our Q1 positive sample, while the rejected categories (X and XX) are described in Sect.~\ref{sc:hard-negatives}.

We further supplemented this sample with candidates identified through complementary Q1 searches: 39 grade A and 29 grade B systems from the bright, low-redshift galaxy search of \citet{Q1-Ecker}; 8 grade A and 26 grade B systems identified by the \texttt{AstroVink} vision transformer \citep{Q1-Vincken, dosovitskiy2020, oquab2023}; and 22 grade A and 28 grade B systems from the search targeting massive elliptical galaxies \citep{Q1-SP052}.

The resulting Q1 sample comprises 319 grade A and 330 grade B lens candidates, providing a total of 649 real-lens systems for our training and validation data sets.

\subsubsection{\label{sc:dr1-lenses} DR1 lenses (200\,deg$^{2}$)}
 Euclid DR1 ultimately encompassed approximately 2000\,deg$^{2}$ of the sky. However, for our training set, our team targeted an initial localised subset of these data covering an area of approximately 200\,deg$^{2}$. This specific footprint was selected and analysed to substantially extend the available training set of real \Euclid lenses, providing the necessary statistical volume to re-train and optimise all lens-finding ML models prior to their large-scale application across the full DR1 survey area.

To assemble an initial comprehensive sample of strong-lens candidates, we generated cutouts from the 200\,deg$^{2}$ footprint extracting all objects meeting the selection criteria established in \citet{DR1-SL01}. We then performed the preliminary search by processing these cutouts through six independent ML algorithms (five from the SLDE, alongside the \texttt{AstroVink} model). The top-ranked candidates from each model's predictions were compiled, yielding an initial pool of 13\,122 candidate cutouts.

These candidates were then subjected to a two-tier visual-inspection process conducted by a group of 14 lensing experts. During the `Stage 1' inspection, the experts performed a rapid, one-by-one visual assessment intended solely to identify and filter out obvious non-lenses and spurious artefacts. This initial inspection yielded a subset of 2245 promising systems. These selected objects then advanced to the `Stage 2' inspection, where they were thoroughly re-evaluated. In this phase, each expert independently assigned a classification grade to the candidates, using the standard grading scheme described in Sect.~\ref{sc:q1-lenses}. These two stages refer to the expert visual-inspection procedure and should not be confused with the Stage 0 and Stage 1 network training scheme presented later. Of this preliminary DR1 sample, 782 originally graded A and B candidates were used in the train3 ablation, while the final fine-tuning set uses our own-derived 953 $\rm \bar{A}$ and $\rm \bar{B}$ objects. This ensemble of grades was later refined and calibrated as described in \citet{DR1-SL02}. The final set from the 200\,deg$^{2}$ is included in the ECLIPSE catalogue presented in \citet{DR1-SL01}.

To construct our training sample, we independently combined the expert grades assigned to the $2245$ preliminary candidates. Specifically, we encoded A, B, C, and X as $3$, $2$, $1$, and $0$, respectively. Their mean scores defined four categories: $\rm \bar{A}$ ($>2.3$), $\rm \bar{B}$ ($1.5$--$2.3$), $\rm \bar{C}$ ($0.5$--$1.5$), and $\rm \bar{X}$ ($<0.5$).
We emphasise that this numerical encoding provides a practical means of combining subjective visual classifications, rather than a calibrated measure of lensing probability. In particular, the numerical spacing between grades is arbitrary, and the resulting mean scores may be sensitive to differences in the grading criteria and confidence of individual inspectors. Consequently, candidates assigned to the same mean-score category may exhibit substantially different degrees of morphological evidence for lensing, motivating the additional visual selection described below.

Finally, to improve the reliability of the training labels, we conducted an additional targeted visual selection of the mean-scored objects. Candidates with similar mean scores may exhibit substantially different features, while disagreement among inspectors does not automatically imply that a system is unsuitable for training. The objective of this additional selection was therefore to minimise potentially misleading or inconsistent labels, rather than to exclude all candidates with uncertain grades. We preferentially retained systems for which the additional inspection supported a sufficiently reliable positive or negative classification, while excluding cases where the available evidence remained inconclusive. This approach aims to limit the impact of label noise on the learned representations \citep{reed2014,zhang2016}, while preserving the morphological diversity needed for the models to generalise to less obvious lensing configurations.

Following this re-evaluation phase, the final composition of the DR1 subset lens data set used in this work is detailed below:
\begin{enumerate}[label=(\roman*)]
    \item Grade $\rm \bar{A}$: 376 confirmed systems (7 initially mean-scored as grade A were rejected as lacking sufficient definitive features).
    \item Grade $\rm \bar{B}$: 742 confirmed systems (38 rejected).
    \item Grade $\rm \bar{C}$: despite their relatively low mean grades, $76$ candidates with sufficiently convincing lensing features were retained as positives, while $351$ systems considered more consistent with non-lensing configurations were assigned to the negative pool. The remaining $733$ inconclusive candidates were excluded. We emphasise that this additional selection also involves subjective visual judgement and cannot entirely eliminate label uncertainty.
    \item Grade $\rm \bar{X}$: 54 confirmed negative systems (3 rejected).
\end{enumerate}
Rejected or inconclusive objects were excluded from the training sample. We note that the preliminary DR1 sample contains a larger fraction of grade B candidates relative to grade A than the Q1 sample. This difference should not necessarily be interpreted as increased contamination, since grade B identifies probable lenses rather than rejected or highly ambiguous systems. Moreover, the two samples were obtained through different ML selection and visual-inspection campaigns, potentially introducing differences in both the populations of selected candidates and the subjective assignment of confidence grades. The relative abundance of grade B systems therefore cannot be used alone as a reliable indicator of sample purity.

\section{\label{sc:network-training} Network implementation and training}

To maximise the efficiency of our lens search and ensure our models can accurately interpret complex morphology within field and cluster environments, we leverage the power of transfer learning through state-of-the-art foundation models \citep{bommasani2021}. Rather than training architectures for our specific task from scratch, which requires unfeasibly large volumes of annotated data, we initialise our networks with weights from models that have already been extensively pre-trained on diverse, large-scale data sets. This pre-training enables the models to learn generalised feature extractors, which are then adapted to the specific task of strong-lens identification. In this work, we employ the \texttt{Zoobot} foundational architecture: a supervised computer vision model specifically trained for galaxy morphology \citep{walmsley2023}. The structural details of this network and our specific modifications are described below.

\subsection{\label{sc:zoobot} Zoobot-Euclid ConvNeXt-Base}
\texttt{Zoobot} \citep{walmsley2023,walmsley2024gz} is a DL framework originally consisting of a Bayesian CNN developed for detailed morphological classification of galaxies, then extended to several different ML architectures. The base models provided by \texttt{Zoobot} operate as foundation models for astronomy, having been pre-trained on up to 92 million volunteer responses gathered through the Galaxy Zoo citizen science project \citep{lintott2008}. Because identifying GGSLs is fundamentally a visual morphological task, relying on the detection of structured anomalies like arcs and rings against typical galactic light profiles, the latent space pre-learned by \texttt{Zoobot} to distinguish standard galaxy morphologies serves as an exceptionally strong prior for lens detection. Indeed, \citet{Pearce-Casey24} and \citet{Q1-SP053} demonstrated that fine-tuning \texttt{Zoobot} architectures yields superior performance in identifying strong lenses in \Euclid field data compared to models trained from scratch. However, those previous \Euclid lens searches prioritised computational speed, using the compact \texttt{ConvNeXt-Nano} (15.6 million parameters) and \texttt{EfficientNet-B0} (5.3 million parameters) variants of \texttt{Zoobot}. 

In this work instead, we chose to fine-tune the considerably deeper and best-performing \texttt{Zoobot} architecture available: \texttt{ConvNeXt-Base}. Comprising approximately $87.5$ million trainable parameters, this deeper model achieved the lowest recorded test loss (19.05) on the original GZ Evo test set, outperforming both the \texttt{Nano} (19.23) and \texttt{EfficientNet-B0} (19.48) variants. These benchmark losses refer to the galaxy-morphology task and should not be interpreted as a controlled comparison of lens-finding performance: we keep the architecture fixed throughout the ablations described in Sect.~\ref{sc:ablation-analysis}.

\texttt{ConvNeXt} \citep{liu2022} is a modernised convolutional architecture that retains the inherent translation equivariance and structural simplicity of traditional \texttt{ResNets} \citep{he2016} while integrating the macro-design choices that drive the success of Vision Transformers (ViTs; \citealt{dosovitskiy2020}), such as large kernel sizes, inverted bottlenecks, and Layer Normalisation \citep{ba2016}. 

Crucially, the specific \texttt{Zoobot} variant used in this work possesses a specific two-stage pre-training history. Initially trained on the extensive GZ Evo data set (about $814\,000$ galaxies), comprising classifications from multiple ground- and space-based surveys (excluding \Euclid), the model was subsequently fine-tuned on the GZ \Euclid data set (approximately $174\,000$ galaxies). This secondary pre-training phase uniquely aligns the network's internal representations with \Euclid's specific instrumental PSF, noise properties, and pixel scales. We emphasise that this pre-training did not involve either strong lensing labels or Q1 and DR1 images; therefore, the model remains an unbiased foundation for our specific task.

To adapt the base \texttt{Zoobot-Euclid} model to our binary classification task (distinguishing GGSLs from non-lenses), we restructured its architecture. In DL, a vision model is typically conceptually divided into two parts: a ‘backbone', which acts as the core feature extractor that detects visual patterns (such as shapes and textures) from the input images, and a ‘classification head', which translates these extracted patterns into final class predictions. 

We retained the \texttt{ConvNeXt-Base} backbone but removed its default head, originally designed for general galaxy morphology classification. Instead, we extract the raw 1024-dimensional feature vector directly from the backbone output and pass it into a newly constructed classification head. This custom head is built as a Multi-Layer Perceptron (MLP), a traditional sequence of fully connected CNN layers that map complex inputs to simpler outputs. Our MLP gradually funnels the high-dimensional features down to a binary decision via progressive dimensionality reduction: 1024 $\to$ 512 $\to$ 128 $\to$ 2, where the final output is a tensor containing the probability of an object to be class 0 (non-GGSL) and class 1 (GGSL). 

To ensure the network trains efficiently and accurately, we incorporated three standard regularising mechanisms within this head: 
\begin{enumerate}[label=(\roman*)]
    \item Layer Normalisation: applied immediately to the incoming feature vector, this standardises the data values with Z-score normalisation, ensuring the network learns smoothly and stably.
    \item Activation Units: we interspersed two Gaussian Error Linear Units (GELU; \citealt{hendrycks2016}) between the size reductions. These mathematical functions introduce non-linearity, allowing the network to learn complex relationships rather than simple weighted averages.
    \item Dropout Layers: applied at rates of 30\% and 20\%, this regularisation technique randomly ‘turns off' a fraction of the network's neurons during each training step. This prevents the network from memorising the specific training examples (overfitting) and forces it to learn generalised visual features.
\end{enumerate}

Furthermore, unlike the approaches adopted in previous searches, where the majority of the pre-trained weights were frozen and only the final layers were allowed to update, we perform a full-network fine-tuning. We initialise the model with the \texttt{Zoobot-Euclid} weights but allow gradients to backpropagate through the entire network during training. While freezing early layers protects low-level feature extractors from catastrophic forgetting when training data are scarce, full-network fine-tuning frequently achieves higher peak performance when the new target data set is sufficiently large, as is the case here \citep{zhai2019, kornblith2018}. By permitting the entire network to dynamically re-arrange its convolutional filters, the model can actively shift its attention away from standard galactic morphology and fully optimise its early-stage edge and texture detectors for the specific physical characteristics of strongly lensed sources and the dense visual crowding inherent to cluster environments.

\subsection{\label{sc:trainingmethodology}Training methodology}
To assess how different representations of the same \Euclid observations influence the performance of DL lens finders, we adopt a common network architecture and training procedure, varying only the preprocessing applied to the input data. This enables a controlled comparison between two complementary approaches: a VIS-only representation that maximises the use of \Euclid's VIS spatial resolution, and a pseudo-RGB representation that incorporates the additional colour information provided by the NISP photometry. The details of each preprocessing pipeline are presented in the following sections, while illustrative examples of the resulting input representations are provided in Appendix~\ref{app:input_representations}.

\subsubsection{\label{sc:visonly-model}VIS-only preprocessing}
Astronomical images natively possess broad dynamic ranges, meaning faint background structures are frequently drowned out by the exceptionally bright cores of foreground galaxies. Due to this contrast, a single global intensity mapping cannot simultaneously represent bright galaxy cores, mid-tone structure, and very faint lensing features with equal fidelity. To optimise the single-band VIS data for our CNN, which expect 3-channel input arrays, we transform the 1-channel raw \texttt{FITS} cutouts into a 3-channel pseudo-image. Rather than redundantly duplicating the data, we apply three mathematically distinct non-linear stretches to the same raw pixel values. This multi-perspective approach allocates the available input dynamic range to different regimes and objectives, mapping complementary morphological information into separate input layers. Specifically, the first channel is engineered for strict morphological characterisation (highlighting boundaries), the second channel offers the best visual tradeoff between faint-feature visibility and bright-core saturation, and the third channel is specifically tailored to expose the morphology of very faint features. By stacking these, we force the network to simultaneously interpret complementary geometric and luminous representations of the exact same physical structure. 

More specifically, the logic behind our stretches implementation is detailed below:

\begin{enumerate}[label=(\roman*)]
    \item Gradient-enhanced stretch: 
    the first channel applies a custom stretch designed to act as an edge-detector and enhancer. We utilise a Sobel filter, a first order differential operator that computes an approximation of the gradient $G(x,y)$ of the image intensity in a direction. By applying the filter along both the $x$ and $y$ pixel axes (obtaining $G_x$ and $G_y$, respectively), we calculate the gradient magnitude as $M= \sqrt{G_x^2(x,y)+G_y^2(x,y)}$, highlighting regions of rapid intensity transition (i.e., edges and structural boundaries). This normalised gradient map is then added back to the original image weighted by a fraction parameter ($\alpha = 0.3$, empirically chosen), and the result is linearly stretched between the 1st and 99.5th percentiles. This operation enhances high-frequency spatial structures, preventing the diffuse light profile of massive early-type galaxies from bleeding into and masking the thin boundaries of nearby lensed arcs.
    
    \item Inverse hyperbolic sine (arcsinh) stretch:
    the second channel applies an arcsinh stretch to produce a numerically well-conditioned representation of the VIS flux field. We normalise the raw array against its 99th percentile, chosen as our brightness reference scale, and we apply the stretch via $f(k) = \mathrm{arcsinh}(k/\beta)$ with $\beta = 0.1$, without further clipping the image as the asinh function already cleverly compresses the dynamic range. Unlike a linear mapping (where a small fraction of very bright pixels can dominate the available range) or a logarithmic mapping (which can overly compress mid-tones), arcsinh allows for dual behaviour: it is approximately linear at low intensities, but transitions to a log-like behaviour for high surface brightness regions. This cleanly allocates the limited numeric dynamic range of the network input more efficiently across faint, intermediate, and bright structures \citep{lupton2004}. 
    
    \item Contrast limited adaptive histogram equalisation (CLAHE) stretch: 
    the third channel utilises the CLAHE method \citep{pizer1987, zuiderveld1994}. Unlike standard global histogram equalisation, which works on the entire image at once, CLAHE divides the image into small tiles and enhances the contrast within each respective tile. Crucially, it imposes a user-defined clip limit for the desired contrast expansion. Therefore, to stabilise the output and prevent the amplification of background noise in blank regions of the sky, we scale the input data to a 16-bit integer array and run the CLAHE algorithm three separate times with increasing clip limits (99.9, 99.5, and 99.0 percentiles). The resulting representations are subsequently blended using a weighted average that favors the most conservative clippings. We implemented this routine to iteratively maximise the visibility of extremely faint features hidden within the varying intensity of the lens galaxy light, while retaining a substantial portion of the original luminous information.
\end{enumerate}

\subsubsection{\label{sc:rgb-model}Pseudo-RGB preprocessing}
For the architecture designed to exploit the multi-band photometric information provided by \Euclid, we combine the high-resolution VIS camera data with the lower-resolution NISP camera data. We construct a 3-channel pseudo-RGB composite by mapping the longer-wavelength NIR data (\HE) to the red channel, the shorter-wavelength NIR data (\JE) to the green channel, and the high-resolution VIS band (\IE) to the blue channel. 

To bridge the gap in resolution and intensity across different instruments, we execute a custom preprocessing pipeline inspired by pan-sharpening techniques. The procedure operates as follows:
\begin{enumerate}[label=(\roman*)]
    \item Channel standardisation: missing or NaN values in each individual band are replaced with the global median of that array. Each band is then independently clipped between the 0.5th and 99.5th percentiles to clear bad pixels, normalised to bounds of $[0, 1]$, and stretched using the $\text{arcsinh}$ mapping ($\beta = 0.1$). 
    \item Luminance substitution: the normalised RGB stack is then transformed into the CIELAB (LAB) colour space, which mathematically decouples geometric brightness (Luminance, L) from colour information (A and B channels). We then select the original asinh stretched map generated strictly from the high-resolution VIS band, and substitute it entirely over the composite's original L channel. 
    \item Colour re-projection: the modified LAB matrix is inverted back into the RGB colour space. We conclude the pipeline by artificially bumping the global colour saturation by a factor of 1.3 (chosen empirically) to further highlight chromatic differences between lens galaxies and star-forming background sources.
\end{enumerate}

This luminance substitution represents the key step of the pseudo-RGB approach, as it physically forces the superior resolution of the VIS instrument in the multi-band image while maintaining the colour gradients contributed by the NISP photometry, optimising the tensor for pattern recognition.

\subsection{\label{sc:fits-vs-jpeg}Information loss: FITS vs. JPEG}

To strictly quantify the amount of information retained by our pipelines, we conducted an information-loss analysis on a representative sample of 1000 \Euclid cutouts, including Q1 and preliminary DR1 lens candidates together with grade X hard negatives. Although most of these objects are also included in the final ablation test set, the two samples are not identical. Transforming astronomical \texttt{FITS} files with high dynamic range floating-point values into standardised inputs for DL models inherently involves data degradation. To evaluate this, we measured the bit depth, defined as $\log_2(N_{\mathrm{unique}})$, where $N_{\mathrm{unique}}$ is the number of distinct discrete pixel intensity levels residing within the physical range. The effective bit depth quantifies the granularity of the image (i.e. how many distinct structural and photometric variations the network can actually resolve).

\begin{figure}[htbp]
    \centering
    \includegraphics[width=0.49\textwidth]{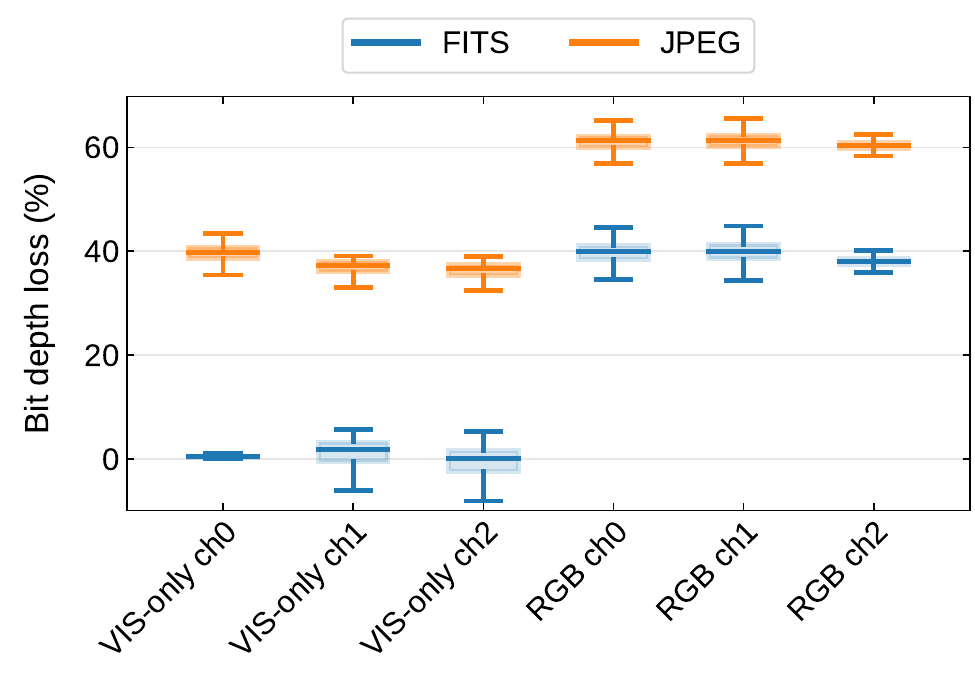} 
    \caption{Distributions of effective bit-depth loss across 1000 test set example cutouts. Boxes indicate the interquartile range (IQR), central lines mark the median, and whiskers extend up to $1.5\times{\rm IQR}$ of the lower and upper quartiles. The boxplots report the relative loss of photometric granularity incurred by different preprocessing strategies compared to the raw VIS \texttt{FITS} base. The underlying uncompressed pipelines (prepr.~FITS, in blue) demonstrate near-zero information loss for the isolated VIS channels, while highlighting the inherent ${\sim}\,40\%$ compromise required to build the pseudo-RGB composite. The orange overlays (prepr.~JPEG) reveal the severe compounding loss in bit-depth resolution introduced if these arrays were bottlenecked through 8-bit JPEG compression for storage.}
    \label{fig:dynamic_range_stats}
\end{figure}

Our VIS-only pipeline was explicitly designed to preserve the natural bit-depth information of \texttt{FITS} files. By applying three distinct mathematical stretches (gradient, asinh, and CLAHE) to the raw VIS images without compressing them, we map astronomical scaling into the network's domain without reducing image granularity. Our analysis confirms this (see Fig.~\ref{fig:dynamic_range_stats}): compared directly to the raw VIS \texttt{FITS} files, the VIS-only preprocessing yields a negligible median loss in bit depth ($0.34\%$, $1.81\%$ and $0.07\%$ for gradient, asinh, and CLAHE respectively). This confirms that the information encoding remains almost uncompromised.

Conversely, our pseudo-RGB methodology, which integrates \IE, \JE, and \HE bands into a 3-channel composite, forces a structural compromise. Generating a coherent RGB stack inherently requires image compression, resulting in a quantifiable loss of photometric granularity. As shown in Fig.~\ref{fig:dynamic_range_stats}, we obtain a median reduction in bit depth of approximately $38\%\mathrm{-}40\%$ per channel relative to the raw \texttt{FITS}. However, despite this information bottleneck, our composition method provides superior feature integration for the CNNs compared to a directly stacked \texttt{FITS} multi-band composite of lower resolution.

\begin{figure*}[htbp]
    \centering
    \includegraphics[width=\textwidth]{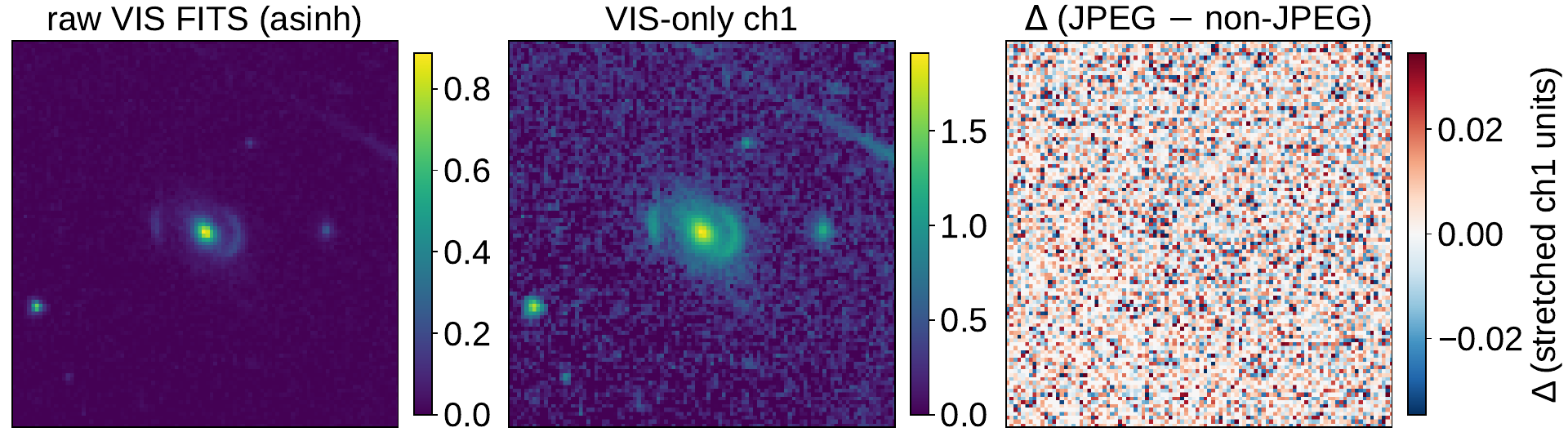} 
    \caption{Visual impact of JPEG compression on continuous photometric distributions for $10\arcsec \times 10\arcsec$ cutouts. The left panel shows the original raw VIS \texttt{FITS} cutout for a grade A lens (scaled via an asinh stretch for visualisation). The middle panel shows the uncompressed, VIS-only network preprocessed \texttt{FITS} output for channel 1, additionally clipped for visualisation (this clipping is not applied to the network input). The right panel presents the pixel-wise difference map ($\Delta$) tracking the absolute residual loss between the uncompressed FITS and its JPEG counterpart. The difference map visually demonstrates the high-frequency quantisation noise and granularity degradation introduced by forcing floating-point arrays into 256 discrete bins.
    }

    \label{fig:jpeg_diff_map}
\end{figure*}

Finally, we used this framework to investigate the impact of pipeline storage formats. Part of the ECLIPSE pipeline relies on converting and storing cutouts as 8-bit JPEGs to reduce processing time. Although JPEG compression roughly preserves the macroscopic contrast span, it ruthlessly degrades the granularity. If our preprocessed inputs were subjected to 8-bit JPEG compression, the VIS-only pre-processed image would suffer a composite median bit-depth loss of approximately $35\%$--40\% across all channels, while the pseudo-RGB pre-processed image would lose in median more than 60\% of its bit depth on all channels. The physical manifestation of this quantisation noise is illustrated in Fig.~\ref{fig:jpeg_diff_map}, where we present a visual comparison between the raw \texttt{FITS} image, the uncompressed image, and the resulting difference map. To the human eye, the compressed and uncompressed panels appear perceptually identical. However, as evidenced by the structured residuals in the difference map, the two images are not quantitatively equivalent. While these subtle signal losses are visually imperceptible to us, they are significant for CNNs, as described in Sect.~\ref{ssc:fits-jpeg-models}.

\subsection{\label{sc:training-configuration}Training configuration}

We trained the two variations of our GGSL-finder model using identical backbone architectures and training procedures, differing only in their input representations. For all experiments, the input consists of $100\times 100$ pixel cutouts, and the task is binary classification ($y=1$ for lenses, $y=0$ for non-lenses). To bridge the domain gap between simulated and empirical data, we initially explored a ten-stage progressive training. In this approach, training began exclusively using simulated lenses as positive class, with each subsequent stage replacing 10\% of the simulated lenses with real lenses, initialising the network with weights from the preceding stage. However, empirical evaluation demonstrated that this gradual transition was unnecessary, as optimal performance was instead achieved by both our VIS-only and pseudo-RGB models using a simplified two-stage transfer learning approach: an initial pre-training phase using only simulated lenses (from here on referred to as Stage 0), followed directly by fine-tuning using only real lenses (Stage 1). Consequently, our final methodology adopts this two-stage strategy. During trainings, we regularise the networks using class-conditional data augmentation which differ from Stage 0 to Stage 1. Stage 0 data augmentation consists of random horizontal and vertical flips of the images applied with equal probability to the positive and negative class. Conversely, positive examples in Stage 1, which represent the minority class, are augmented using the full dihedral group $D_4$ (comprising all eight possible distinct rotations and reflections of one image) to enforce strict rotational invariance for lens configurations and provide more augmented versions of the lenses, while negative examples receive only random horizontal and vertical flips. The Stage 0 networks are initialised with weights pre-trained on the GZ \Euclid data set, while Stage 1 initialisation proceeds directly from the converged Stage 0 checkpoint. Details on the two stages training sets are given below, while the overall composition is reported in Table~\ref{tab:dataset_composition}.

\subsubsection{\label{sc:stage0}Stage 0: simulation-driven pre-training}
Stage 0 uses a balanced training set of $N_\mathrm{train}=21\,392$ cutouts ($10\,096$ lenses, $11\,296$ non-lenses). The positive class is entirely simulated and consists of $5011$ cluster-lens simulations, plus $2585$ and $2500$ field-lens simulations (S1 and S2 sets, respectively). The negative class consists of $8800$ real non-lenses from the hard negatives sample (grade ‘XX', see Sect.~\ref{sc:hard-negatives}) and $2496$ artefact cutouts (see Sect.~\ref{sc:artefacts}). Throughout Stage 0 we monitor generalisation on a fixed real-lens validation set of $N_\mathrm{val}=943$ cutouts ($127$ lenses, $816$ non-lenses), composed of $69$ Q1 lenses (see Sect.~\ref{sc:q1-lenses}), $58$ lenses from the 200\,$\rm deg^2$ DR1 area (see Sect.~\ref{sc:dr1-lenses}), $745$ hard negatives (grade ‘X', see Sect.~\ref{sc:hard-negatives}) and $71$ grade $\rm \bar{C}$+$\rm \bar{X}$ (see Sect.~\ref{sc:dr1-lenses}). 

\subsubsection{\label{sc:stage1}Stage 1: real-lens fine-tuning}
Stage 1 fine-tunes on $N_\mathrm{train}=6150$ cutouts including only real data ($1478$ lenses, $4672$ non-lenses). Lenses comprise $953$ objects from the 200\,$\rm deg^2$ DR1 area (see Sect.~\ref{sc:dr1-lenses}) and $525$ lenses from Q1 area (see Sect.~\ref{sc:q1-lenses}). Non-lenses comprise $4200$ grade ‘X' (see Sect.~\ref{sc:hard-negatives}), $280$ non-lenses from grades $\rm \bar{C}$ (see Sect.~\ref{sc:dr1-lenses}) and $192$ artefacts (see Sect.~\ref{sc:artefacts}). The validation set is the same used in Stage 0 ($N_\mathrm{val}=943$).

\renewcommand{\arraystretch}{1.2}
\begin{table}[ht]
    \centering
    \caption{Summary of data-set compositions across training stages and validation.}
    \label{tab:dataset_composition}
    \begin{tabular}{lccc}
        \hline\hline
        \textbf{Data source} & \textbf{Stage 0} & \textbf{Stage 1} & \textbf{Valid.} \\
        \hline
        \multicolumn{4}{l}{\textit{Lenses}} \\
        Cluster-lens simulations & 5011 & -- & -- \\
        Field-lens simulations (S1) & 2585 & -- & -- \\
        Field-lens simulations (S2) & 2500 & -- & -- \\
        200\,$\rm deg^2$ DR1 area lenses & -- & 953 & 58 \\
        Q1 area lenses & -- & 525 & 69 \\
        \hline
        \textbf{Total lenses} & \textbf{10\,096} & \textbf{1478} & \textbf{127} \\
        \hline
        \multicolumn{4}{l}{\textit{Non-lenses}} \\
        Hard negatives (grade `XX') & 8800 & -- & -- \\
        Hard negatives (grade `X') & -- & 4200 & 745 \\
        DR1 non-lenses ($\rm \bar{C}, \bar{X}$) & -- & 280 & 71 \\
        Artefacts & 2496 & 192 & -- \\
        \hline
        \textbf{Total non-lenses} & \textbf{11\,296} & \textbf{4672} & \textbf{816} \\
        \hline\hline
        \textbf{Total objects} & \textbf{21\,392} & \textbf{6150} & \textbf{943} \\
        \hline\hline
    \end{tabular}
\end{table}

\subsubsection{\label{sc:optaugandloss}Optimisation and loss function}

Prior to training and inference, we apply a channel-wise standardisation (Z-scoring) to the images after the stretch pipeline, such that each input channel $x$ is scaled as 
\begin{equation}
\hat{x}=\frac{x-\mu}{\sigma}\,,
\end{equation}
where $\mu$ and $\sigma$ are the mean and standard deviation of that channel computed directly from the corresponding training split. Standardising the features to zero mean and unit variance is a critical preprocessing step for deep neural networks; it ensures that all input channels operate on a similar numerical scale \citep{LeCun1998}. 

During training, the network weights, i.e. the learnable parameters controlling how input pixels and intermediate features are combined, are iteratively updated to minimise a predefined loss function measuring the discrepancy between predicted and true labels. Training is performed over successive epochs, each corresponding to one full pass through the training set, with weight updates computed on smaller data subsets. The best-performing weight configurations on the validation set are saved as checkpoints, and training is stopped once validation performance no longer improves, reducing the risk of overfitting.

The network weights are optimised using the AdamW algorithm \citep{loshchilov2017}. We configure AdamW with its default exponential decay rates ($\beta_1=0.9$, $\beta_2=0.999$, $\epsilon=10^{-8}$) and a weight decay of $10^{-3}$, processing the training data in batches of 100 images at a time. 

The learning rate, a hyperparameter that dictates the magnitude of the step size the optimiser takes when updating the model's weights, is initially set to $5\times 10^{-5}$ in Stage 0. It is subsequently lowered to $1\times10^{-5}$ in Stage 1 to ensure a smooth and stable fine-tuning process. We actively modulate this step size using a \texttt{ReduceLROnPlateau} scheduler: if the validation metric fails to improve for 5 consecutive epochs, the scheduler decays the learning rate by a factor of 0.5 (down to a minimum of $5\times 10^{-8}$).

To guide the optimiser, we utilise two distinct loss functions depending on the model and the training stage. In Stage 0 (for both models) and in Stage 1 (for the pseudo-RGB model), we employ the standard cross-entropy (CE) loss. In practice, the overall CE loss for a batch of $N$ images evaluates the negative log-likelihood of the softmax probabilities, formulated as
\begin{equation}
\mathrm{CE}(x, y) = - \frac{1}{N} \sum_{n=1}^N \ln \left( \frac{\exp(x_{n, y_n})}{\sum_{c=1}^C \exp(x_{n, c})} \right)\,,
\end{equation}
where $C=2$ is the number of classes (lens or non-lens), $x_{n,c}$ represents the raw and unnormalised network output (logit) for the $n$-th example in class $c$, and $y_n \in \{0, 1\}$ is the true class index for that example. This formulation effectively quantifies the deviation between the predicted probability and the true discrete label, equally penalising misclassifications across the data set. 

However, standard CE loss can be quickly overwhelmed by heavily imbalanced data sets (as our Stage 1 set), where a massive number of easily classified examples from the majority class can numerically dominate the overall gradient and stall learning. To counteract this, in Stage 1 for the VIS-only model we adopt a focal loss (FL; \citealt{lin2017}), formulated as
\begin{equation}
\mathrm{FL}(p_{t}) = \underbrace{\alpha_{t}}_{\text{Class weight}} \, \times \, \underbrace{(1-p_{t})^\gamma}_{\text{Difficulty weight}} \, \underbrace{-\ln(p_{t})}_{\text{Standard CE loss}}\,,
\end{equation}
where $p_{t}$ is the network's estimated softmax probability for the true class ${t}\in \{0,1\}$, $\alpha_{t}$ is a static class-specific weight parameter and $\gamma>0$ is the focusing parameter, which defines how fast the loss of well-classified examples (where $p_{t} \rightarrow 1$) should be pushed towards zero. The FL acts as a dynamically scaled variation of cross-entropy, where the modulating factor $(1-p_{t})^\gamma$ down-weights the loss contributed by simple and well-classified examples. This mechanism forces the optimiser's attention almost exclusively toward “difficult'' to classify cases. The $\alpha_{t}$ parameter provides an additional explicit weighting mechanism to balance generic class frequencies. For our VIS-only fine-tuning, we configure the focusing parameter $\gamma=2$ with class weights $\alpha_{y=1}=0.6$ and $\alpha_{y=0}=0.4$, slightly favouring class 1 (lenses). Empirically, we observed a difference in the optimal loss function between the two network variants during Stage 1 fine-tuning: while the VIS-only model benefited considerably from the application of FL, the pseudo-RGB model consistently achieved better performance using the standard CE loss.

We trained both models for 30 epochs in Stage 0. Convergence was reached in 12 epochs for the VIS-only model and 16 epochs for the pseudo-RGB model. For Stage 1 fine-tuning, models were trained for 40 epochs: the VIS-only model converged within 22 epochs while the pseudo-RGB model reached convergence in just 11 epochs. 

Conventionally, CNN training is halted (and the optimal epoch selected) by identifying the weights checkpoint that yields the lowest absolute validation loss. However, relying strictly on the raw validation loss can be misleading for strongly imbalanced classification tasks: in an imbalanced scenario, the model can achieve a low validation loss simply by pushing the predicted probabilities of the majority class (the non-lenses) extremely close to zero, even if its ability to confidently distinguish and rank the rare minority class (the lenses) remains mediocre. 

To circumvent this, we determine our final model checkpoints by tracking predictive ranking metrics rather than raw loss. Specifically, we select the epoch that maximises the area under the Precision-Recall curve (PR-AUC) evaluated on the fixed validation set. Recall is defined as the fraction of lenses that are correctly identified, whereas Precision is the fraction of predicted lenses that are, in fact, true lenses. To contextualise this choice, we must consider the astronomical objective: we aim to maximise the number of discovered lenses (measured in terms of completeness, or ‘Recall'), while ensuring that the resulting catalogue is not flooded with false positives (measured in terms of purity, or ‘Precision') even after scoring several tens of millions of objects.  

A common alternative in ML is maximising the area under the receiver operating characteristic curve (ROC-AUC), defined as
\begin{equation}
    {\rm AUC} = \int_{0}^{1} {\rm TPR(FPR)}\,{\rm dFPR}\,,
\end{equation} where TPR (true positive rate, or Recall) and FPR (false-positive rate) are calculated from the ranked scores. For rare-object searches, however, ROC curves present a mathematical blind spot: because the FPR is obtained dividing by the total number of non-lenses, a massive amount of class 0 objects will artificially shrink this rate to near-zero sizes. Consequently, a model could output hundreds of false positives, while still maintaining an overly optimistic ROC curve \citep{davis2006,saito2015}. 

\section{\label{sc:ablation-analysis}Ablation analysis}

\begin{figure*}
    \centering
    \includegraphics[width=\textwidth]{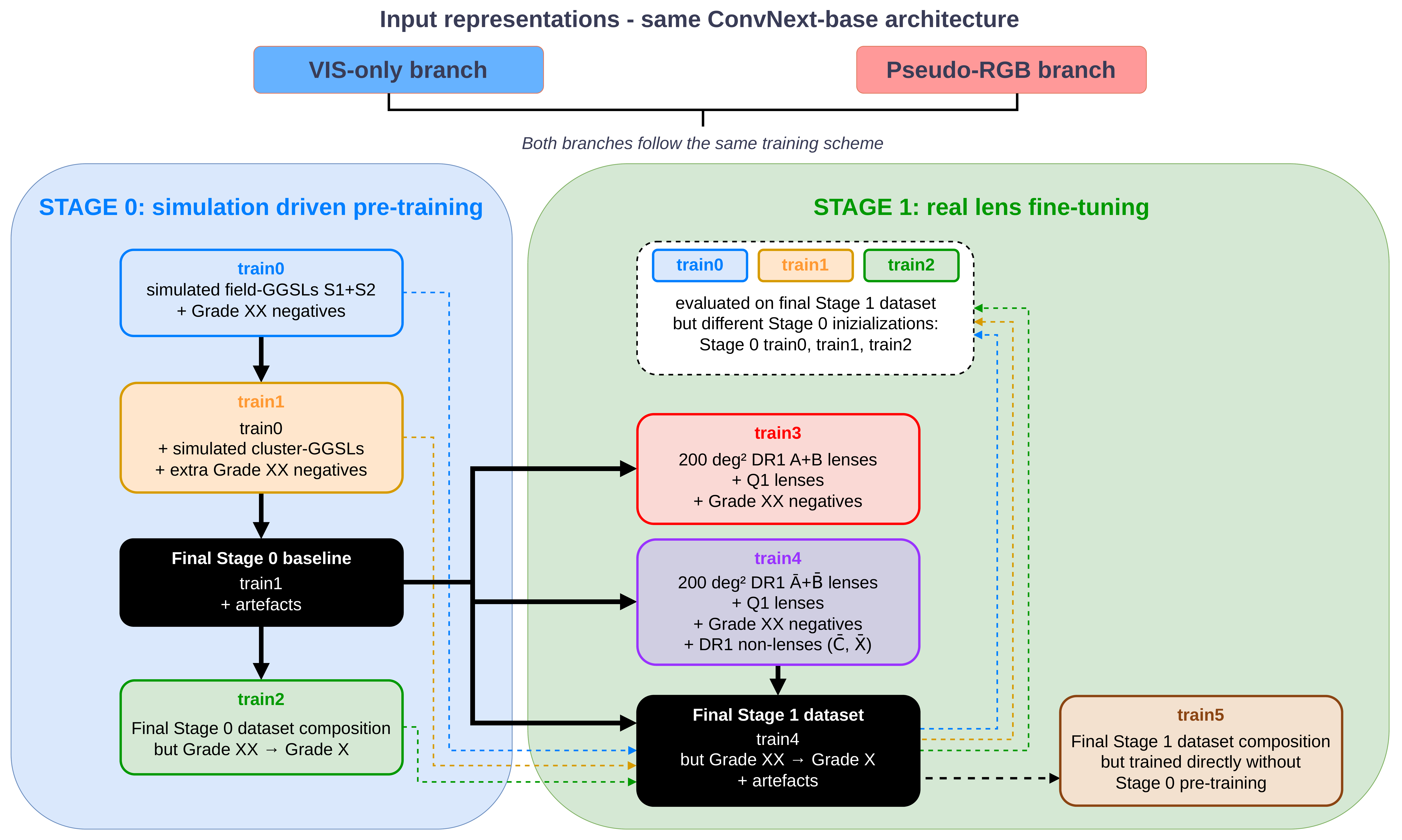}
    \caption{Schematic overview of the staged training strategy adopted for the ConvNeXt-base models. Box colours are matched to the corresponding training configurations shown in Fig.~\ref{fig:ablation_stage10_pr}.
    }
    \label{fig:training_scheme}
\end{figure*}

We now quantify how the staged training data set and the composition of the training sets affect the final ranking quality of the VIS-only and pseudo-RGB models.

The ablation study is organised into two ladders. The first one varies only the initial Stage 0 training set while keeping the later Stage 1 fine-tuning data set fixed. The second one starts from the Stage 0 configuration used for the final models and changes the Stage 1 training data set, including a direct-from-scratch Stage 1 run. A schematic overview of the ablation study is given in Fig.~\ref{fig:training_scheme}.

\subsection{Ablation metrics}
\label{ssc:ablation-metrics}

For each model in the ablation tables we report a small set of complementary ranking and forecast quantities to assess both the intrinsic ordering quality and the expected operational yield when the classifier is deployed. All reported PR‑based quantities are computed using per‑sample weights that approximate DR1 survey class balance: positive examples are left unweighted ($w_1=1$), while each of the $N_{\rm neg}=19\,999$ negative examples is assigned a weight $w_0=3300$, such that $w_0N_{\rm neg}\simeq6.6\times10^7$, approximately reproducing the number of objects scored over the DR1 footprint. The negative test sample consists of randomly selected DR1 objects visually inspected through SW during the ECLIPSE campaign, while the positive sample comprises the $15\,070$ ECLIPSE grade A and grade B systems. 
The resulting weighted PR metrics therefore account for the extreme class imbalance encountered at survey scale, while retaining a computationally feasible and defined test set for comparing the different model configurations. Infact, the reason we adopt this common test set for the ablation analysis is because evaluating all the 22 tested models configurations over the complete DR1 footprint would require the repeated execution of the full ECLIPSE data processing and inference pipeline, involving excessive computational costs for an analysis of this scope. The weighted test set metrics should therefore be interpreted as comparative measures under a DR1-like class prevalence, rather than as direct measurements of performance over the full survey. Full-survey statistics are instead presented in Sect.~\ref{sc:model-ensemble} for the final models actually deployed in the DR1 lens-search pipeline. 

The ablation metrics are defined as follows:
\begin{enumerate}[label=(\roman*)]

\item ROC‑AUC: the area under the ROC curve (see Sect.~\ref{sc:optaugandloss}) summarises rank separability across thresholds, representing the probability that our classifier gives a higher score to a positive sample than to a negative one. Due to its definition, ROC-AUC is insensitive to constant per‑class rescaling of sample counts (i.e., ROC shape is invariant to the class weights).

\item Weighted average precision ($\rm AP_{w}$): the area under the PR curve computed while accounting for the class weights ($w_0,w_1$). Concretely, $\rm AP_{w}$ is the average Precision produced by integrating $P$ as a function of Recall on the DR1-weighted PR curve. $\rm AP_w$ then ignores True Negatives (TN) and reflects the expected Precision of the ranked candidate lists at survey scale, measuring how well a model balances making highly accurate positive predictions with actually finding all available positive cases.

\item Normalised high-Recall partial area ($\mathrm{AP}_{\rm w,\mathrm{HR}}$): the average weighted precision $\rm AP_w$ over the high-$R$ interval $R\in[0.75,1.00]$, defined as
\begin{equation}
\mathrm{AP}_{\rm w,\mathrm{HR}} = \frac{1}{1-0.75}\int_{0.75}^{1.00} P_{\rm w}(R)\, {\rm d}R\,,
\end{equation}
where $P_{\rm w}(R)$ is the weighted Precision as a function of weighted Recall. This is the most relevant summary metric for our study, because it isolates the regime in which the classifier must remain highly pure while recovering most of the lenses. Unlike the full $\rm AP_{w}$, which averages performance across the entire Recall range and is therefore influenced by the easier low-Recall region, $\mathrm{AP}_{{\rm w},\mathrm{HR}}$ concentrates on the part of the curve that matters most for candidate generation. We therefore use it as the primary quantity for comparing ablations and for motivating the choice of the final models.

\item Precision at top-\textit{K} (P@\textit{K}) and True Positives at top-\textit{K} (TP@\textit{K}): these two columns summarise the expected Precision and the expected number of True Positives (TP) among the top-\textit{K} scored cutouts. We set $K=70\,000$ as this is approximately the number of candidates from our ensemble model selected for the final SW and GJ visual inspections \citep{DR1-SL02}. P@\textit{K} is the weighted Precision among the top-\textit{K} entries (i.e., TP divided by weighted total selected), and TP@\textit{K} is the corresponding weighted TP count (weights act as a count proxy, so TP@\textit{K} is interpretable as the expected number of TP at survey scale contained in the top‑\textit{K} candidate list). These fixed-K weighted forecasts should be distinguished from the directly measured $K=68\,021$ recovery over the complete ECLIPSE score catalogue reported in Sect.~\ref{sc:model-ensemble}.

\item Precision at fixed Recall ($\text{P@R}\in{0.50, 0.75, 0.90}$): these are the weighted Precision values read from the weighted PR curve at fixed Recall levels $R\in{0.50,0.75,0.90}$. We obtain these by interpolating the weighted PR curve so that ${\rm P@R}=r$ is the Precision corresponding to Recall $r$. We focus on relatively high-$R$ targets because the operational goal is to recover the majority of lenses while keeping the candidate list as pure as possible. Reporting $P$ at multiple high-$R$ points makes explicit the trade‑off between purity and completeness.
\end{enumerate}

Given the reduced number of negatives available in our test set, the ROC curves are close to saturation across all figures, so the weighted PR curves are the more discriminative visual diagnostic. In the discussion below, percentage changes are interpreted primarily through $\mathrm{AP}_{{\rm w},\mathrm{HR}}$, since this is the metric that best reflects the operating regime used to select the final models, complemented by precision at fixed Recall and by top-$K$ recovery forecasts. To assess the sensitivity of the reported metrics to the assumed survey prevalence, we repeated the evaluation using negative-class weights of $w_0=3000$ and $3600$, corresponding to variations of approximately $\pm\,9\%$ around the adopted value, while keeping $w_1=1$ fixed. Although the absolute weighted Precision and average Precision values change as expected, the relative ordering of all Stage 1 configurations in $\mathrm{AP}_{{\rm w},\mathrm{HR}}$ remains unchanged for both VIS-only and Pseudo-RGB models. The final Stage 1 models also retain the highest $\text{P@\textit{K}}$ across the tested weights. These results indicate that our main conclusions regarding high-Recall ranking performance are robust to moderate variations in the assumed class prevalence.

 All models computed metrics can be found in Table~\ref{tab:ablation_stage0_metrics_paper} and Table~\ref{tab:ablation_stage10_metrics_paper} for Stage 0 and Stage 1 trainings respectively.

\begin{table*}[t]
\centering
\caption{Baseline Stage-0 ablation metrics on the DR1 test set. }
\label{tab:ablation_stage0_metrics_paper}
\begin{tabular}{llcccccccc}
\hline\hline
Train & Modality & ROC-AUC & $\rm AP_{w}$ & $\rm AP_{w,\mathrm{HR}}$ & P@\textit{K} & TP@\textit{K} & $\rm P@R=0.50$ & $\rm P@R=0.75$ & $\rm P@R=0.90$ \\
\hline
final  & pseudo-RGB & 0.9790 & 0.2681 & 0.0112 & 0.1035 & 7248  & 0.0900 & 0.0293 & 0.0061 \\
train0 & pseudo-RGB & 0.9660 & 0.1498 & 0.0059 & 0.0816 & 5715  & 0.0419 & 0.0161 & 0.0030 \\
train1 & pseudo-RGB & 0.9574 & 0.3437 & 0.0037 & 0.1140 & 7977 & 0.1489 & 0.0129 & 0.0017 \\
train2 & pseudo-RGB & 0.9721 & 0.4052 & 0.0070 & 0.1094 & 7660 & 0.1317 & 0.0191 & 0.0036 \\
\hline
final & VIS     & 0.9858 & 0.3648 & 0.0179 &  0.1324 & 9271 & 0.2018 & 0.0547 & 0.0088  \\
train0 & VIS &  0.9678 & 0.2773 & 0.0101 & 0.1267 & 8872 & 0.1714 & 0.0330 & 0.0042 \\
train1 & VIS    & 0.9747 & 0.2947 & 0.0131 & 0.1043 & 7300 &  0.0900 & 0.0358 & 0.0060 \\
train2 & VIS & 0.9890 & 0.2918 & 0.0198 & 0.1281 & 8968 &  0.1853 & 0.0499 & 0.0122 \\
\hline\hline
\end{tabular}
\par\smallskip
\tablefoot{All values are computed with sample weights $w_0=3300$ for class 0 and $w_1=1$ for class 1. $\rm AP_{w}$ is the weighted average Precision over the full Recall range, while $\rm AP_{w,\mathrm{HR}}$ is the normalised weighted partial area under the PR curve restricted to the high-$R$ interval $R\in[0.75,1.00]$. $\text{P@R}\in{0.50, 0.75, 0.90}$ are the weighted Precision values interpolated from the weighted PR curve at Recall $0.50$, $0.75$, and $0.90$. ROC-AUC is reported for completeness. P@\textit{K} is the weighted precision among the top \textit{K}$=70\,000$ scored cutouts (as a forecasted candidate list, with weights acting as count proxies), and TP@\textit{K} is the corresponding weighted number of TP.}
\end{table*}

\subsection{\label{sc:stage0-models}Stage 0 models}

\begin{figure*}[t]
\centering
\includegraphics[width=0.49\textwidth]{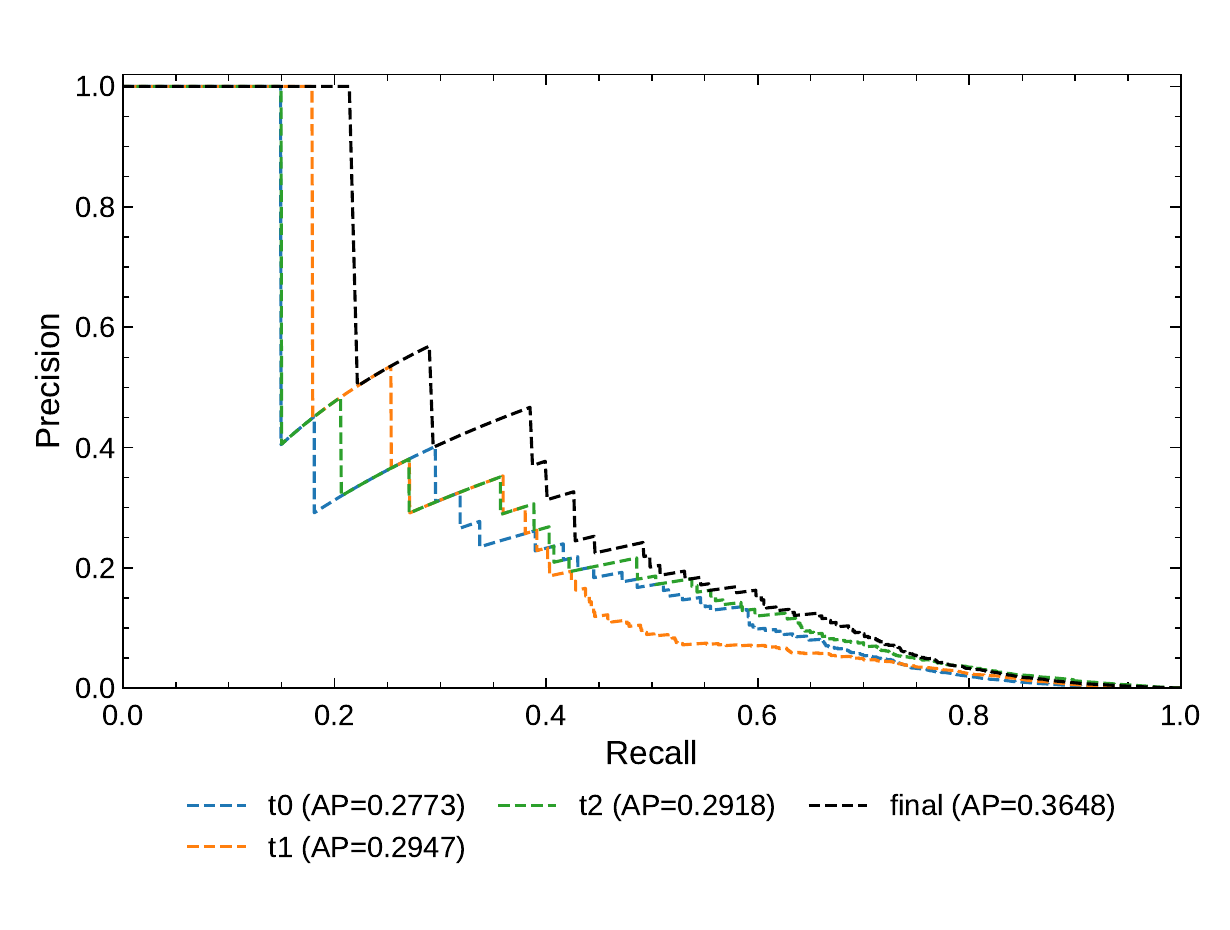}
\includegraphics[width=0.49\textwidth]{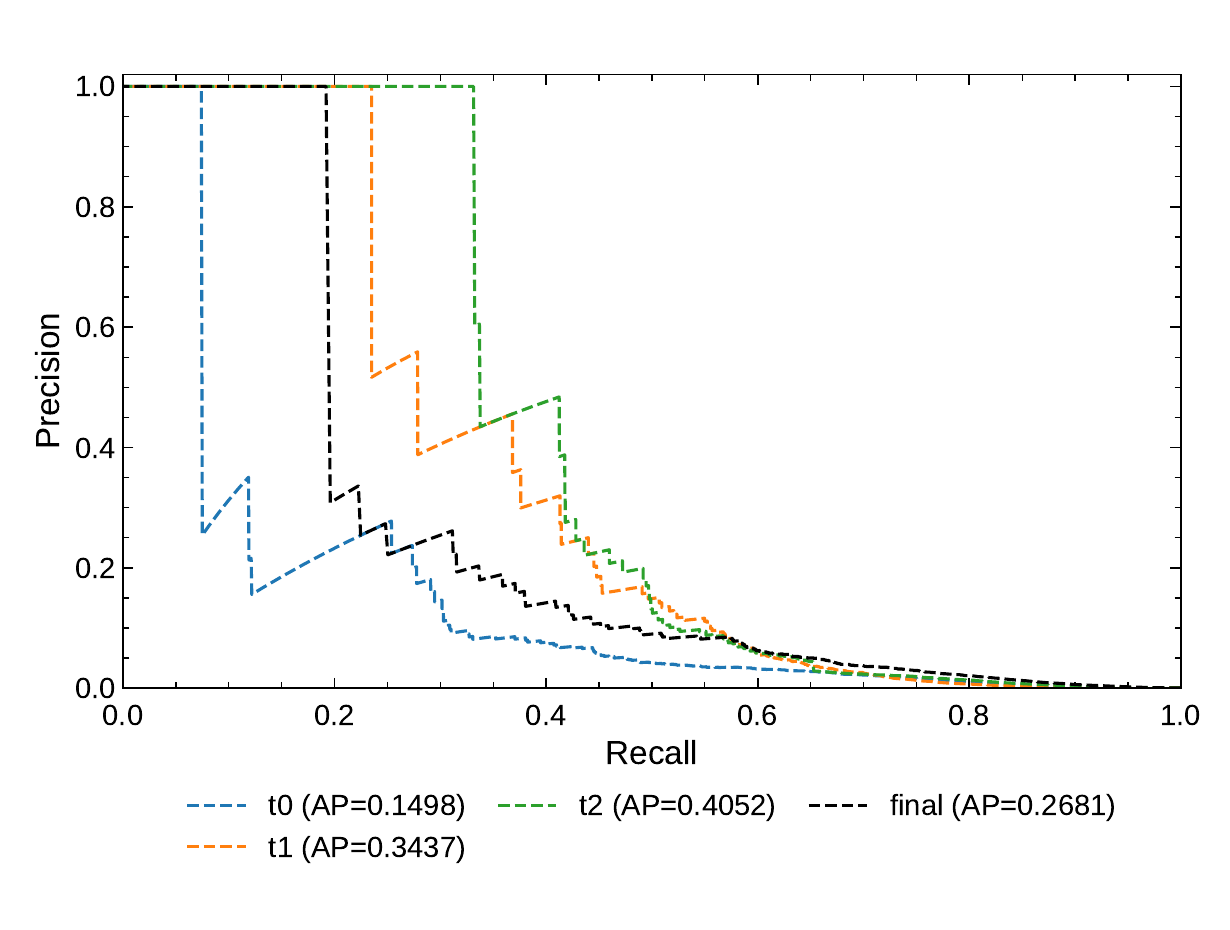} \\
\includegraphics[width=0.49\textwidth]{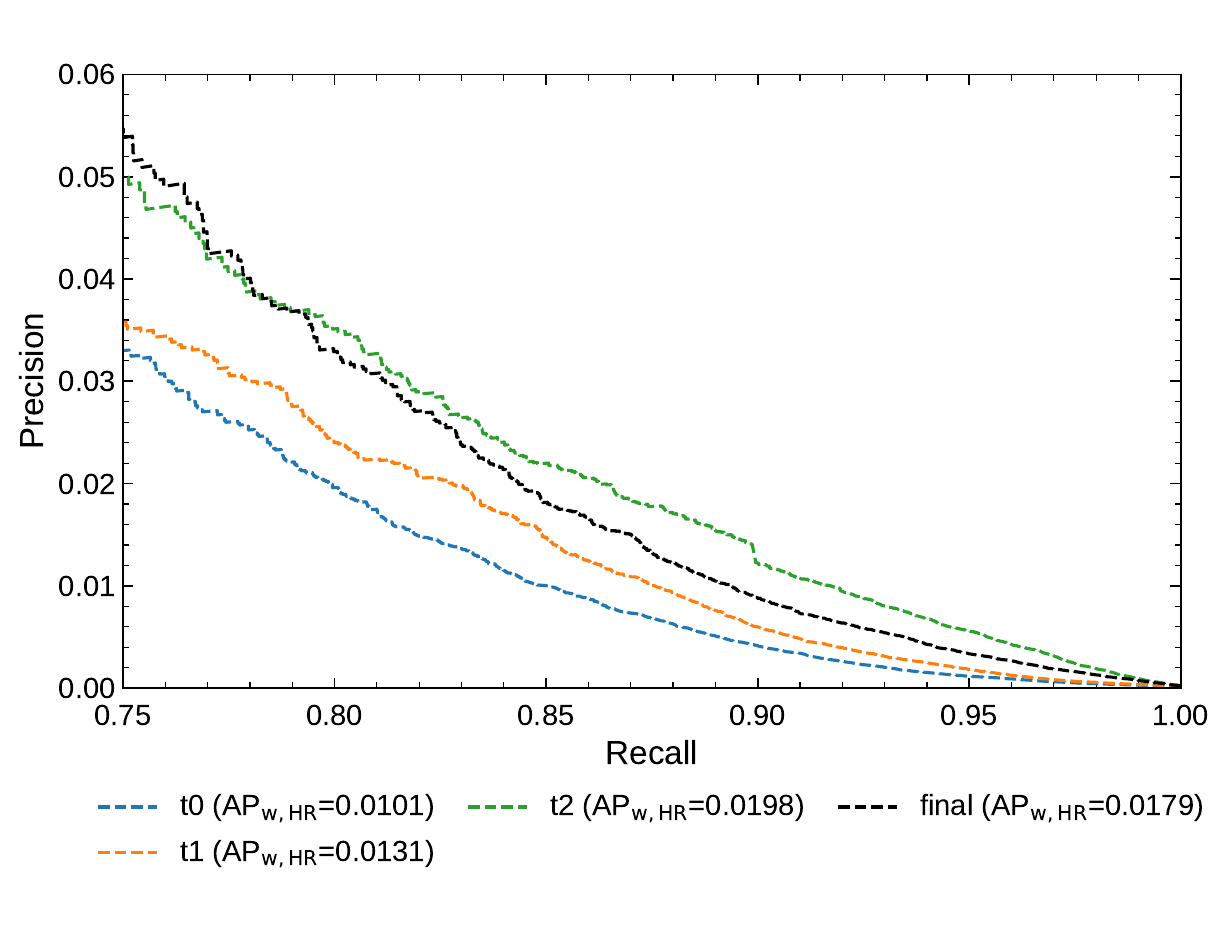}
\includegraphics[width=0.49\textwidth]{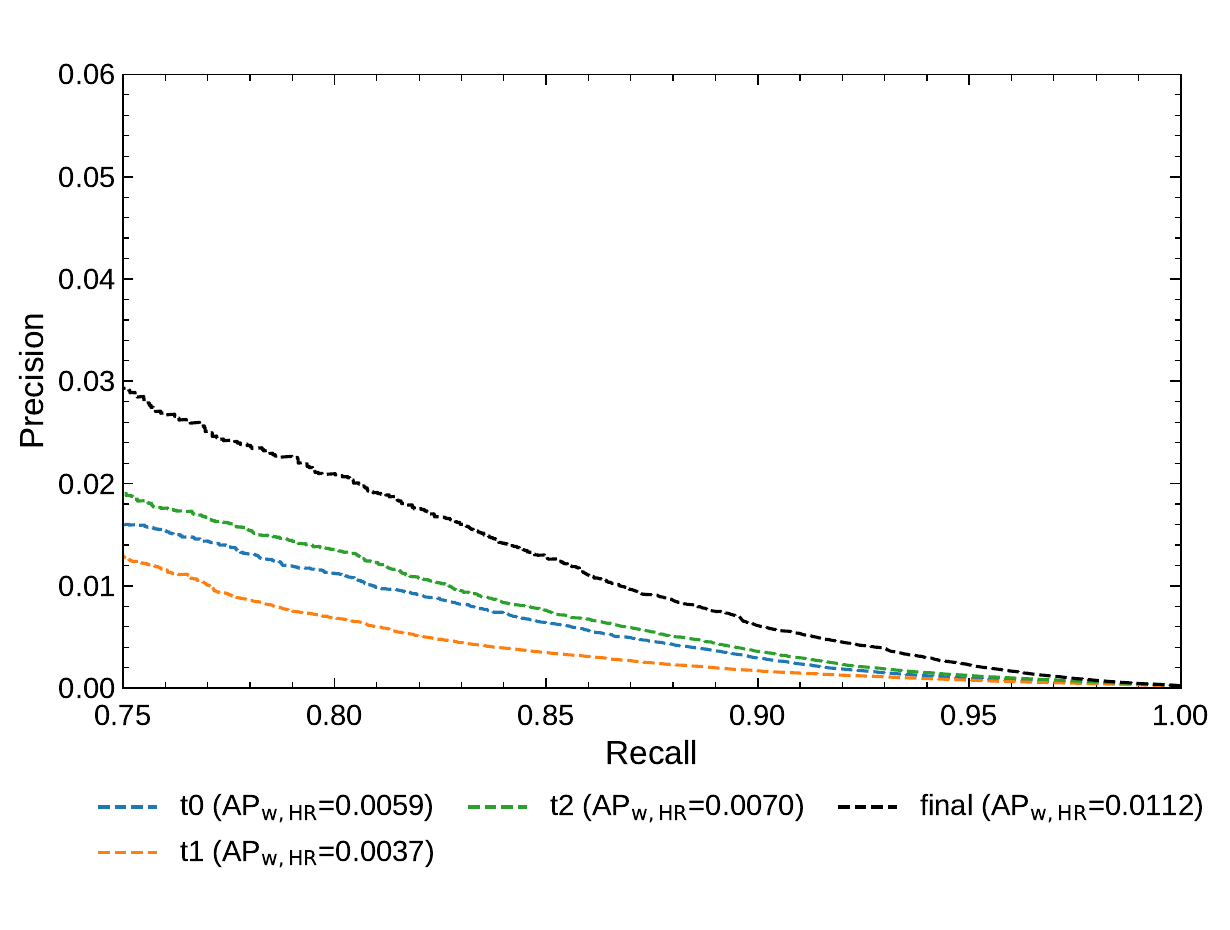}
\caption{\emph{Upper panels}: baseline Stage 0 weighted PR curves for the VIS-only model family (left) and the Pseudo-RGB family (right). Legends report the weighted average Precision. \emph{Lower panels}: same as the upper panels but zoomed between $R=0.75$ and $R=1.00$. Legends report the normalised weighted average Precision for $R \in [0.75,1.00]$. The high-$R$ zoom shows how some training sets preserve Precision once Recall rises above ${\sim}\,0.75$, which is exactly the regime that matters for candidate generation.}
\label{fig:ablation_baseline_pr}
\end{figure*}

\begin{figure*}[t]
\centering
\includegraphics[width=0.49\textwidth]{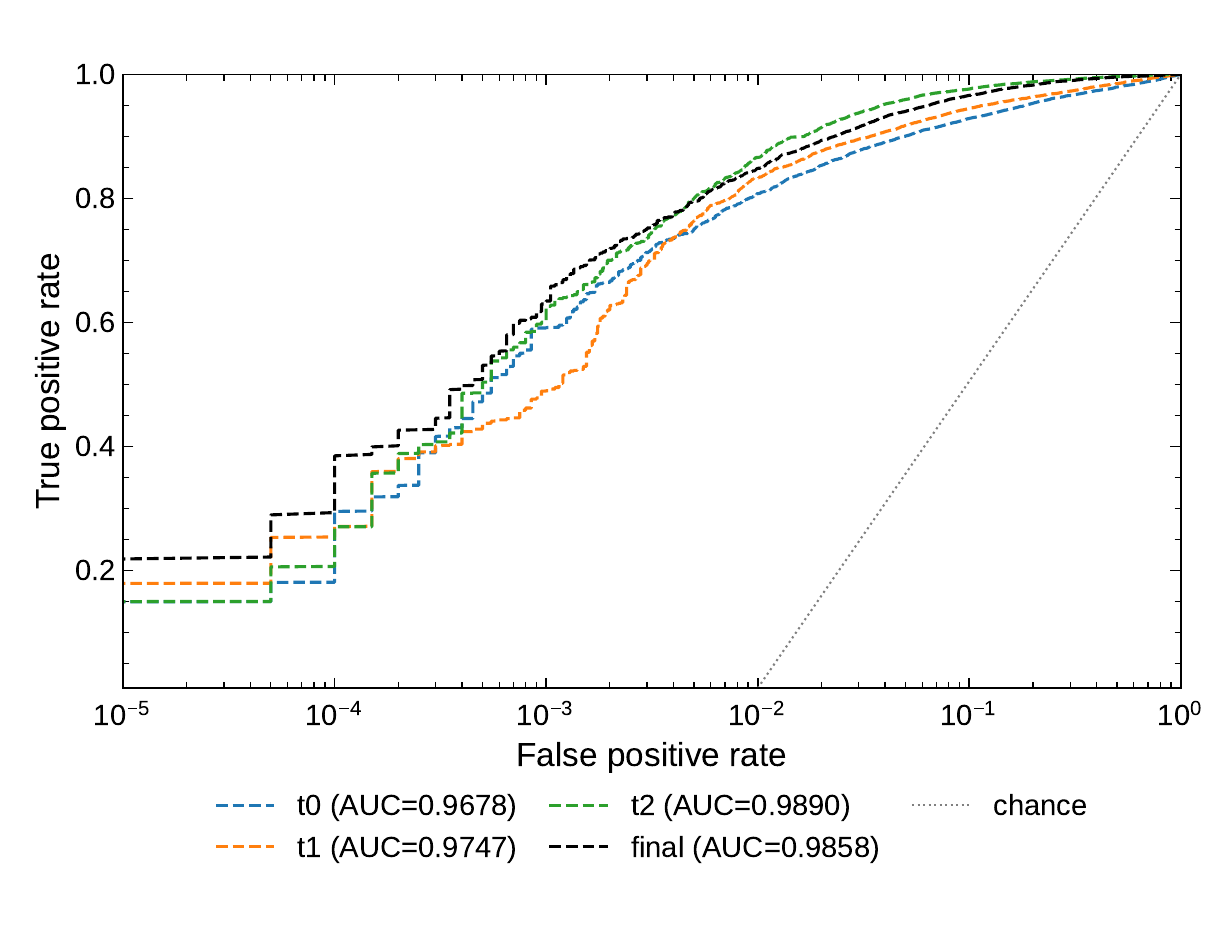}
\includegraphics[width=0.49\textwidth]{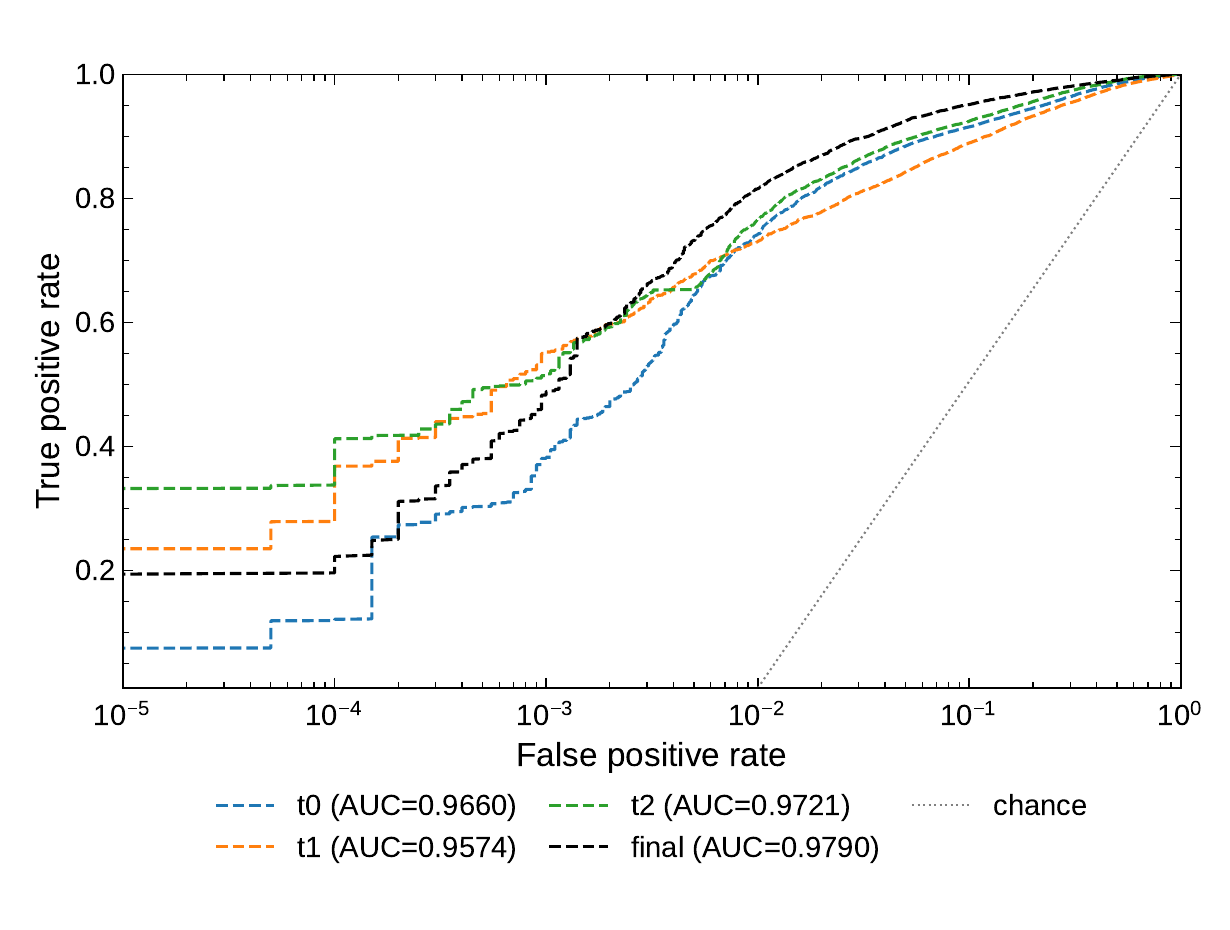}
\caption{Baseline Stage 0 ROC curves for the VIS-only model family (left) and the Pseudo-RGB family (right). The curves are concentrated near the upper-left corner, which means the ranking problem is already fairly strong even before the later Stage 1 refinement. The more interesting separation between ablations happens in the low-FPR region, where small changes in the leftmost part of the curve correspond to non-trivial changes in the FP burden at survey scale.}
\label{fig:ablation_baseline_roc}
\end{figure*}

\begin{figure*}[t]
\centering
\includegraphics[width=0.49\textwidth]{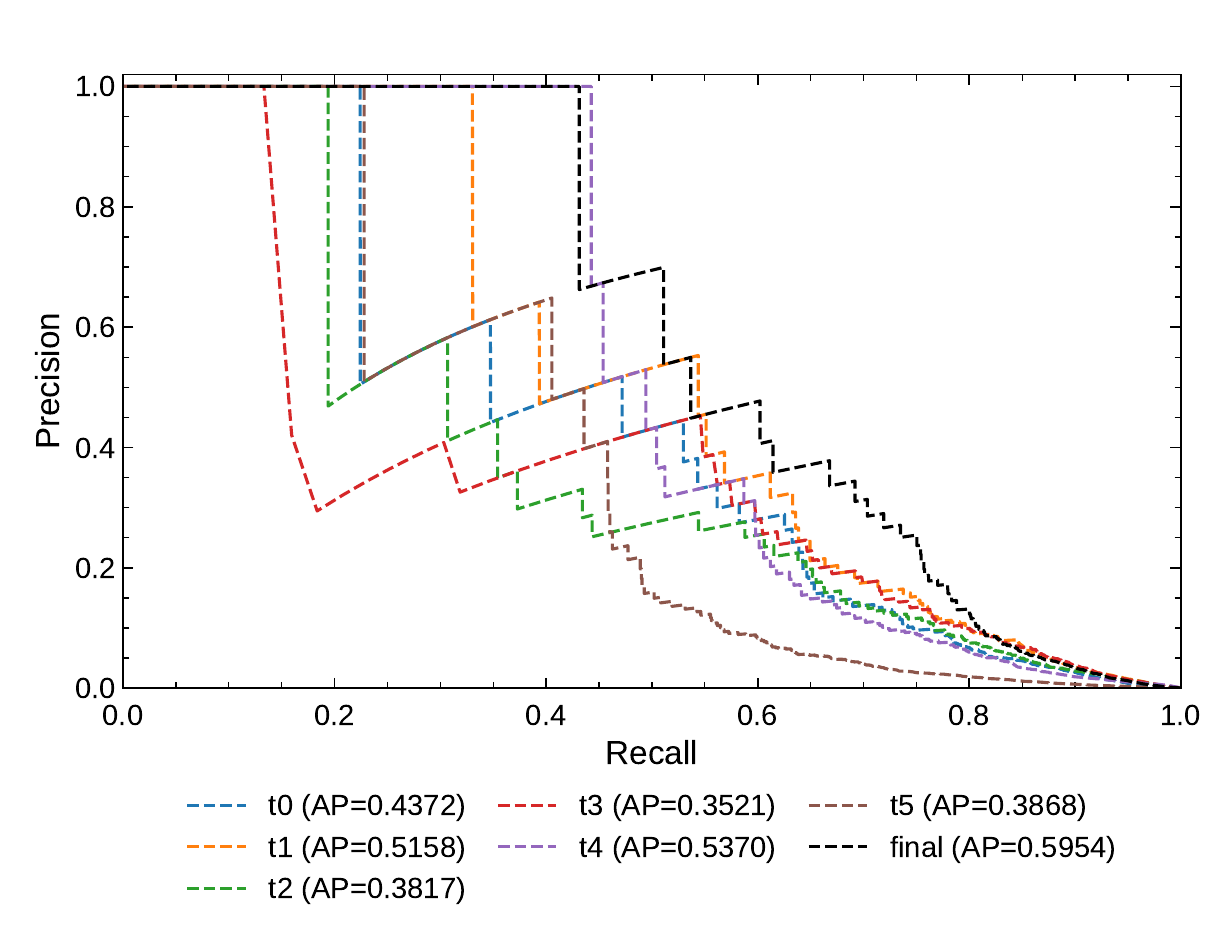}
\includegraphics[width=0.49\textwidth]{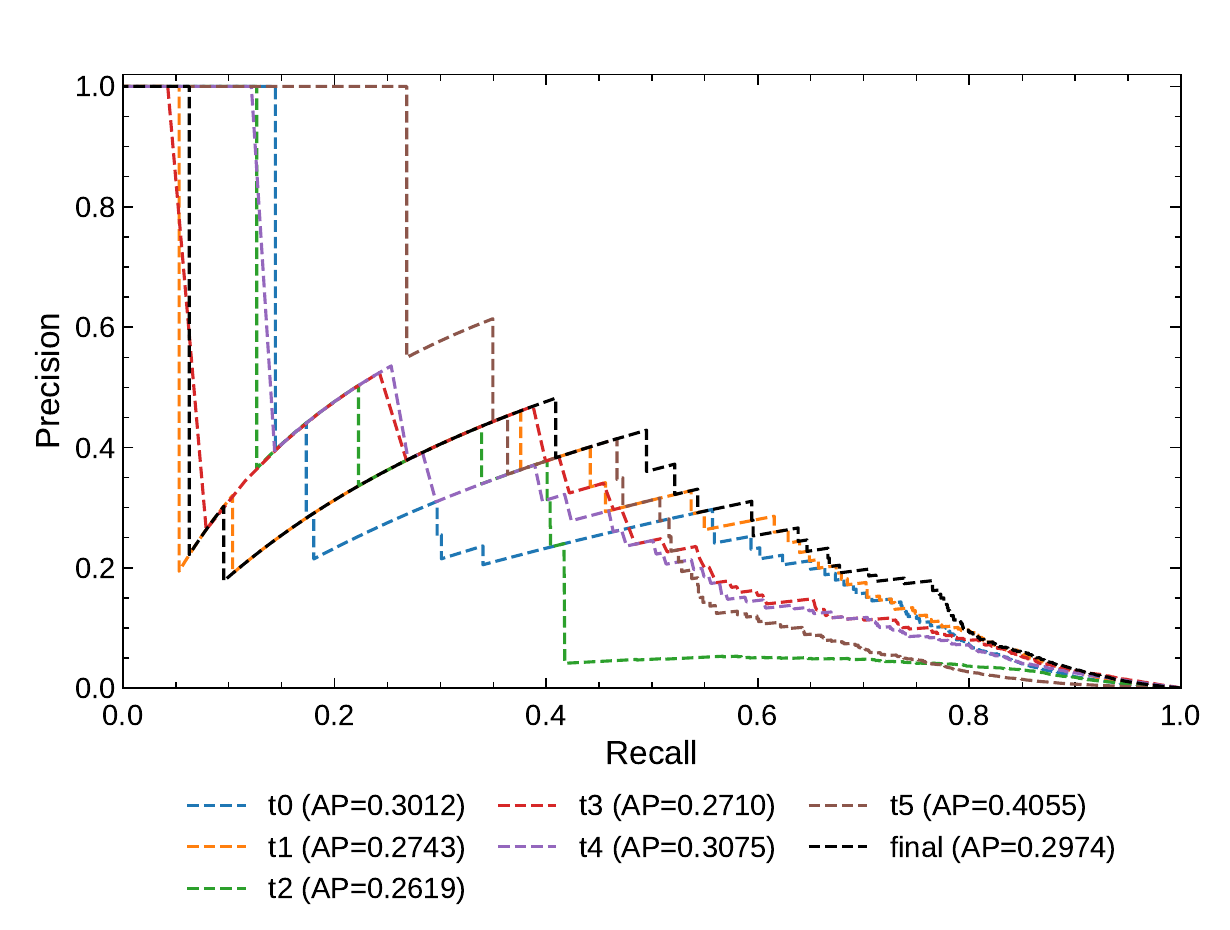} \\
\includegraphics[width=0.49\textwidth]{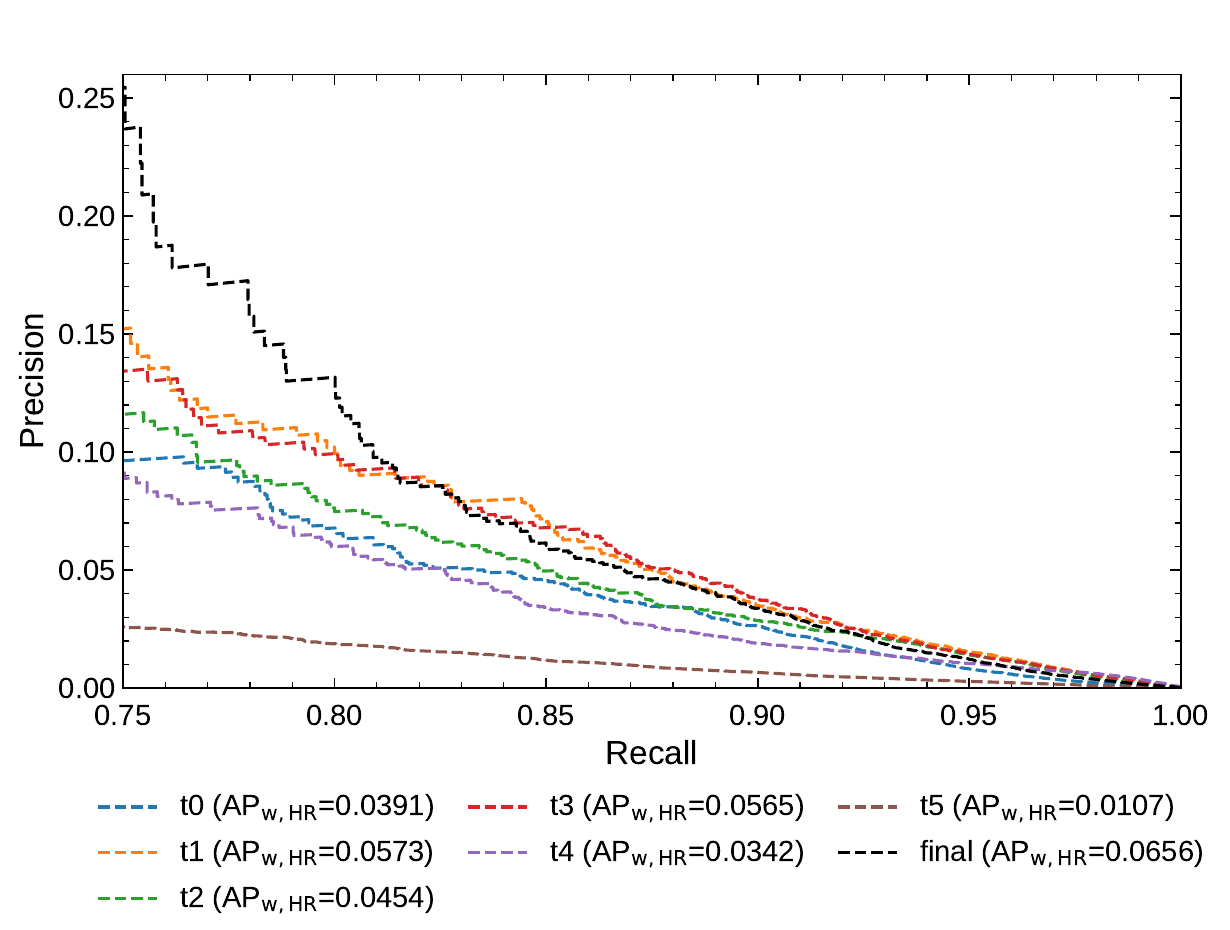}
\includegraphics[width=0.49\textwidth]{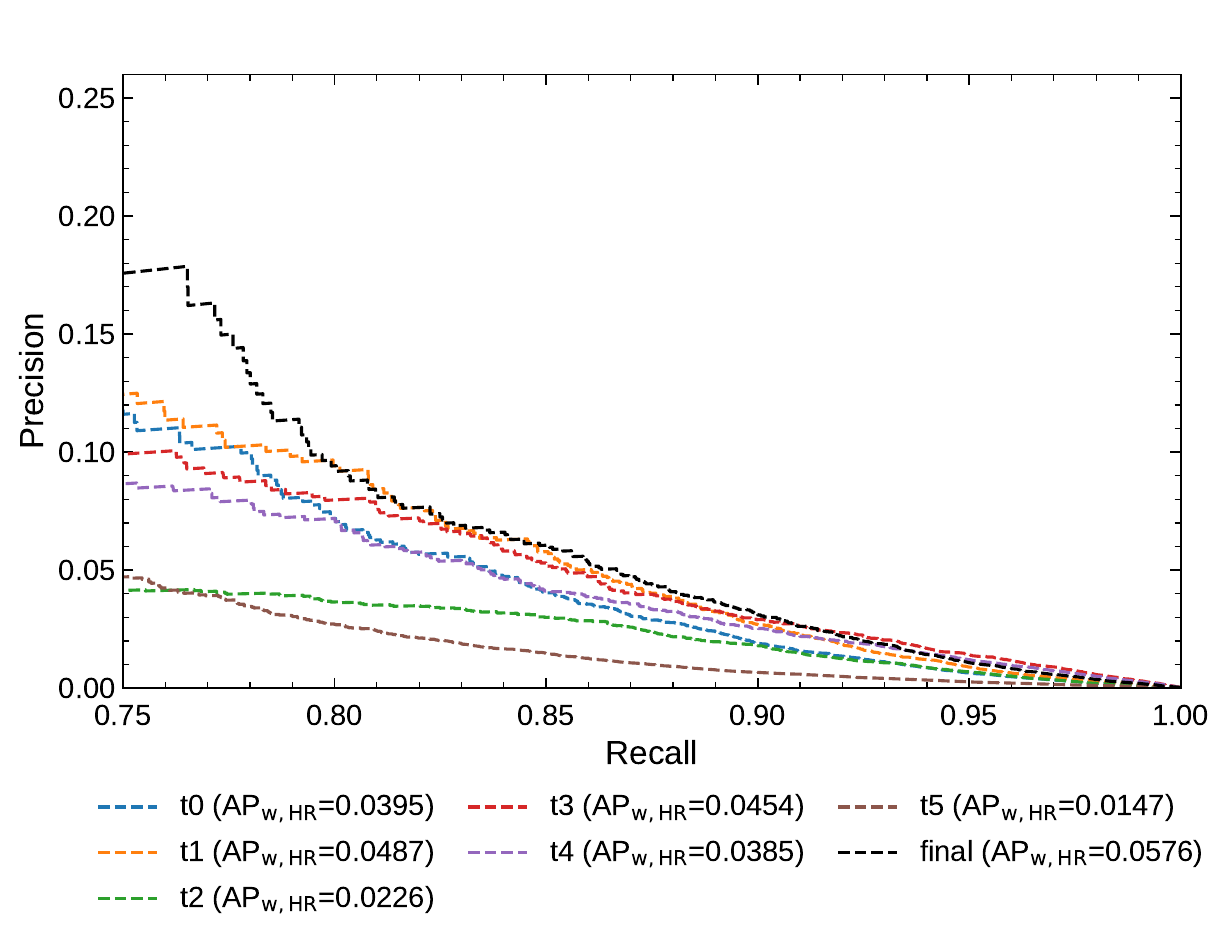} 
\caption{\emph{Upper panels}: Stage 1 weighted PR curves for the VIS-only model family (left) and the Pseudo-RGB family (right). Legends report the weighted average Precision. The curves separate most clearly at high Recall, where the practical operating point is determined by how much precision can be retained while still recovering a large fraction of the lenses. \emph{Lower panels}: same as the upper panels but zoomed between $R=0.75$ and $R=1.00$. Legends report the normalised weighted average Precision for $R \in [0.75,1.00]$. The high-$R$ zoom shows that VIS is comparatively robust under several data-set changes, while the Pseudo-RGB branch can degrade sharply if the Stage 1 training set is altered.}
\label{fig:ablation_stage10_pr}
\end{figure*}

\begin{figure*}[t]
\centering
\includegraphics[width=0.49\textwidth]{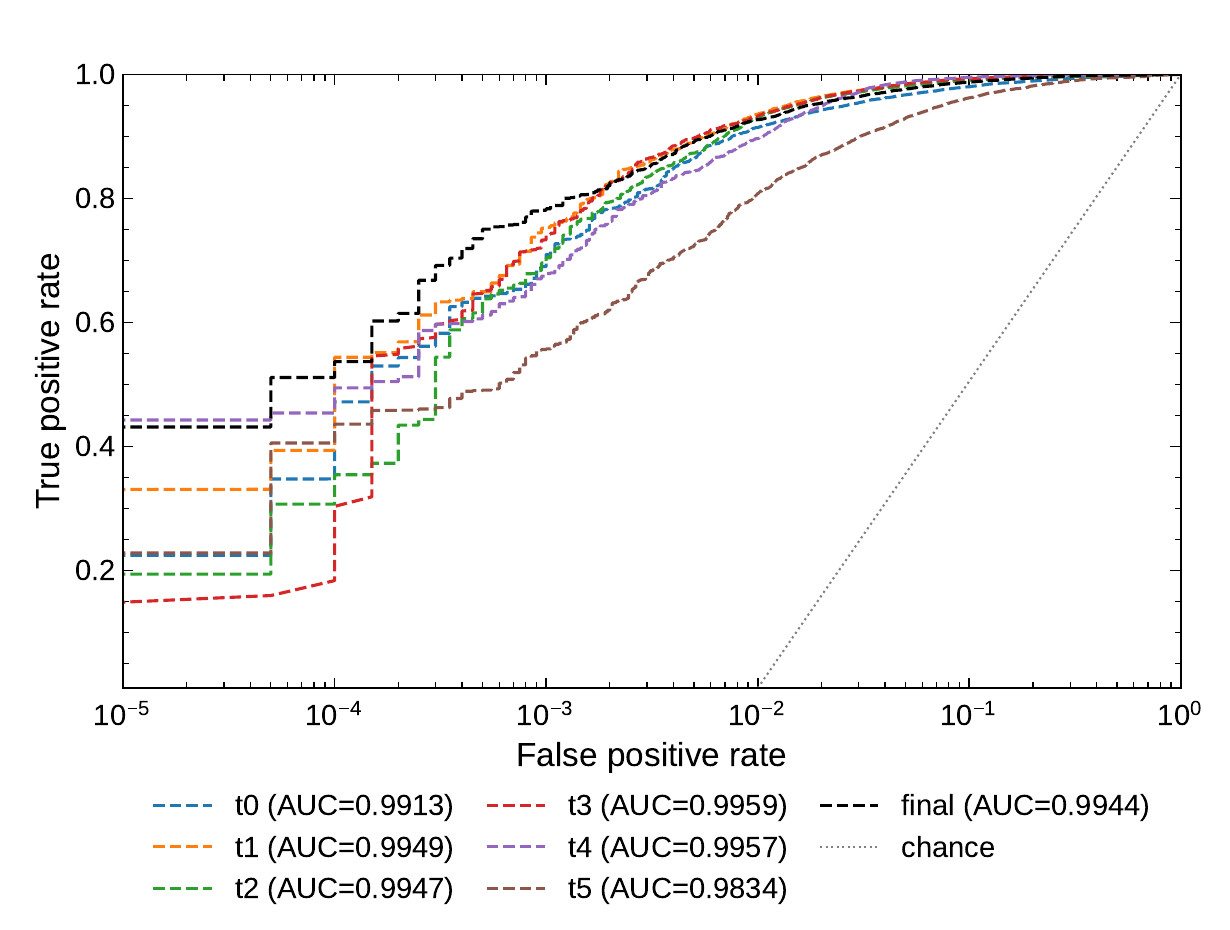}
\includegraphics[width=0.49\textwidth]{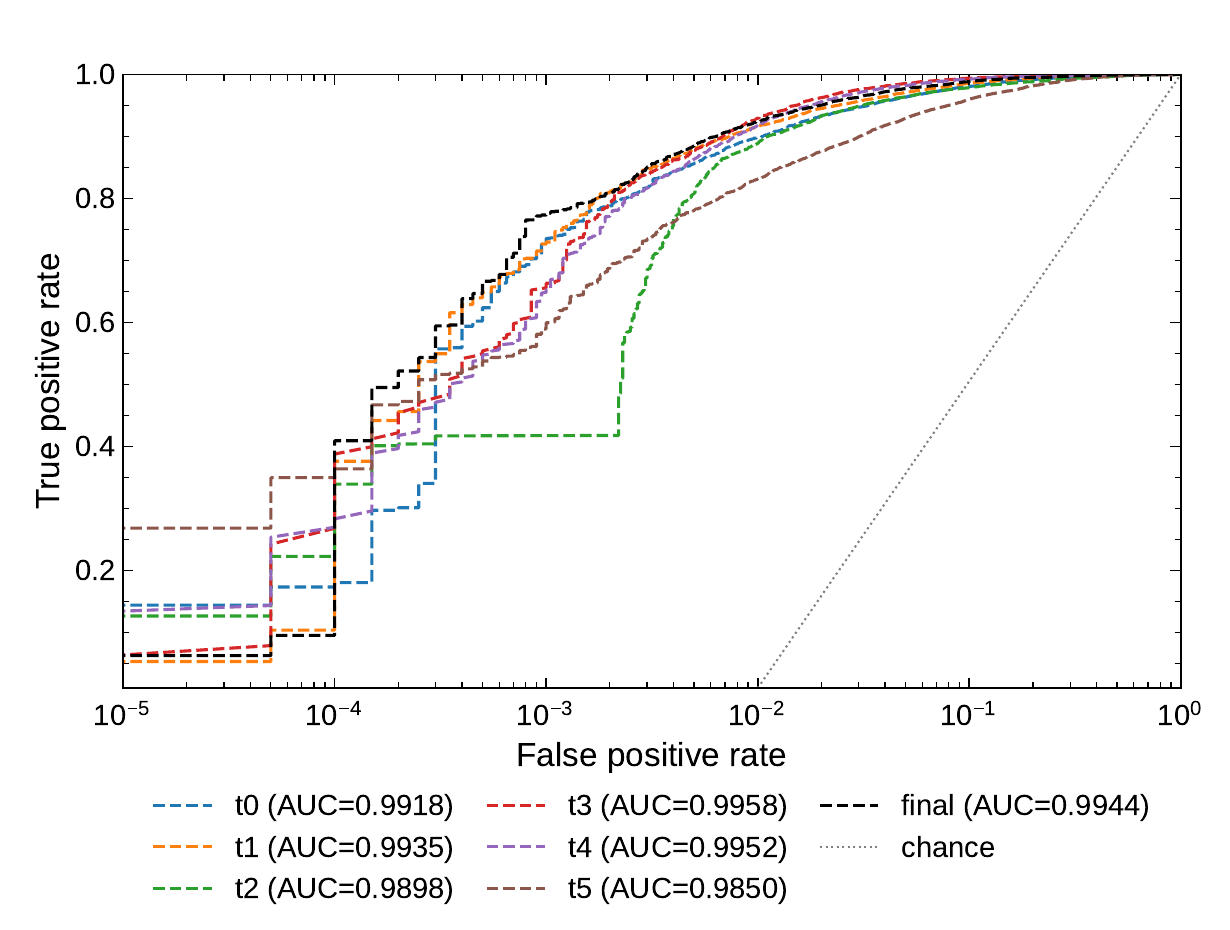}
\caption{Stage 1 ROC curves for the VIS-only model family (left) and the pseudo-RGB family (right). As in the baseline case, the ROC curves are highly compressed near the upper-left corner, which means that ROC-AUC alone does not separate the ablations cleanly. The low-FPR zoom is the more informative version of this figure because it exposes how much false-positive mass is accumulated before the curve reaches the high-TPR regime.}
\label{fig:ablation_stage10_roc}
\end{figure*}

The Stage 0 ablations isolate the effect of the initial training pool before any Stage 1 fine-tuning is applied. Train0 uses S1 and S2 simulations as positives and grade XX objects as negatives. Train1 adds cluster-GGSL simulations and an equal number of negatives to preserve class balance. The final Stage 0 baseline models (see Table~\ref{tab:dataset_composition}) add artefacts on top of this training data set and are the configurations carried forward as the reference starting point for the Stage 1 study. Train2 then partially replaces the grade XX negatives in the final Stage 0 baseline set with the harder grade X negatives, making the discrimination task more realistic and substantially more demanding. A detailed summary of Stage 0 ablation set compositions is given in Table~\ref{tab:stage0sets}.

\renewcommand{\arraystretch}{1.2}
\begin{table}[ht]
    \centering
    \caption{Summary of Stage 0 ablation-set compositions}
    \label{tab:stage0sets}
    \begin{tabular}{lccc}
        \hline\hline
        \textbf{Data source} & \textbf{train0} & \textbf{train1} & \textbf{train2} \\
        \hline
        \multicolumn{4}{l}{\textit{Lenses}} \\
        Cluster-lens simulations & -- & 5011 & 5011 \\
        Field-lens simulations (S1) & 2585 & 2585 & 2585 \\
        Field-lens simulations (S2) & 3737 & 3737 & 3737 \\
        200\,$\rm deg^2$ DR1 area lenses & -- & -- & -- \\
        Q1 area lenses & -- & -- & -- \\
        \hline
        \textbf{Total lenses} & \textbf{6322} & \textbf{11\,333} & \textbf{11\,333} \\
        \hline
        \multicolumn{4}{l}{\textit{Non-lenses}} \\
        Hard negatives (grade `XX') & 6322 & 11\,333 & 4600 \\
        Hard negatives (grade `X') & -- & -- & 4200 \\
        DR1 non-lenses ($\rm \bar{C}, \bar{X}$) & -- & -- & -- \\
        Artefacts & -- & -- & 2496 \\
        \hline
        \textbf{Total non-lenses} & \textbf{6322} & \textbf{11\,333} & \textbf{11\,296} \\
        \hline\hline
        \textbf{Total objects} & \textbf{12\,644} & \textbf{22\,666} & \textbf{22\,629} \\
        \hline\hline
    \end{tabular}
\end{table}

\subsubsection{train0: baseline VIS-only and pseudo-RGB models}
The baseline train0 configuration is the simplest Stage 0 setting, as reflected in its low $\mathrm{AP}_{{\rm w},\mathrm{HR}}$. With the simple grade XX negative pool, the pseudo-RGB branch reaches only $\mathrm{AP}_{{\rm w},\mathrm{HR}}=0.0059$, while the VIS-only branch reaches $0.0101$. These values are $47.3\%$ and $43.5\%$ below the final Stage 0 references, respectively. The weighted PR curves (see Fig.~\ref{fig:ablation_baseline_pr}) nevertheless show substantial room for improvement: $\rm AP_{w}$ is $0.1498$ for RGB and $0.2773$ for VIS. Relative to the final baseline reference, train0 is $44.1\%$ lower and $23.9\%$ lower in $\rm AP_{w}$ for the Pseudo-RGB and VIS-only models respectively.

For Pseudo-RGB, the precision at $R=0.75$ is only $0.0161$, showing that the candidate list becomes heavily contaminated once the model is asked to recover most of the lenses. VIS behaves somewhat better ($\rm P@R_{0.75} = 0.0330$), but the high-$R$ area is still clearly below the final baseline. 

Visually, the ROC curves (see Fig.~\ref{fig:ablation_baseline_roc}) already show that even the baseline train0 models retain a substantial ranking performance, so our conclusion is not that train0 fails outright, but that the training pool is less efficient in producing a sharply ordered ranking once the candidate list becomes long, in the low-FPR region. We interpret this as an indication that the grade XX negative pool is insufficient in preserving purity once Recall is pushed toward the operating regime of interest.

\subsubsection{train1: adding simulated cluster-GGSLs}
Adding cluster-GGSL simulations and an equal number of negatives to train0 increases sample diversity, with model-dependent effects.

For Pseudo-RGB, $\mathrm{AP}_{{\rm w},\mathrm{HR}}$ decreases to $0.0037$, $37.3\%$ below train0 and $66.9\%$ below the final baseline. Nevertheless, train1 achieves the highest Precision among the top $K=70\,000$ candidates of all Stage 0 Pseudo-RGB configurations ($\text{P@\textit{K}}=0.1140$), recovering the most true positives ($\text{TP@\textit{K}}=7977$). By contrast, $\mathrm{AP}_{\rm w}$ increases from 0.1498 to 0.3437 ($+\,129.4\%$). This contrasting behaviour shows that even if the additional cluster simulations are effective at prioritising a substantial fraction of lenses within the first $70\,000$ candidates, train1 Precision deteriorates more rapidly than the other ablations as Recall approaches unity.

VIS-only responds differently to the cluster-GGSL simulations. Its high-$R$ weighted average precision increases from $\mathrm{AP}_{{\rm w},\mathrm{HR}}=0.0101$ to $0.0131$, $29.7\%$ above train0 but $26.8\%$ below the final Stage 0 reference. The overall weighted average precision also moderately increases from $\mathrm{AP}_{\rm w}=0.2773$ to $0.2947$ ($+\,6.3\%$). The additional cluster simulations therefore benefit both models, but in different regions of the PR curve.

This contrast suggests that the two representations respond differently to the morphological complexity of cluster environments. For the VIS-only model, exposure to more diverse lens configurations may improve its ability to recognise lensing features in crowded environments, particularly at high completeness. Conversely, the Pseudo-RGB model can exploit colour information to distinguish lenses from contaminants, potentially explaining its substantial improvement in $\mathrm{AP}{\rm w}$. However, its decreased high-$R$ performance indicates that the additional simulations do not improve discrimination uniformly across the PR curve. This reinforces the need to evaluate the operating regime relevant to DR1. Nevertheless, since train1 simultaneously increases the number of training examples and introduces cluster examples (while keeping the same label ratio of train0), this ablation does not disentangle the effects of sample size and morphological diversity: a size-matched experiment would be required for a stricter causal attribution.

\subsubsection{Final Stage 0 baseline models}
The final Stage 0 baselines add artefacts to train1 and serve as the reference initialisations for Stage 1 (train3 and train4, see Sect.~\ref{sc:stage1-models}). Although train2 is the strongest Stage 0 choice in $\mathrm{AP}_{{\rm w},\mathrm{HR}}$ for VIS and remains competitive for Pseudo-RGB (see Sect.~\ref{sc:train2}), the final Stage 0 models are retained because they are the most balanced and operationally robust starting point for the Stage 1 experiments, as exploiting grade X hard negatives already from Stage 0 does not translate into an higher performance in Stage 1 (see Table~\ref{tab:ablation_stage10_metrics_paper}, where train2 results after Stage 1 training do not show competitive advantages over other ablations in $\mathrm{AP}_{{\rm w},\mathrm{HR}}$ and $\rm AP_{w}$, especially with respect to the final Stage 0 baselines).

For VIS, the final baseline is the strongest Stage 0 configuration in $\rm AP_{w}$ ($0.3648$) and is better than train2 precision at both top-\textit{K} and high-$R$ (except at Recall $>0.90$, where the difference becomes more substantial). For Pseudo-RGB, the final baseline is more conservative than train2 in $\rm AP_{w}$ but gives the best $\mathrm{AP}_{{\rm w},\mathrm{HR}}$ among the Stage 0 Pseudo-RGB runs (see Table~\ref{tab:ablation_stage0_metrics_paper}), and also provides a balanced trade-off: $\text{P@\textit{K}}=0.1035$ (5.4\% lower than train2), and the best weighted Precision at $R=0.90$ of $0.0061$. 

\subsubsection{\label{sc:train2}train2: introducing grade X hard negatives in Stage 0 final trainset}

Train2 investigates the effect of replacing the grade XX negatives in the final Stage 0 training set (see Sect.~\ref{sc:stage0}) with grade X hard negatives, exposing the models to more lens-like contaminants. The effects depend on the representation and Recall regime.

For Pseudo-RGB, train2 achieves the highest Stage 0 $\mathrm{AP}_{\rm w}$ ($0.4052$), $51.1\%$ above the final baseline reference. It also improves Precision among the top-$K$ candidates, reaching $\text{P@\textit{K}}=0.1094$, approximately $5.7\%$ above the final baseline reference. However, performance deteriorates at high Recall: its $\mathrm{AP}_{{\rm w},\mathrm{HR}}$ remains $37.5\%$ below the final baseline value (0.0070 versus 0.0112). Indeed, Fig.~\ref{fig:ablation_baseline_pr} shows that train2 maintains the highest Precision over much of the low and intermediate Recall range, reaching perfect Precision up to $R\simeq0.35$, but at higher Recall the final baseline becomes more competitive, particularly beyond $R\simeq0.65$, and at $R=0.90$ the final baseline reaches a precision of $0.0061$, compared with $0.0036$ for train2. These results indicate that hard-negative training strongly improves the early ranking of Pseudo-RGB candidates, but does not translate into a corresponding improvement in the high-$R$ regime.

The VIS-only branch behaves differently: train2 reaches $\mathrm{AP}_{{\rm w},\mathrm{HR}}=0.0198$, the highest value among the Stage 0 VIS configurations and $10.6\%$ above the final baseline reference. This gain is clearest at high-$R$ (see Fig.~\ref{fig:ablation_baseline_pr}), where train2 begins to outperform the final baseline at $R\simeq0.8$ and maintains higher precision towards $R=1$. At $R=0.90$, its Precision reaches $0.0122$, compared with $0.0088$ for the final baseline, corresponding to a $38.6\%$ improvement. Its $\text{P@\textit{K}}=0.1281$ also remains close to the highest Stage 0 value of $0.1324$, although its $\mathrm{AP}_{\rm w}$ is $20\%$ lower than that of the final baseline (0.2918 against 0.3648). Thus, unlike Pseudo-RGB, VIS-only benefits mainly at high-$R$ rather than in overall ranking.

These results show that grade X hard negatives can improve discrimination, but their benefits depend on the representation and operating regime. For Pseudo-RGB, the improvement is concentrated at low and intermediate Recall and in the top-ranked candidate sample, whereas for VIS-only it becomes significant at high Recall. However, given the limited amount of grade X objects available for training (and validation), the best strategy we found was to exclude grade X from Stage 0, and to exploit those in Stage 1.

\subsection{\label{sc:stage1-models}Stage 1 models}

\renewcommand{\arraystretch}{1.2}
\begin{table}[ht]
    \centering
    \caption{Summary of Stage 1 ablation-set compositions.}
    \label{tab:stage10sets}
    \begin{tabular}{lccc}
        \hline\hline
        \textbf{Data source} & \textbf{train3} & \textbf{train4} & \textbf{train5} \\
        \hline
        \multicolumn{4}{l}{\textit{Lenses}} \\
        200\,$\rm deg^2$ DR1 A+B lenses & 782 & -- & -- \\
        200\,$\rm deg^2$ DR1 $\rm \bar{A}$+$\rm \bar{B}$ lenses & -- & 953 & 953 \\
        Q1 area lenses & 525 & 525 & 525 \\
        \hline
        \textbf{Total lenses} & \textbf{1307} & \textbf{1478} & \textbf{1478} \\
        \hline
        \multicolumn{4}{l}{\textit{Non-lenses}} \\
        Hard negatives (grade `XX') & 4200 & 4200 & -- \\
        Hard negatives (grade `X') & -- & -- & 4200 \\
        DR1 non-lenses ($\rm \bar{C}$) & -- & $280^{*}$ & 280 \\
        Artefacts & -- & -- & 192 \\
        \hline
        \textbf{Total non-lenses} & \textbf{4200} & \textbf{4480} & \textbf{4672} \\
        \hline\hline
        \textbf{Total objects} & \textbf{5507} & \textbf{5958} & \textbf{6150} \\
        \hline\hline
    \end{tabular}
    \tablefoot{Train5 is reported for completeness, as it represents the final Stage 1 training set. Train5 results are therefore relative to the non-use of Stage 0 pre-training, as described in Sect.~\ref{sc:train5}. $^{*}$For the Pseudo-RGB train4 configuration, the 280 additional grade $\rm \bar{C}$ negatives were excluded after preliminary experiments revealed a substantial degradation in performance. The VIS-only train4 configuration retains these objects.}
\end{table}

\begin{table*}[t]
\centering
\caption{Stage 1 ablation metrics on the DR1 test set.}
\label{tab:ablation_stage10_metrics_paper}
\begin{tabular}{llcccccccc}
\hline\hline
Train & Modality & ROC-AUC & $\rm AP_{w}$ & $\rm AP_{w,\mathrm{HR}}$ & P@\textit{K} & TP@\textit{K} & $\rm P@R=0.50$ & $\rm P@R=0.75$ & $\rm P@R=0.90$ \\
\hline
final & pseudo-RGB & 0.9944 & 0.2974 & 0.0576 & 0.1642 & 11\,491 & 0.3625 & 0.1758 & 0.0312 \\
train0 & pseudo-RGB & 0.9918 & 0.3012 & 0.0395 & 0.1514 & 10\,600  & 0.2749 & 0.1201 & 0.0191 \\
train1 & pseudo-RGB & 0.9935 & 0.2743 & 0.0487 & 0.1514 & 10\,600 & 0.3127 & 0.1245 & 0.0270 \\
train2 & pseudo-RGB & 0.9898 & 0.2619 & 0.0226 & 0.0895 & 6264 & 0.0481 & 0.0414 & 0.0180 \\
train3 & pseudo-RGB & 0.9958 & 0.2710 & 0.0454 & 0.1402 & 9812 & 0.2453 & 0.0992 & 0.0290 \\
train4$^{*}$ & pseudo-RGB & 0.9952 & 0.3075 & 0.0385 & 0.1361 & 9524 & 0.2453 & 0.0866 & 0.0254 \\
train5 & pseudo-RGB & 0.9850 & 0.4055 & 0.0147 & 0.1246 & 8720 & 0.3127 & 0.0471 & 0.0067 \\
\hline
final & VIS & 0.9944 & 0.5954 & 0.0656 & 0.1672 & 11\,704 &  0.6946 & 0.2544 & 0.0338 \\
train0 & VIS & 0.9913 & 0.4372 & 0.0391 & 0.1475 & 10\,324 & 0.4313 & 0.0964 & 0.0262 \\
train1 & VIS & 0.9949 & 0.5158 & 0.0573 & 0.1582 & 11\,074 & 0.5321 & 0.1523 & 0.0350 \\
train2 & VIS & 0.9947 & 0.3817 & 0.0454 & 0.1466 & 10\,263 & 0.2749 & 0.1160 & 0.0286 \\
train3 & VIS & 0.9959 & 0.3521 & 0.0565 & 0.1538 & 10\,768 &  0.4313 & 0.1343 & 0.0379 \\
train4 & VIS & 0.9957 & 0.5370 & 0.0342 & 0.1437 & 10\,062 & 0.4313 & 0.0912 & 0.0189 \\
train5 & VIS & 0.9834 & 0.3868 & 0.0107 & 0.1191 & 8338 & 0.1594 & 0.0260 & 0.0067 \\
\hline\hline
\end{tabular}

\par\smallskip
\tablefoot{All values are computed with sample weights $w_0=3300$ for class 0 and $w_1=1$ for class 1. Same as Table~\ref{tab:ablation_stage0_metrics_paper}. $^{*}$The train4 Pseudo-RGB metrics correspond to the configuration in which the 280 grade $\rm \bar{C}$ negative examples were excluded from the training set, unlike the VIS-only configuration.}
\end{table*}

The Stage 1 ablations study how fine-tuning set composition affects performance, starting from the Stage 0 baselines (see Sect.~\ref{sc:stage0-models}). Train3 replaces the curated final Stage 1 positives with the original grade A and B lenses and uses grade XX negatives. Train4 instead adopts the curated positive sample (see Sect.~\ref{sc:dr1-lenses}), testing its effect together with variations in the negative pool. The final models use the training set described in Sect.~\ref{sc:stage1}, including artefacts and grade X hard negatives, while train5 omits Stage 0 entirely. We also fine-tune the train0, train1, and train2 Stage 0 baselines on the identical final Stage 1 set to assess the effect of initialisation.
Table~\ref{tab:stage10sets} summarises the data sets, while Table~\ref{tab:ablation_stage10_metrics_paper} and Figs.~\ref{fig:ablation_stage10_pr} and~\ref{fig:ablation_stage10_roc} report the results.

\subsubsection{train0, train1, train2: effect of Stage 0 initialisation on Stage 1 fine-tuning}

These configurations share the final Stage 1 fine-tuning set but differ in Stage 0 initialisation (described in Sect.~\ref{sc:stage0-models}), isolating its effect on subsequent performance.

For Pseudo-RGB, better Stage 0 performance does not necessarily translate into better Stage 1 results. Train0 reaches $\mathrm{AP}_{\rm w}=0.3012$, marginally exceeding the final Stage 1 reference ($0.2974$) by $1.3\%$. However, its high-$R$ performance is considerably weaker, with $\mathrm{AP}_{{\rm w},\mathrm{HR}}=0.0395$, $31.4\%$ below the final reference ($0.0576$). Train1, which includes additional cluster-GGSL simulations during Stage 0, reaches $\mathrm{AP}_{\rm w}=0.2743$ ($7.8\%$ below the final reference) and $\mathrm{AP}_{{\rm w},\mathrm{HR}}=0.0487$ ($15.5\%$ below). Interestingly, train1 high-$R$ partial area is better than train0, despite the opposite behaviour observed when evaluating the Stage 0 models alone. Thus, the benefits of cluster simulations also depend on subsequent fine-tuning. Train2 instead deteriorates after fine-tuning. Despite achieving the highest $\mathrm{AP}_{\rm w}$  among the Stage 0 Pseudo-RGB configurations, its Stage 1 performance falls to $\mathrm{AP}_{\rm w}=0.2619$ and $\mathrm{AP}_{{\rm w},\mathrm{HR}}=0.0226$, corresponding to reductions of $11.9\%$ and $60.8\%$ respectively, relative to the final Stage 1 reference. The performance drop is also evident at fixed Recall: at $R=0.75$, train2 reaches a Precision of only $0.0414$, compared with $0.1758$ for the final reference. Its $\text{P@\textit{K}}=0.0895$ is also substantially lower than the final configuration ($0.1642$), resulting in $6264$ recovered lenses among the top-$K$ candidates, compared with $11\,491$ for the final model. These results indicate that introducing grade X hard negatives already during Stage 0 does not provide a valid initialisation for the subsequent Pseudo-RGB fine-tuning.

The VIS-only branch shows a similar trend. Train0 reaches $\mathrm{AP}_{\rm w}=0.4372$ and $\mathrm{AP}_{{\rm w},\mathrm{HR}}=0.0391$, respectively $26.6\%$ and $40.4\%$ below the final Stage 1 reference. Adding cluster-GGSL simulations in train1 improves both metrics to $\mathrm{AP}_{\rm w}=0.5158$ and $\mathrm{AP}_{{\rm w},\mathrm{HR}}=0.0573$, reducing the gaps relative to the final reference to $13.4\%$ and $12.7\%$, respectively. Train1 also achieves $\text{P@\textit{K}}=0.1582$, compared with $0.1475$ for train0, recovering $7.3\%$ additional lenses within the top-$K$ candidates. Unlike Pseudo-RGB, the improvement associated with cluster simulations is therefore maintained after Stage 1 fine-tuning, although the final baseline remains the best-performing configuration overall. Train2 also loses its Stage 0 advantage in VIS-only. Although it achieved the highest $\mathrm{AP}_{{\rm w},\mathrm{HR}}$ among the Stage 0 VIS-only configurations, after fine-tuning it reaches only $\mathrm{AP}_{\rm w}=0.3817$ and $\mathrm{AP}_{{\rm w},\mathrm{HR}}=0.0454$, corresponding to reductions of $35.9\%$ and $30.8\%$ relative to the final Stage 1 reference. Its $\text{P@\textit{K}}=0.1466$ is also lower than both train1 and the final model, while its $\rm P@R_{0.75}$ remains limited to $0.1160$, compared with $0.2544$ for the final reference. Therefore, introducing grade X negatives during Stage 0 does not improve subsequent fine-tuning, even for the VIS-only branch, where they had been particularly beneficial at high Recall.

These ablations show that Stage 0 initialisation affects the final ranking even with identical fine-tuning data. The additional cluster-GGSL simulations generally provide a more ideal starting point than the simpler train0 configuration, particularly for VIS-only, while introducing grade X hard negatives during Stage 0 does not yield corresponding benefits after Stage 1. One possible explanation is the reuse of the same grade X hard negatives across both training stages, which may limit the diversity of new examples analysed during fine-tuning. However, this interpretation remains theoretical, as the present ablations do not isolate the effect of sample reuse from that of the training-set composition. Overall, the results favour the final Stage 0 baseline (which represent a balanced rather than maximally hard initialisation) as the most effective starting point for the two-stage training procedure, highlighting that optimising pre-training performance alone is insufficient to guarantee optimal performance after fine-tuning.

\subsubsection{train3: A+B graded lenses and grade XX negatives}
This configuration uses grade A and B lenses (from the original 200\,$\rm deg^2$ search, see Sect.~\ref{sc:dr1-lenses}) together with the grade XX negative pool as Stage 1 training set. Both branches perform worse than their final Stage 1 counterparts.

For Pseudo-RGB, $\mathrm{AP}_{\rm w}=0.2710$, representing an $8.9\%$ reduction relative to the final Stage 1 reference. The high-$R$ performance also decreases, with $\mathrm{AP}_{{\rm w},\mathrm{HR}}=0.0454$, approximately $21.2\%$ below the final reference. This difference is reflected also in $\rm P@R_{0.75}$, where train3 reaches $0.0992$ compared with $0.1758$ for the final model, corresponding to a $43.6\%$ reduction. Similarly, $\text{P@\textit{K}}=0.1402$, compared with $0.1642$ for the final reference, resulting in $14.6\%$ less recovered lenses among the top-$K$ candidates. The VIS-only branch instead exhibits a more pronounced drop in overall ranking performance. Its $\mathrm{AP}_{\rm w}=0.3521$ is $40.9\%$ below the final Stage 1 reference, while $\mathrm{AP}_{{\rm w},\mathrm{HR}}=0.0565$ remains comparatively close to the final value of $0.0656$, representing a $13.9\%$ reduction. Despite this relatively modest difference in the high-$R$ partial area, $\rm P@R_{0.75}$ decreases from $0.2544$ to $0.1343$, corresponding to a $47.2\%$ reduction. Train3 also achieves $\text{P@\textit{K}}=0.1538$, which compared with $0.1672$ for the final model gives $10,768$ recovered lenses within the top-$K$ candidates instead of $11,704$ ($8\%$ decrease).

The PR curves in Fig.~\ref{fig:ablation_stage10_pr} illustrate these differences more clearly than the ROC curves alone in Fig.~\ref{fig:ablation_stage10_roc}. Although train3 retains substantial discriminative power, its Precision is generally lower than that of the final models across the operating regimes most relevant for candidate selection, especially for the Pseudo-RGB model. These differences favour the final Stage 1 training set, although the effects of positive-sample curation and harder negative examples cannot be isolated. However, since train3 differs from the final configuration in both its positive and negative samples, their individual contributions cannot be isolated from this comparison alone, motivating the subsequent train4 ablation. 

\subsubsection{train4: removing positive confounders from train3}

Train4 investigates the impact of refining the Stage 1 training sample by replacing the original grade A and B lenses used in train3 with the curated $\rm \bar{A}$ and $\rm \bar{B}$ sample (see Sect.~\ref{sc:dr1-lenses}), while retaining the same grade XX negative pool. For the VIS-only branch, the negative sample also includes 280 additional $\rm \bar{C}$ objects, whereas these were excluded from the final Pseudo-RGB train4 configuration, as discussed below. This ablation therefore allows us to investigate how modifications to the positive training sample affect the two representations, while also providing insights into their sensitivity to difficult negative examples.

For Pseudo-RGB, the refined training sample improves the overall $\mathrm{AP}_{\rm w}$, which increases from $0.2710$ in train3 to $0.3075$ in train4, corresponding to a $13.5\%$ improvement and exceeding the final Stage 1 reference by $3.4\%$. However, this improvement does not extend to the high-$R$ regime, where $\mathrm{AP}_{{\rm w},\mathrm{HR}}$ decreases from $0.0454$ to $0.0385$, representing a $15.2\%$ reduction relative to train3. Similarly, $\text{P@\textit{K}}$ decreases slightly from $0.1402$ to $0.1361$, with $9524$ recovered lenses among the top-$K$ candidates, compared with $9812$ for train3 ($-\,2.9\%$). These results indicate that refining the positive training sample can improve the overall ranking performance of the Pseudo-RGB branch, although it does not necessarily improve its ability to maintain Precision at high-$R$. Interestingly, the initial version of train4, which actually included the 280 additional grade $\rm \bar{C}$ objects in the negative training pool, exhibited a substantially poorer performance for Pseudo-RGB, reaching only $\mathrm{AP}_{\rm w}=0.0838$ and $\mathrm{AP}_{{\rm w},\mathrm{HR}}=0.0216$. Removing these objects corresponded to improvements of $267.0\%$ and $78.2\%$ for these two metrics, respectively. The corresponding $\text{P@\textit{K}}$ also increased from $0.0967$ to $0.1361$. This strong sensitivity to a relatively small modification of the negative training sample suggests that the Pseudo-RGB branch may be particularly vulnerable to extremely difficult negative labels, especially when the latter make up just a small part of the training set. In particular, some grade $\rm \bar{C}$ candidates may retain colour or morphological features associated with true lensing systems, potentially introducing conflicting information during fine-tuning. Although the precise nature of this behaviour remains uncertain, the result highlights the importance of carefully selecting negative examples when training \Euclid multi-band lens finders.

The VIS-only branch shows a different response. Even if including the few grade $\rm \bar{C}$ objects, its overall $\mathrm{AP}_{\rm w}$ increases from $0.3521$ in train3 to $0.5370$ in train4, corresponding to a $52.5\%$ improvement and approaching the final Stage 1 reference ($0.5954$). However, the high-$R$ performance worsens, with $\mathrm{AP}_{{\rm w},\mathrm{HR}}$ decreasing from $0.0565$ to $0.0342$, representing a $39.5\%$ reduction relative to train3. Similarly, $\text{P@\textit{K}}$ decreases from $0.1538$ to $0.1437$, corresponding to $-6.5\%$ recovered lenses among the top-$K$ candidates. Unlike Pseudo-RGB, the VIS-only branch retains relatively strong overall performance despite the inclusion of the additional grade $\rm \bar{C}$ negatives, suggesting a greater robustness to this particular modification of the training sample. To further investigate the impact of the additional grade $\rm \bar{C}$ negatives, we also repeated the VIS-only train4 experiment after removing these 280 objects, as previously done for Pseudo-RGB. Again, the VIS-only branch behaves differently: while $\mathrm{AP}_{\rm w}$ decreases from $0.5370$ to $0.3822$ ($-28.8\%$), the high-$R$ performance improves substantially, with $\mathrm{AP}_{{\rm w},\mathrm{HR}}$ increasing from $0.0342$ to $0.0548$ ($60.2\%$ increase). Similarly, $\text{P@\textit{K}}$ increases from $0.1437$ to $0.1512$, corresponding to $5.1\%$ additional recovered lenses among the top-$K$ candidates. At $R=0.90$, Precision nearly doubles, increasing from $0.0189$ to $0.0356$. These results indicate that unlike the Pseudo-RGB model, the inclusion of the additional grade $\rm \bar{C}$ negatives benefits the overall ranking performance of VIS-only, but at the expense of Precision at high-$R$. Thus, although the two branches respond differently in terms of global ranking performance, both benefit from excluding these objects in the high-$R$ regime.

Taken together, these experiments reveal a strong dependence of the training outcome on both the input representation and the composition of the negative sample. While the inclusion of just a few grade $\rm \bar{C}$ negatives substantially degrades Pseudo-RGB performance across all evaluated metrics, it improves the overall $\mathrm{AP}_{\rm w}$ of VIS-only at the expense of high-$R$ performance. Notably, removing these objects improves $\mathrm{AP}_{{\rm w},\mathrm{HR}}$ for both branches, suggesting that the inclusion of a small number of particularly challenging negatives within a predominantly grade XX training pool may adversely affect the recovery of difficult lenses at high-$R$. Conversely, the final Stage 1 configurations, which employ a more uniformly challenging grade X negative pool, achieve stronger high-$R$ performance, suggesting that the effectiveness of hard-negative mining depends not only on the difficulty of individual examples, but also on their balance within the training set.

\subsubsection{Final Stage 1 models}
The final Stage 1 models use the training set described in Sect.~\ref{sc:stage1}, similar to train4 but with a few artefact additions and the replacement of grade XX with grade X hard negatives examples. As shown in Table~\ref{tab:ablation_stage10_metrics_paper}, these configurations perform best across the survey-relevant metrics. For VIS-only, the final Stage 1 model achieves the highest $\mathrm{AP}_{\rm w}$ ($0.5954$) and $\mathrm{AP}_{{\rm w},\mathrm{HR}}$ ($0.0656$) among all Stage 1 configurations. It also reaches the highest $\text{P@\textit{K}}=0.1672$, corresponding to $11\,704$ recovered lenses among the top $K=70\,000$ candidates, and the highest $\rm P@R_{0.50}$ and $\rm P@R_{0.75}$, with values of $0.6946$ and $0.2544$, respectively. For Pseudo-RGB, the final Stage 1 model achieves $\mathrm{AP}_{\rm w}=0.2974$ and the highest $\mathrm{AP}_{{\rm w},\mathrm{HR}}$ among its Stage 1 configurations ($0.0576$). It also provides the highest $\text{P@\textit{K}}=0.1642$, recovering $11,491$ lenses among the top-$K$ candidates, and reaches the highest $\rm P@R_{0.75}$ and $\rm P@R_{0.90}$, with values of $0.1758$ and $0.0312$, respectively. Although some ablations achieve higher overall $\mathrm{AP}_{\rm w}$ or slightly better ROC-AUC values, the final models consistently provide the most favourable trade-off between completeness and contamination at the operating points relevant for the \Euclid DR1 lens search. Interestingly, while the VIS-only model substantially outperforms Pseudo-RGB in $\mathrm{AP}_{\rm w}$ ($0.5954$ versus $0.2974$), the gap narrows considerably in the high-$R$ regime ($\mathrm{AP}_{{\rm w},\mathrm{HR}}=0.0656$ versus $0.0576$), indicating greater competitiveness at high completeness and motivating their combination (see Sect.~\ref{sc:model-ensemble}).

The PR curves in Fig.~\ref{fig:ablation_stage10_pr} show that the final Stage 1 training set preserves Precision much better than the other alternatives, and this is especially visible in the high-$R$ regime beyond $0.75$. Particularly for the VIS-only branch, the low-FPR region in the ROC plot (see Fig.~\ref{fig:ablation_stage10_roc}) also confirms that the improvement is not confined to one isolated threshold: the model stays strong across a broad range of operating points.

\begin{figure*}[t]
\centering
\includegraphics[width=0.49\textwidth]{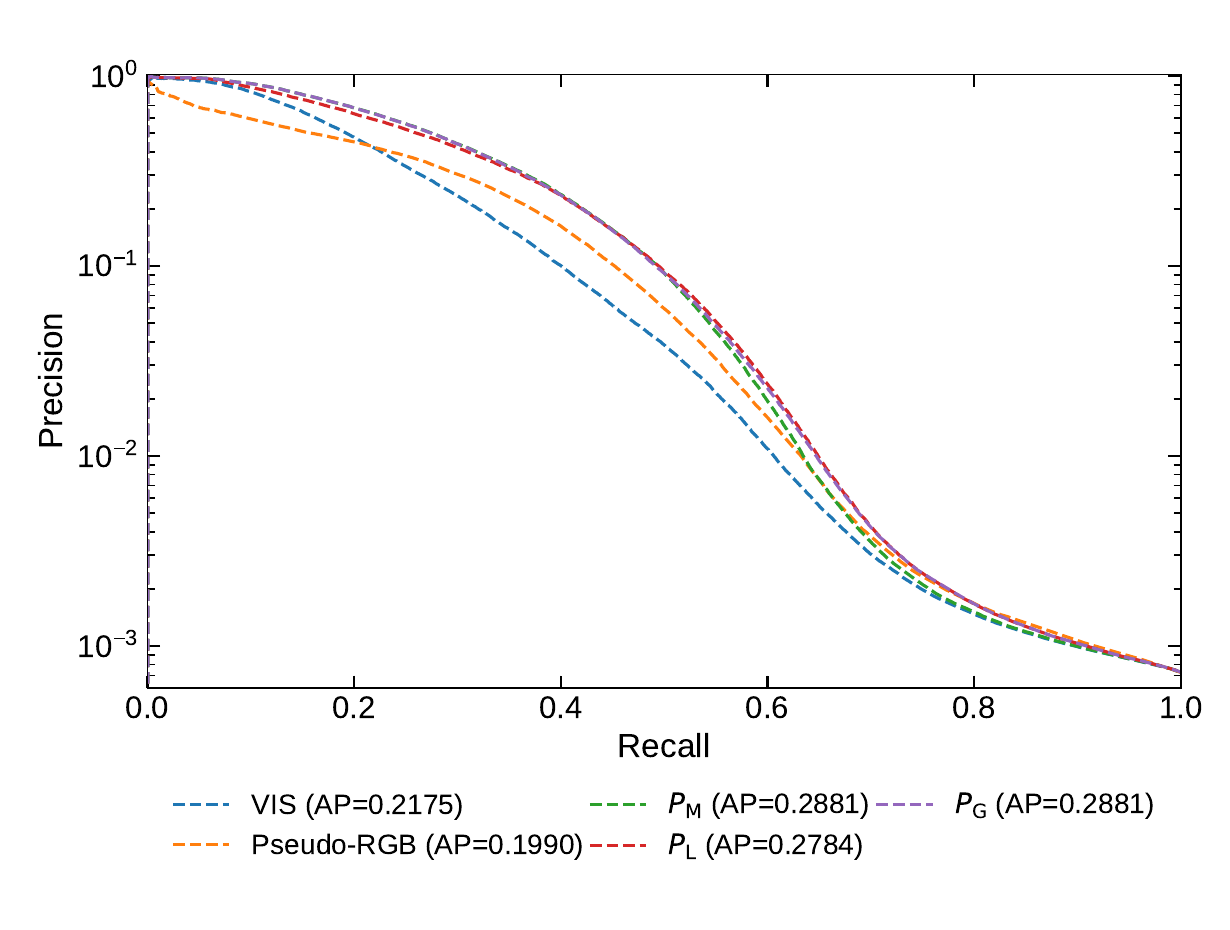}
\includegraphics[width=0.49\textwidth]{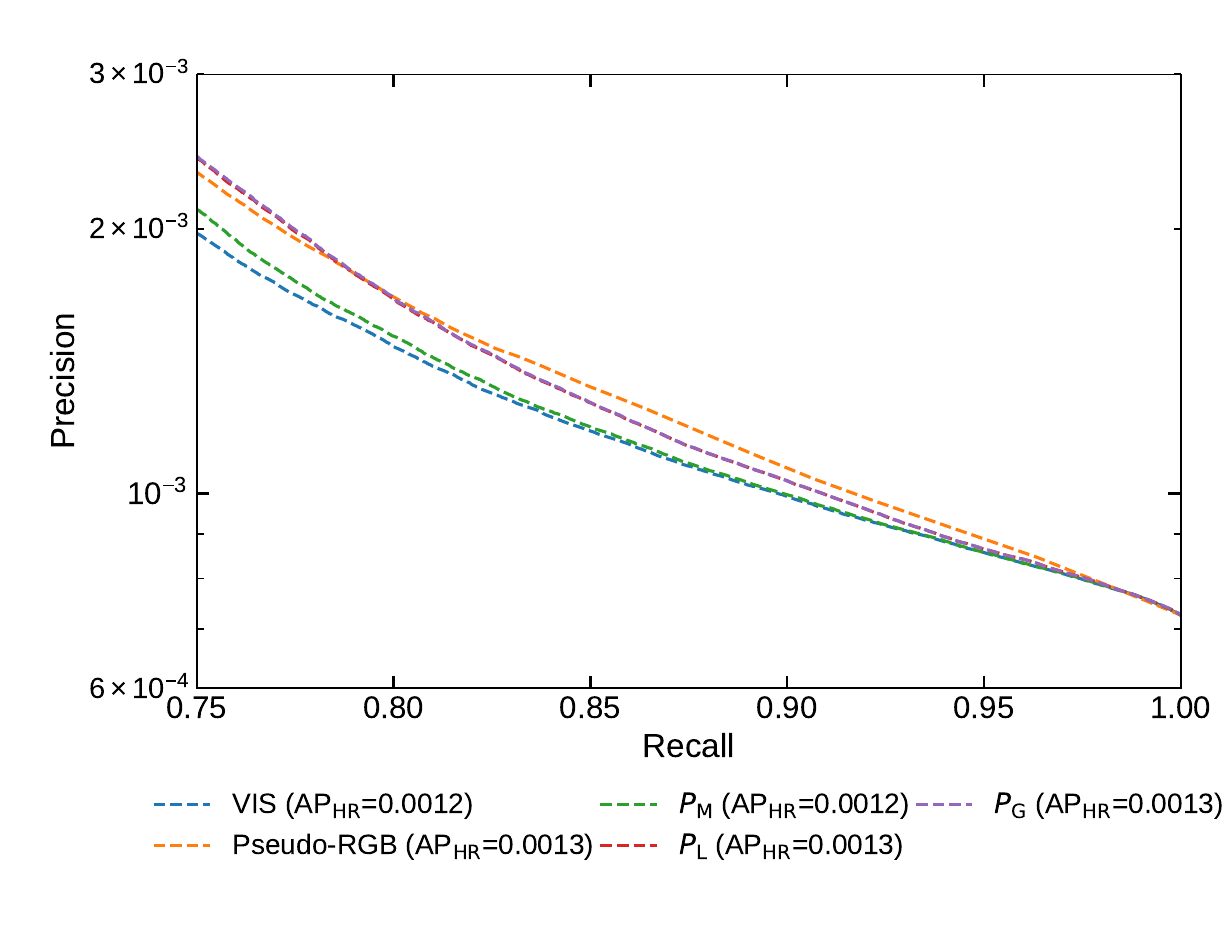}
\caption{
PR curves obtained on the full DR1 sample for the final VIS-only model, Pseudo-RGB model, and three score-level ensemble combinations: mean probability $P_{\rm M}$, geometric mean probability $P_{\rm G}$, and logit mean probability $P_{\rm L}$. The left panel shows the full PR curve and the legend reports the average Precision. The right panel zooms into the high-$R$ interval $R\in[0.75,1.00]$, with legend reporting the normalised average Precision for $R \in [0.75,1.00]$.}
\label{fig:ensemble_pr}
\end{figure*}

\begin{figure}[t]
\centering
\includegraphics[width=\columnwidth]{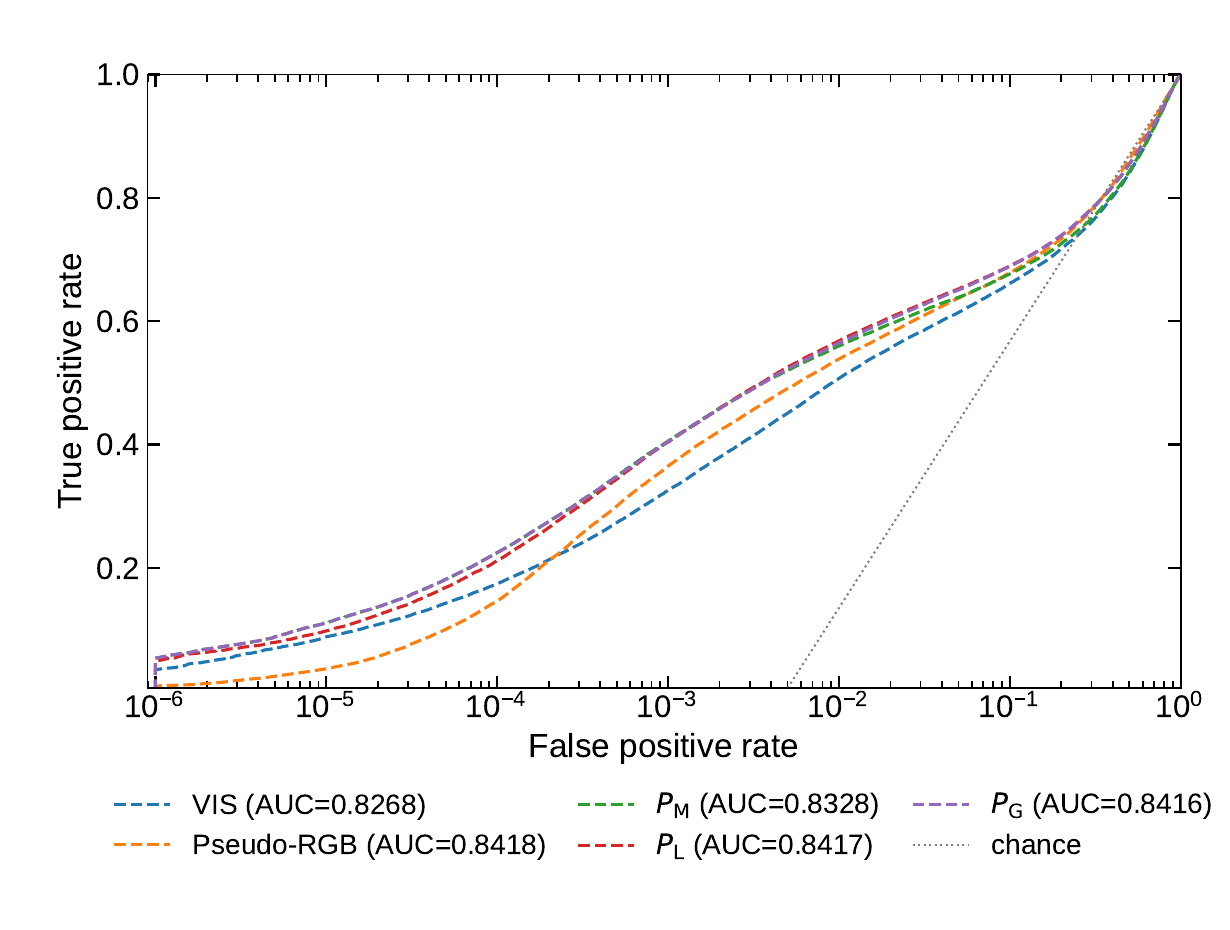}
\caption{
ROC curves obtained on the full DR1 sample for the final single models and ensemble scores: mean probability $P_{\rm M}$, geometric mean probability $P_{\rm G}$, and logit mean probability $P_{\rm L}$. The FPR axis is logarithmic to emphasise the low-FPR regime relevant for survey-scale candidate generation.}
\label{fig:ensemble_roc}
\end{figure}

\subsubsection{\label{sc:train5}train5: direct training on Stage 1 data set}

The final ablation removes the Stage 0 pre-training entirely and trains directly on the final Stage 1 training set. This is the clearest test of whether the two-stage approach is actually useful. The results again reveal a strong dependence on the input representation.

For VIS-only, omitting Stage 0 degrades all metrics. $\mathrm{AP}_{\rm w}$ decreases to $0.3868$, representing a $35.0\%$ reduction relative to the final Stage 1 reference. The degradation is even more pronounced at high-$R$, where $\mathrm{AP}_{{\rm w},\mathrm{HR}}=0.0107$, corresponding to an $83.7\%$ reduction relative to the final model. Similarly, $\rm P@R_{0.75}$ drops from $0.2544$ to $0.0260$, while $\rm P@R_{0.90}$ decreases from $0.0338$ to $0.0067$. The top-$K$ performance also worsens, with $\text{P@\textit{K}}=0.1191$ compared with $0.1672$ for the final reference, corresponding to $8338$ recovered lenses instead of $11,704$ among the top-$K$ candidates (corresponding to a $28.8\%$ decrease). These results demonstrate that Stage 0 pre-training is particularly important for the VIS-only branch, substantially improving its ability to prioritise true lenses throughout the candidate ranking.

Interestingly, for the Pseudo-RGB branch train5 achieves $\mathrm{AP}_{\rm w}=0.4055$, the highest among all Stage 1 Pseudo-RGB configurations and $36.3\%$ above the final reference. However, this apparent improvement does not extend to the high-$R$ regime, where $\mathrm{AP}_{{\rm w},\mathrm{HR}}=0.0147$, representing a $74.5\%$ reduction relative to the final model. Moreover, $\rm P@R_{0.75}$ decreases from $0.1758$ to $0.0471$, while $\rm P@R_{0.90}$ falls from $0.0312$ to $0.0067$. Similarly, $\text{P@\textit{K}}=0.1246$ remains below the final reference ($0.1642$), resulting in $-24.1\%$ recovered lenses compared with the final configuration. Thus, although direct training on real \Euclid data produces a Pseudo-RGB model with strong overall ranking performance, it is considerably less effective at maintaining Precision as Recall approaches unity.

These results demonstrate that Stage 0 pre-training plays a fundamental role in achieving robust high-$R$ performance for both image representations. While the VIS-only branch benefits from pre-training across the entire PR curve, the Pseudo-RGB branch exhibits a trade-off between global ranking performance and high-$R$ Precision. One interpretation is that the simulated Stage 0 sample provides a broader range of lens configurations than the relatively limited real-lens fine-tuning set, improving the models’ ability to recognise less common or more challenging systems. Overall, the substantial degradation observed in $\mathrm{AP}_{{\rm w},\mathrm{HR}}$ and $\text{P@\textit{K}}$ confirms that the two-stage training strategy is preferable for the high-$R$ candidate-selection regime adopted in the \Euclid DR1 search.

\section{\label{sc:model-ensemble}Model ensemble}

The final candidate-ranking method used in this work is based on the combination of two independently trained classifiers: a VIS-only model, trained on the single-band \Euclid-VIS cutouts, and a Pseudo-RGB model, trained on colour-composite images built from the available VIS and NISP bands. The VIS-only branch primarily exploits lensing morphology, including arcs, tangential elongation, and the contrast between deflectors and lensed features. Pseudo-RGB additionally exploits colour differences between deflectors, sources, and contaminants, potentially resolving morphologically ambiguous configurations. More importantly, by incorporating the NISP bands it can identify candidate lenses whose lensed features are faint or absent in VIS imaging but detectable in the NIR. An example is presented by Euclid Collaboration: Harvey et al. (in prep.), who report a $z_{\rm phot}=7.95$ VIS-dropout lens candidate missed by all the networks participating in the SLDE and the following Q1 searches, but assigned a score greater than 0.99 by our Pseudo-RGB model, boosting its final logit-mean ensemble score to approximately 0.95 (despite receiving only modest support from the VIS-only branch). This demonstrates that NIR information is not merely complementary to morphology, but can reveal an additional population of systems inaccessible to optical-only searches.

The single-model performances of the final Stage 1 networks on our ablation test set are reported in detail in Table~\ref{tab:ablation_stage10_metrics_paper}. We briefly summarise here the quantities most relevant for the ensemble construction. The final VIS-only model reaches a weighted average precision of $\rm AP_w=0.5954$ and a normalised high-$R$ partial area of $\mathrm{AP}_{{\rm w},\mathrm{HR}}=0.0656$. The final Pseudo-RGB model reaches $\rm AP_w=0.2974$ and $\mathrm{AP}_{{\rm w},\mathrm{HR}}=0.0576$. The VIS-only model is therefore the stronger individual classifier on this test set, especially in the high-$R$ regime that is most relevant for survey candidate generation. This does not imply, however, that the Pseudo-RGB model is uninformative: its ranking is different from that of the VIS-only model, and this difference is precisely what motivates the ensemble.

To combine the two networks, we tested three simple score-level fusion rules. Given the lens probabilities predicted by the VIS-only and Pseudo-RGB models, $p_{\rm VIS}$ and $p_{\rm RGB}$, we define
\begin{equation}
p_{\rm mean} = \frac{p_{\rm VIS}+p_{\rm RGB}}{2}\,,
\end{equation}
\begin{equation}
p_{\rm geo}
=
\sqrt{p_{\rm VIS}\,p_{\rm RGB}}
=
\exp\!\left[
\frac{1}{2}
\left(
\ln p_{\rm VIS}
+
\ln p_{\rm RGB}
\right)
\right]\,,
\end{equation}
\begin{equation}
p_{\rm logit} =
\sigma \left\{
\frac{1}{2}
\left[
\mathrm{logit}(p_{\rm VIS})+
\mathrm{logit}(p_{\rm RGB})
\right]
\right\}\,,
\end{equation}
where $\sigma(x)$ is the sigmoid function. The arithmetic mean averages the probabilities directly, the geometric mean penalises objects for which one model assigns a low probability, and the logit mean averages the models in log-odds space. The latter is less sensitive to the compression of probabilities near zero and one, and is therefore a natural choice when the two networks produce calibrated but differently distributed confidence scores.

As the VIS-only and Pseudo-RGB classifiers were combined into the ensemble model deployed in the ECLIPSE strong lens search, both were applied to the full DR1 sample, comprising approximately $67$ million scored objects. As a consequence, unlike the weighted test set evaluation adopted for the ablation study (Sect.~\ref{sc:ablation-analysis}), the following results are computed directly from the full survey predictions, without applying class-prevalence weights. This allows us to study how the individual models and their combinations rank the strong lensing systems identified in the ECLIPSE catalogue within the complete population of scored objects.

Figure~\ref{fig:ensemble_pr} shows the resulting PR curves for the two individual models and the three score-fusion strategies. The VIS-only and Pseudo-RGB branches achieve overall $\mathrm{AP}$ values of $0.2175$ and $0.1990$, respectively. Combining their predictions substantially improves the ranking performance, with the arithmetic and geometric means reaching $\mathrm{AP}=0.2881$, while the logit mean achieves $\mathrm{AP}=0.2784$. These correspond to improvements of approximately $32.5\%$ and $28.0\%$, respectively, relative to the VIS-only model. The ensemble therefore provides a significant gain in the recovery of known DR1 lenses across the full ranked sample, demonstrating the practical value of combining the complementary information encoded by the two image representations.

Notably, the relative performance of the individual branches differs from that observed in the weighted ablation test set. While VIS-only substantially outperformed Pseudo-RGB in $\mathrm{AP}_{\rm w}$ ($0.5954$ versus $0.2974$), their performance becomes considerably more comparable on the full DR1 sample. This difference likely reflects the distinct distributions of the two evaluation populations, as the weighted test set reproduces the approximate survey class prevalence but cannot fully capture the diversity of contaminants encountered across the complete DR1 footprint. The full-survey evaluation therefore provides an important complementary perspective on the operational behaviour of the models.

The high-$R$ portion of the PR curves (right panel of Fig.~\ref{fig:ensemble_pr}) reveals that although the ensembles provide the strongest overall ranking performance, their advantage becomes less pronounced as Recall approaches unity. In particular, the Pseudo-RGB branch becomes slightly more competitive than VIS-only and the ensemble combinations at $R\gtrsim0.8$, reaching $\mathrm{AP}_{\rm HR}\simeq0.0013$, compared with $\sim0.0012$ for VIS-only. Nevertheless, the general low Precision values in this regime reflect the strong contamination associated with attempting to recover nearly all known lenses from tens of millions of objects. Moreover, these results are based on only two of the five ML algorithms contributing to the ECLIPSE catalogue, whose complementary selection capabilities further improve the overall lens recovery (Euclid Collaboration: Harvey et al. in prep.).

To assess the practical impact of these differences on the DR1 search, we compared the models at the true fixed candidate-selection depth of $K=68\,021$ objects, corresponding to the sample subsequently inspected through SW. This number results from applying a logit-mean ensemble score threshold of approximately $0.73$ to the full DR1 sample, and was calibrated on an independent $50$\,deg$^2$ test set to achieve an estimated Precision of $0.08$ (see \citealt{DR1-SL02} for details). Although the final ECLIPSE catalogue contains $15\,070$ unique grade A and B lenses, only $14\,237$ could be associated with at least one cutout in the score catalogue used for this analysis. This difference arises from the fact that the various ML pipelines were executed independently, using the same input cutout selections but different data coverage and processing configurations. We therefore report the recovery fraction relative to the matched lens population, which provides a more direct measure of the ranking performance for systems actually present in our scored sample. Table~\ref{tab:dr1_topk_recovery} reports the number of recovered grade A and B lens detections and unique lens systems for each model, together with the corresponding Recall relative to the $14\,237$ unique systems matched in the ECLIPSE lens catalogue. At this selection depth, the logit-mean ensemble achieves the highest Recall, recovering $11\,256$ unique lenses ($79.1\%$), compared with $10\,395$ ($73.0\%$) for Pseudo-RGB and $9650$ ($67.8\%$) for VIS-only. The ensemble therefore recovers $1606$ additional unique lenses relative to VIS-only and $861$ relative to Pseudo-RGB, corresponding to improvements of $11.3\%$ and $6\%$ in Recall, respectively. Interestingly, Pseudo-RGB also outperforms VIS-only by approximately $5.0\%$, despite the latter achieving substantially higher overall $\rm AP_w$ on the controlled ablation test set. These results demonstrate that the complementary information encoded by the two branches translates into a substantial improvement in the recovery of known DR1 lenses at the actual survey selection depth, further supporting the adoption of the logit-mean ensemble for the ECLIPSE candidate-ranking pipeline.

\renewcommand{\arraystretch}{1.2}
\begin{table}[ht]
    \centering
    \caption{Recovery of DR1 grade A and B lenses among the top $K=68\,021$ candidates independently ranked by each model over the full DR1 survey.}
    \label{tab:dr1_topk_recovery}
    \begin{tabular}{lccc}
        \hline\hline
        \textbf{Model} & \textbf{Detections} & \textbf{Unique lenses} & \textbf{Recall (\%)} \\
        \hline
        VIS-only & 12\,201 & 9650 & 67.8 \\
        Pseudo-RGB & 13\,275 & 10\,395 & 73.0 \\
        Ensemble $P_{\rm L}$ & 14\,528 & 11\,256 & 79.1 \\
        \hline\hline
    \end{tabular}
    \tablefoot{Recall is computed relative to the $14\,237$ unique grade A and B lens systems with a matched cutout in the final ECLIPSE catalogue. Multiple cutouts associated with the same physical lens are counted only once when computing the reported Recall.}
\end{table}

The ROC curves in Fig.~\ref{fig:ensemble_roc} provide a complementary view of the full-survey ranking. The individual VIS-only and Pseudo-RGB models reach ROC-AUC values of $0.8268$ and $0.8418$, respectively, while the arithmetic, geometric, and logit means achieve $0.8328$, $0.8416$, and $0.8417$. Although these values are lower than those obtained on the controlled test set, they reflect a substantially broader and more heterogeneous population of scored objects. Importantly, the ensemble combinations remain competitive in the low-FPR region, where even small improvements in the recovery of strong lenses can significantly reduce the number of contaminants requiring subsequent visual inspection.

Finally, the difference between the weighted test-set metrics and the full-DR1 results should be interpreted with some caution. The weighted evaluation accounts for the expected survey prevalence but is based on a finite sample of visually inspected negatives, whereas the full-DR1 analysis includes the actual distribution of objects encountered by the search pipeline. Furthermore, the ECLIPSE lens catalogue was constructed through a combination of several ML-algorithms and a subsequent visual inspection, and therefore does not represent an independent, complete census of all the GGSLs present in the survey. The full-survey metrics here presented are therefore best interpreted as measures of the recovery and ranking of the identified DR1 lens population, rather than as unbiased estimates of the intrinsic classification performance.

For the final DR1 candidate ranking, we adopted the logit-mean ensemble score. Although the arithmetic and geometric means achieve marginally higher overall AP values, the logit mean provides competitive performance across the relevant operating regimes while combining the two models in log-odds space. Importantly, this approach allows a highly confident prediction from one branch to compensate for a less confident prediction from the other, rather than strongly penalising candidates receiving a low score from either model, as occurs with the geometric mean. This property is particularly valuable when combining complementary image representations, as systems poorly recognised in VIS imaging may still receive high scores from the Pseudo-RGB branch thanks to the additional NIR information. Such behaviour is especially relevant for VIS-dropout lenses, whose identification may rely almost entirely on the NISP bands. The relatively small differences between the ensemble strategies, compared with their improvement over the individual classifiers, suggest that combining complementary image representations is more important than the precise choice of score-fusion rule.

We further assess branch complementarity through the dependence of relative ranking on lens properties. To this end, we use the \texttt{PyAutoLens} modelling results presented by \citet{DR1-SL03} and the available photometric redshifts from \citet{DR1-SL04}. The analysis is restricted to catalogue systems with the required measurements.

\begin{figure*}[t]
\centering
\includegraphics[width=\textwidth]{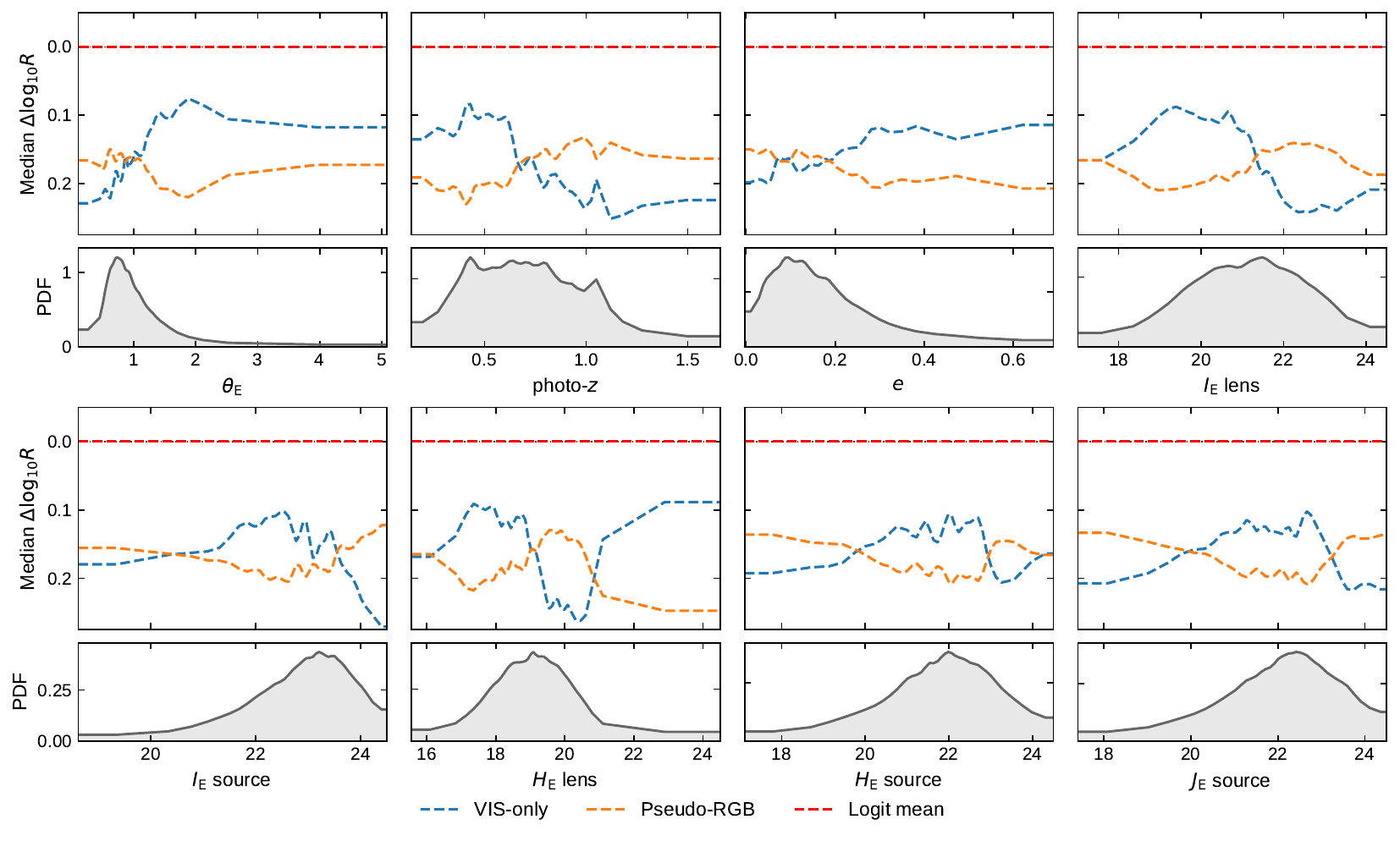}
\caption{
Relative median rank with respect to the logit-mean ensemble,
\(\Delta \log_{10}R=\log_{10}R_{\rm model}-\log_{10}R_{\rm logit}\),
as a function of lens properties. The upper row shows Einstein radius $\theta_\sfont{E}$, photometric redshift photo-$z$, ellipticity $e$, and the \IE-band magnitude of the lens. The bottom row shows the \IE-band magnitude of the source, the \HE-band magnitude of the lens and the source, and the \JE-band magnitude of the source. Negative values indicate that the individual model ranks systems better than the logit-mean reference, while positive values indicate worse ranking. The crossings between the VIS-only and pseudo-RGB curves show that the two branches encode different property-dependent selection functions. Each upper sub-panel shows the median $\Delta \log_{10}R$ as a function of the corresponding property, while the lower sub-panel gives the smoothed probability density function (PDF) of the systems along the $x$-axis.}
\label{fig:rank_bias_delta}
\end{figure*}

Figure~\ref{fig:rank_bias_delta} shows the median relative logarithmic rank difference,
\begin{equation}
\Delta \logten R =
\logten R_{\rm model} -
\logten R_{\rm logit},
\end{equation}
where $R_{\rm model}$ and $R_{\rm logit}$ are the ranks assigned by an individual network and the logit-mean ensemble, respectively, calculated over the full DR1 sample of approximately $67$ million scored cutouts. For each unique lens system, we retain the best-ranked cutout independently for each model, allowing the comparison to reflect the actual ability of each classifier to prioritise lenses within the survey. Since lower ranks correspond to stronger candidates, negative values of $\Delta \logten R$ indicate that an individual model ranks a lens better than the ensemble, while positive values indicate the opposite.

The most striking result is that both single-model curves remain predominantly above zero across the investigated parameter space, indicating that the logit-mean ensemble generally assigns more favourable ranks to the known lenses than either branch alone. Typical median differences of $\Delta\logten R\simeq0.1$–$0.25$ correspond to single-model ranks approximately $30\%$ to $80\%$ larger than those assigned by the ensemble. This improvement is consistent with the higher recovery achieved by the logit-mean combination at the operational selection depth of $K=68\,021$ objects (see Table~\ref{tab:dr1_topk_recovery}), demonstrating that the ensemble advantage extends across the whole range of lens properties rather than being confined to a particular subset of systems.

Beyond this overall improvement, the relative behaviour of the two branches varies systematically with lens properties. In particular, the VIS-only model becomes more competitive at intermediate Einstein radii ($\theta_\sfont{E}\simeq \text{\ang{;;1.5}}$--$\text{\ang{;;2}}$), where the larger angular separation between lensed features and the foreground deflector may facilitate morphological identification. Conversely, the Pseudo-RGB branch generally shows better relative rankings for smaller $\theta_\sfont{E}$, where colour information may help distinguish compact lensing configurations that are less easily recognised through morphology alone. A similar difference emerges as a function of the lens photometric redshift, with the Pseudo-RGB branch becoming increasingly competitive at higher redshifts, consistent with the additional wavelength information provided by the NISP bands.

The magnitude-dependent trends provide further evidence of this complementarity. For relatively faint VIS-band deflectors, the Pseudo-RGB branch generally assigns better ranks than VIS-only, suggesting that multi-band information may compensate for weaker morphological signatures in the optical image. Particularly interesting is the behaviour as a function of the NISP-band source magnitudes: Pseudo-RGB becomes comparatively more effective for several populations of sources bright in \HE and \JE, consistent with its ability to exploit NIR emission that is not directly accessible to the VIS-only classifier. This trend is especially relevant for systems with faint or undetected VIS counterparts, such as the VIS-dropout candidate discussed by Euclid Collaboration: Harvey et al. (in prep.), and highlights the potential of NIR-assisted lens searches for future joint \Euclid and \textit{Roman} analyses. 

These results demonstrate that the two image representations encode different property-dependent ranking preferences, which overall reinforce the results obtained on the full DR1 sample: combining morphological and NIR information improves the prioritisation of known strong lenses while reducing the dependence on either image representation alone. The individual VIS-only and Pseudo-RGB scores are therefore retained alongside the final logit-mean score as useful diagnostics for identifying systems whose detection is driven predominantly by one of the two wavelength regimes.

\subsection{\label{ssc:fits-jpeg-models}FITS-trained and JPEG-trained models}

\begin{figure*}[t]
\centering
\includegraphics[width=0.49\textwidth]{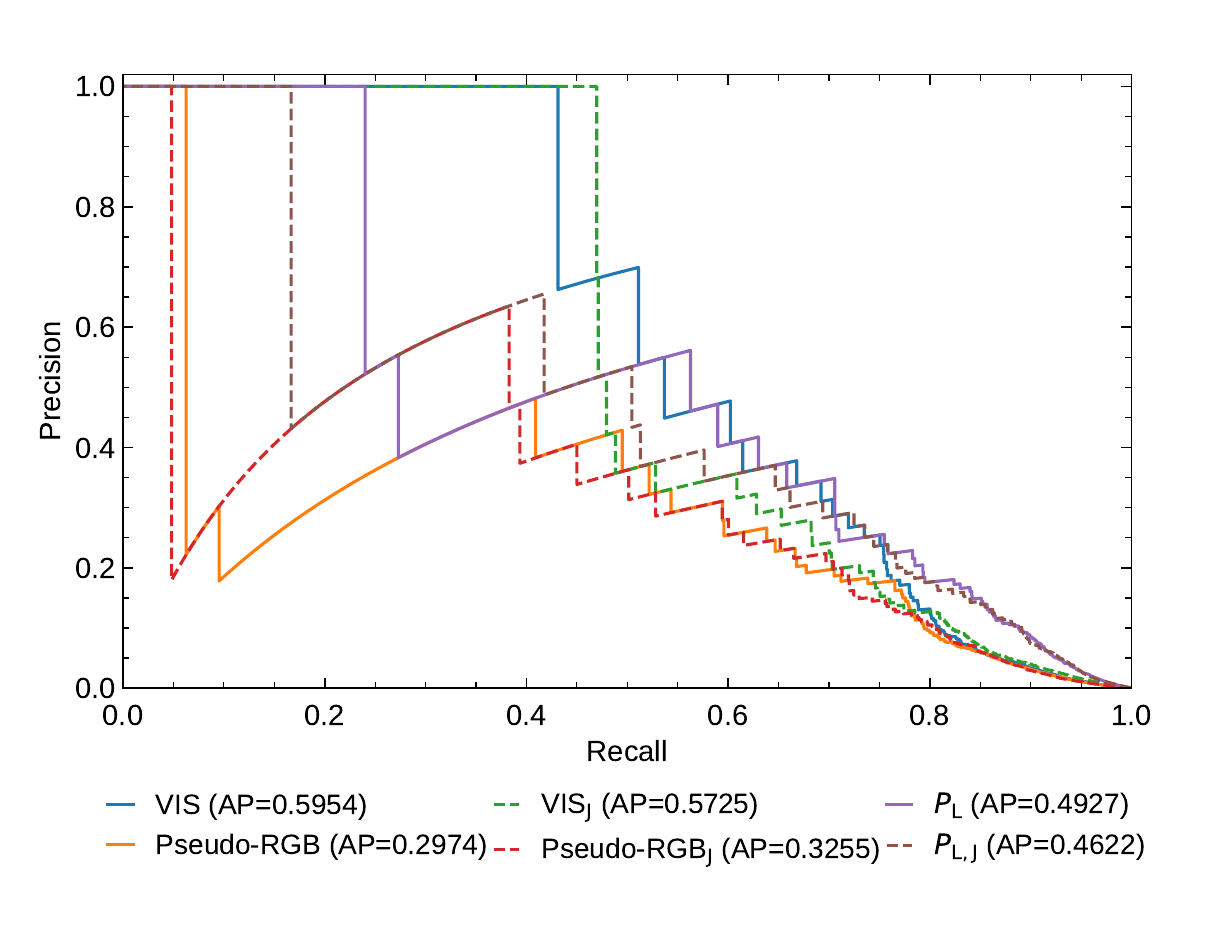}
\includegraphics[width=0.49\textwidth]{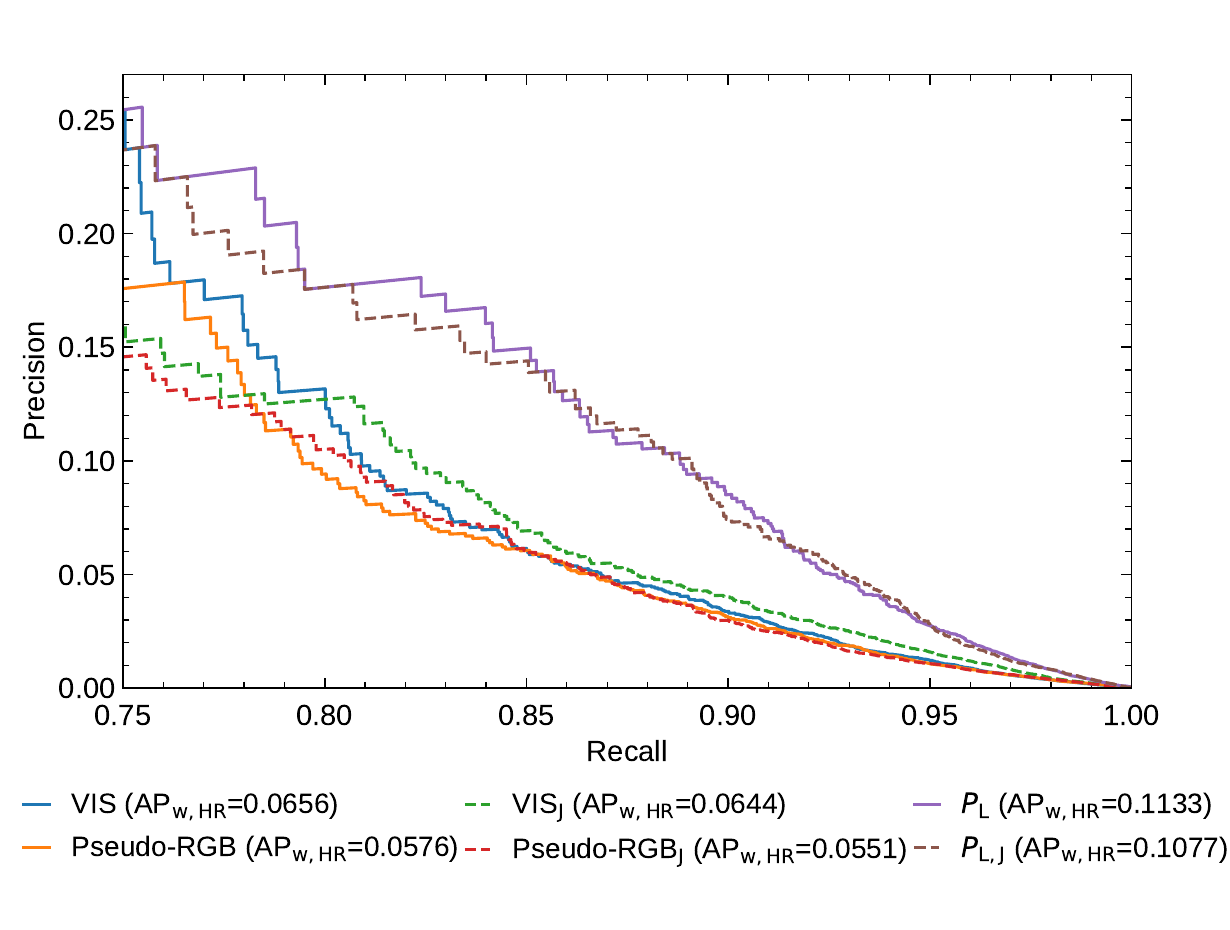}
\caption{
Comparison of weighted PR curves for models trained on FITS-preprocessed inputs and JPEG-compressed versions of the same products. The left panel shows the full weighted PR curves, while the right panel focuses on the high-$R$ interval $R\in[0.75,1.00]$. Legends report the corresponding weighted average Precision and normalised high-$R$ partial areas, respectively. Although the JPEG-trained models remain competitive, the FITS-trained logit-mean ensemble achieves higher overall and high-recall weighted average Precision.
}
\label{fig:fits_jpeg_pr}
\end{figure*}

To assess whether the information loss quantified in Sect.~\ref{sc:fits-vs-jpeg} affects classification, we compare models trained on preprocessed FITS images with equivalent models trained on JPEG-compressed versions of the same inputs.

Figure~\ref{fig:fits_jpeg_pr} shows the weighted PR curves for the individual models and their corresponding logit-mean ensembles, evaluated on the ablation test set. For VIS-only, $\mathrm{AP}_{\rm w}$ decreases moderately from $0.5954$ for the FITS-trained model to $0.5725$ for its JPEG counterpart, corresponding to a $3.8\%$ reduction. Conversely, the JPEG-trained Pseudo-RGB model achieves $\mathrm{AP}_{\rm w}=0.3255$, approximately $9.4\%$ above the FITS-trained reference ($0.2974$). The effect of JPEG compression therefore differs between representations, with both formats remaining competitive in terms of global $\mathrm{AP}_{\rm w}$. At survey-relevant operating points, the trend is more consistent. In the high-$R$ interval, the FITS-trained VIS-only model reaches $\mathrm{AP}_{{\rm w},\mathrm{HR}}=0.0656$, compared with $0.0644$ for JPEG ($-\,1.8\%$), while Pseudo-RGB decreases from $0.0576$ to $0.0551$ ($-\,4.3\%$). Although these differences are relatively modest, the precision among the top $K=70\,000$ candidates also favours FITS for both branches. For VIS-only, $\text{P@\textit{K}}$ decreases from $0.1672$ to $0.1604$ when using JPEGs, corresponding to $-\,4.1\%$ weighted recovered lenses. Similarly, Pseudo-RGB decreases from $\text{P@\textit{K}}=0.1642$ to $0.1554$, corresponding to $-\,5.3\%$ weighted true positives. Thus, FITS-trained models retain an advantage at the DR1 selection depth despite comparable overall $\mathrm{AP}_{\rm w}$.

The ensemble comparison reinforces this interpretation. The FITS-trained logit-mean ensemble reaches $\mathrm{AP}_{\rm w}=0.4927$ and $\mathrm{AP}_{{\rm w},\mathrm{HR}}=0.1133$, compared with $0.4622$ and $0.1077$ for the JPEG-trained ensemble. These correspond to improvements of $6.6\%$ and $5.2\%$, respectively, in favour of FITS.

Overall, these results indicate that JPEG compression does not erase the main lensing information required for successful classification, as JPEG-trained models remain competitive and can even achieve higher overall $\mathrm{AP}_{\rm w}$ (although this advantage is observed only on the ablation test set and may not generalise to the full DR1 survey). Nevertheless, the FITS-preprocessed representations consistently provide better high-$R$ performance and higher top-$K$ precision for both individual branches, while the FITS-trained ensemble also outperforms its JPEG counterpart. Although the present experiment does not isolate which specific image features are responsible for these differences, the results are consistent with the advantages of preserving the original information content of astronomical images. Given the importance of maintaining completeness while limiting contamination on a survey scale, we therefore favoured FITS-based preprocessing for the final ECLIPSE lens-finding pipeline.

\section{\label{sc:discussion-and-conclusion}Discussion and conclusions}

We performed a systematic ablation study of the factors that govern the performance of DL lens finders on \Euclid data. Rather than focusing exclusively on network architecture, we examined how input representation, training set composition, negative-sample selection, training strategy, and score-fusion strategies affect the final survey-scale performance. By combining a controlled evaluation set based on the ECLIPSE lens catalogue with the actual deployment of our final models over approximately $67$ million DR1 cutouts, we demonstrate that some of these practical choices have a substantially larger impact than others and are particularly relevant for future large-scale strong lens searches. 

A first important result concerns the relative contribution of spatial and spectral information. On the weighted ablation test set, the final VIS-only model substantially outperforms Pseudo-RGB in weighted average Precision ($\mathrm{AP}_{\rm w}=0.5954$ versus $0.2974$), demonstrating the efficiency of the high-resolution VIS imaging in the identification of characteristic lensing morphologies. However, this difference becomes much smaller in the high-Recall regime, where the two branches achieve $\mathrm{AP}_{{\rm w},\mathrm{HR}}=0.0656$ and $0.0576$, respectively. More importantly, the relative behaviour changes when the models are evaluated on the complete DR1 population. At the operational selection depth of $K=68\,021$ candidates, Pseudo-RGB recovers $10\,395$ unique grade A and B lenses, compared with $9650$ for VIS-only, despite its lower overall performance on the ablation test set. This demonstrates that the effectiveness of different image representations depends on both the evaluation population and the operating regime, highlighting the importance of validating lens finders under realistic survey conditions. Moreover, the complementary behaviour of the two branches provides a strong motivation for their combination. On the full DR1 sample, the logit-mean ensemble achieves $\mathrm{AP}=0.2784$, compared with $0.2175$ for VIS-only and $0.1990$ for Pseudo-RGB. More importantly, at $K=68\,021$, the ensemble recovers $11\,256$ unique grade A and B lenses, corresponding to $79.1\%$ of the total lensing systems matched to our entire DR1 sample. This represents $1606$ additional unique lenses relative to VIS-only and $861$ relative to Pseudo-RGB, corresponding to improvements of $11.3\%$ and $6\%$ in catalogue recovery, respectively. These gains demonstrate that combining complementary representations can provide substantial improvements in Recall at a fixed candidate-inspection budget, rather than merely increasing integrated classification metrics. We stress, however, that these recovery fractions refer to the identified DR1 catalogue, which was itself constructed through a combined ML and visual-inspection pipeline, and should not be interpreted as unbiased estimates of the completeness of the whole underlying lens population \citep{DR1-SL02}.

The analysis of global candidate rankings as a function of lens properties further illustrates the origin of this complementarity. Across the entire investigated parameter space, the ensemble assigns better ranks to known lenses than both individual branches, with typical median differences of $\Delta\logten R\simeq0.1$--$0.25$, corresponding to individual-model ranks approximately $30\%$--$80\%$ times larger than those of the ensemble. Nevertheless, the relative ranking preferences vary with Einstein radius, lens photometric redshift, ellipticity, and lens/source magnitudes. In particular, Pseudo-RGB becomes comparatively more competitive for several populations of sources bright in the NISP bands, consistent with the additional colour information available to this branch. This is particularly relevant for systems whose lensed features are faint or absent in VIS imaging. The results therefore highlight that NIR information is not merely an alternative representation of the same lensing morphology, but can provide access to systems that optical-only searches may fail to prioritise.

A second major finding concerns the importance of training set composition and, in particular, the selection of negative examples. The Stage 0 ablations demonstrate that replacing relatively simple grade XX negatives with harder grade X contaminants can improve the ranking, although the gains depend strongly on the image representation and Recall regime. For example, the VIS-only train2 configuration improves $\mathrm{AP}_{{\rm w},\mathrm{HR}}$ by $10.6\%$ relative to the final Stage 0 baseline, while Pseudo-RGB achieves a $51.1\%$ improvement in overall $\mathrm{AP}_{\rm w}$ but loses performance at high-$R$. However, these advantages do not persist after Stage 1 fine-tuning: using grade X negatives already during Stage 0 produces less competitive final models, particularly for Pseudo-RGB. This demonstrates that increasing the difficulty of the negative training set is not universally beneficial, and that the optimal composition depends on the stage at which the examples are introduced. The Stage 1 ablations reinforce this conclusion and reveal an additional sensitivity to the reliability and balance of negative labels. In particular, introducing only 280 grade $\rm \bar{C}$ objects into a predominantly grade XX negative pool substantially degrades the Pseudo-RGB performance, while the VIS-only branch exhibits a different response, improving in $\mathrm{AP}_{\rm w}$ but losing performance at high-$R$. Removing these objects improves $\mathrm{AP}_{{\rm w},\mathrm{HR}}$ by $78.2\%$ for Pseudo-RGB and $60.2\%$ for VIS-only relative to the corresponding configurations containing them. The final Stage 1 models instead, which incorporate these challenging examples alongside the harder grade X negative pool, achieve the best high-$R$ performance. Although the underlying mechanism cannot be isolated with the present ablations, these findings suggest that the effectiveness of hard-negative mining depends not only on the difficulty of individual examples, but also on the composition of the surrounding negative sample. For future lens searches, this highlights the importance of carefully curating negative training data rather than simply increasing their difficulty.

Another important result is the benefit of the two-stage training strategy. Removing Stage 0 pre-training causes a substantial degradation in high-$R$ performance, reducing $\mathrm{AP}_{{\rm w},\mathrm{HR}}$ by $83.7\%$ for VIS-only and $74.5\%$ for Pseudo-RGB relative to the final configurations. Interestingly, the Pseudo-RGB model trained directly on the Stage 1 data achieves the highest $\mathrm{AP}_{\rm w}$ among its ablations ($0.4055$), despite its substantially poorer high-$R$ performance. This provides a particularly clear example of how optimising global classification metrics can favour models that are less suitable for high-completeness candidate generation. The results suggest that simulated pre-training remains valuable for exposing the networks to a broad diversity of lens configurations, while subsequent fine-tuning on real observations is essential for adapting the learned representations to the contaminants encountered in the survey. Finally, for the two-stage strategy investigated here, a particularly useful practical choice is to reserve the grade-X negatives for fine-tuning on real observations rather than introducing them during simulation-based pre-training.

The comparison between FITS-trained and JPEG-trained models provides a further practical consideration. Although JPEG compression reduces the effective photometric dynamic range of the input images, its impact on overall classification performance is relatively modest and not always negative. For example, JPEG-trained Pseudo-RGB achieves a $9.4\%$ higher overall $\mathrm{AP}_{\rm w}$ than its FITS-trained counterpart on the ablation test set. Nevertheless, FITS-trained models retain a consistent advantage in the high-$R$ regime, with improvements of $1.8\%$ and $4.3\%$ in $\mathrm{AP}_{{\rm w},\mathrm{HR}}$ for VIS-only and Pseudo-RGB, respectively. The FITS-trained logit-mean ensemble also improves $\mathrm{AP}_{\rm w}$ by $6.6\%$ and high-$R$ performance by $5.2\%$ relative to its JPEG-trained counterpart. These results suggest that preserving the original information content of astronomical images is preferable when computationally feasible, although the benefits are smaller than those associated with training-set composition and model ensembling.

Taken together, these findings provide several practical lessons for the development of future survey-scale strong lens finders. First, the evaluation metric should reflect the intended operational use of the classifier: for candidate-ranking pipelines followed by visual inspection, high-$R$ performance and Recall at a fixed inspection budget are more informative than global classification metrics alone. Second, realistic simulations and carefully selected hard negatives are both essential, but their contribution depends on how they are incorporated into the training procedure. Third, combining complementary image representations can provide substantial gains even when one individual model appears considerably stronger on a controlled test set. Finally, preserving the information content of the input data remains desirable, although improvements in preprocessing alone cannot compensate for limitations in training set design. The results therefore reinforce the idea that training set design and input representation should be treated as a first-order scientific problem rather than a secondary implementation detail. Our conclusions are established for a fully fine-tuned ConvNeXt-Base architecture. Whether the same sensitivity to hard-negative composition and input representation persists for smaller and computationally cheaper networks remains to be tested.

As \Euclid progresses from the 63\,deg$^2$ Q1 release to the ${\sim}\,2000$\,deg$^2$ DR1 footprint and ultimately to the full ${\sim}\,14\,000$\,deg$^2$ survey, the challenge will increasingly shift from classification to ranking: identifying a small number of true lenses among hundreds of millions of galaxies. The methodology developed here is specifically optimised for that operating regime. By combining realistic simulations, extensive hard-negative mining, high dynamic-range image processing, and complementary model ensembles, we establish a scalable framework that has been deployed as one of the ML components of the ECLIPSE DR1 strong-lens search (\citealt{DR1-SL01,DR1-SL02}, Euclid Collaboration: Vincken et al. in prep.) and that can be readily extended to future \Euclid data releases. More broadly, the demonstrated complementarity between optical and NIR information suggests that similar approaches will be valuable also for forthcoming wide-field surveys such as \textit{Roman} and LSST, but remains to be established through instrument-specific training and validation. These developments will be essential for efficiently discovering the expected population of ${\sim}\,10^5$ strong lenses and for enabling the wide range of cosmological, dark matter, galaxy evolution, and high-redshift astrophysical studies that such samples will make possible.

\begin{acknowledgements}
\AckEC
\AckDatalabs
\AckQone
\AckDRone
This work has made use of CosmoHub \citep{tallada2020,carretero2017}, developed by PIC (maintained by IFAE and CIEMAT) in collaboration with ICE-CSIC. It received funding from the Spanish government (grant EQC2021-007479-P funded by MCIN/AEI/10.13039/501100011033), the EU NextGeneration/PRTR (PRTR-C17.I1), and the Generalitat de Catalunya. 

AN gratefully acknowledges the support of the Belgian Federal Science Policy Office (BELSPO) for the provision of financial support in the framework of the PRODEX Programme of the European Space Agency (ESA) under contract number 4000143347. 

MG acknowledges support from the Polish National Agency for Academic Exchange grant No. BPN/BEK/2024/1/00325/DEC/1. 

This project has received funding from the European Research Council (ERC) under the European Union’s Horizon 2020 research and innovation programme (LensEra: grant agreement No 945536). TEC is funded by the Royal Society through a University Research Fellowship.

Numerical computations were done on the Sciama High Performance Compute (HPC) cluster which is supported by the ICG, SEPNet and the University of Portsmouth.

NL is supported through a graduate studentship from the UKRI STFC and the University of Portsmouth, and acknowledges travel support provided by Euclid:UK through STFC grant APP91126 and by STFC for UK participation in the Rubin LSST through grant ST/X001334/1.

PH is funded by the Royal Society (RF/ERE/210033) and acknowledges travel support provided by Euclid:UK through STFC grant APP91126.

PB acknowledges financial support from the Italian Space Agency (ASI) through contract ``Euclid - Phase E'', and INAF Grants ``The Big-Data era of cluster lensing'' and ``Probing Dark Matter and Galaxy Formation in Galaxy Clusters through Strong Gravitational Lensing''.
\end{acknowledgements}

%
%

\bibliography{Euclid, DR1,my_cit} 
%
%

%

\clearpage
\appendix
\onecolumn

\section{Examples of simulated strong lenses}
\label{app:simulated_lenses}

\begin{figure}[H]
\centering
\includegraphics[width=\textwidth,height=0.80\textheight,keepaspectratio]{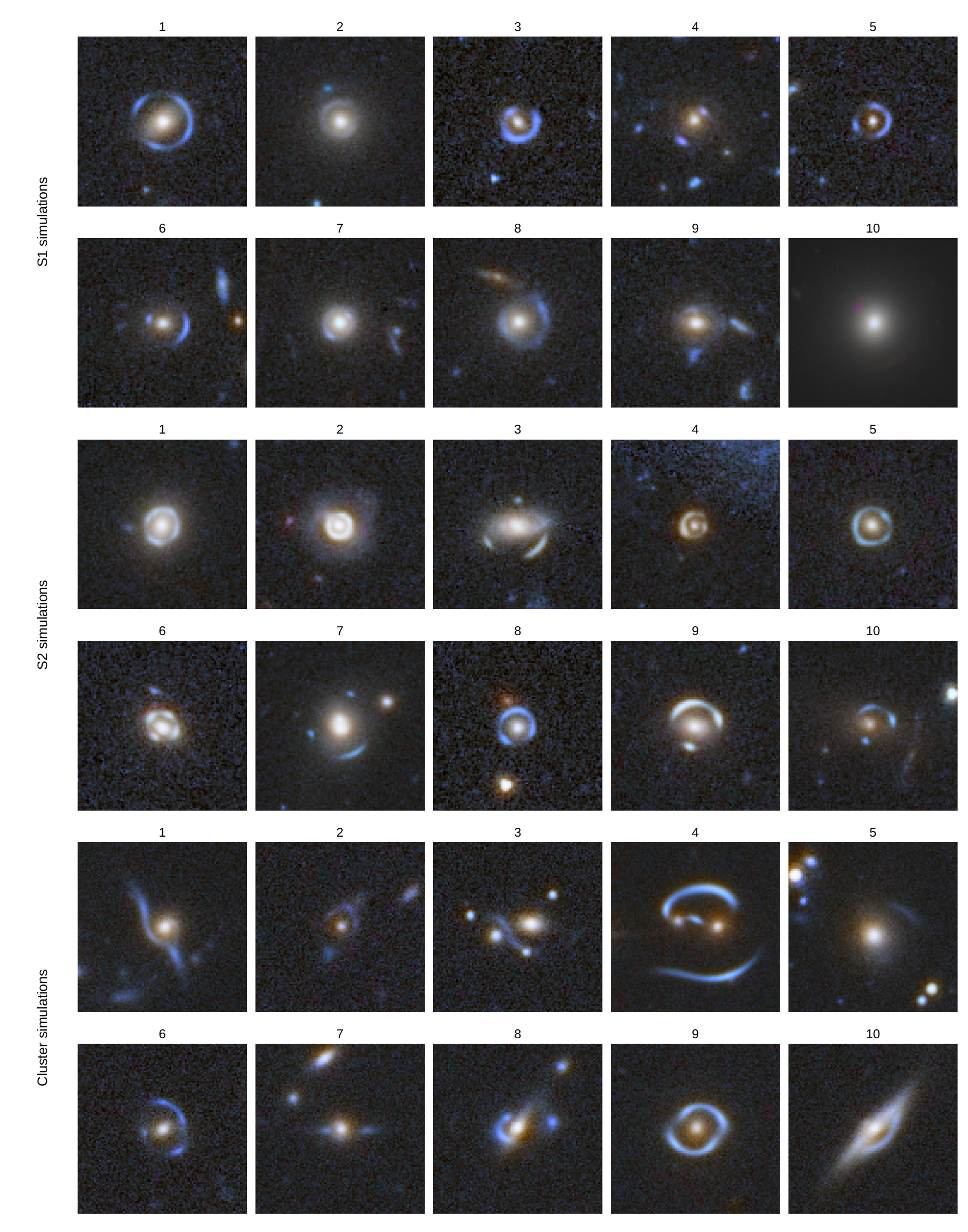}
\caption{Examples of simulated strong lensing systems used in this work. The first two rows show S1 field-GGSL simulations, while the following two rows show S2 field-GGSL simulations. The bottom two rows show cluster-GGSL simulations, with ten examples per population. The images are displayed as VIS and NISP colour composites using an MTF-like nonlinear stretch, and all panels show $10\arcsec \times 10\arcsec$ cutouts. This rendering is applied for visualisation only and does not represent the preprocessing adopted for network training.}
\label{fig:appendix_simulations}
\end{figure}

\noindent\begin{minipage}{0.48\textwidth}
Figure~\ref{fig:appendix_simulations} shows representative examples of the three simulated lens populations described in Sect.~\ref{sc:simulated-ggsl-events}, including the S1 and S2 field-GGSL simulations and the cluster-GGSL simulations.
\end{minipage}

\clearpage

\section{Input image representations}
\label{app:input_representations}

\begin{figure}[H]
\centering
\includegraphics[width=\textwidth,height=0.69\textheight,keepaspectratio]{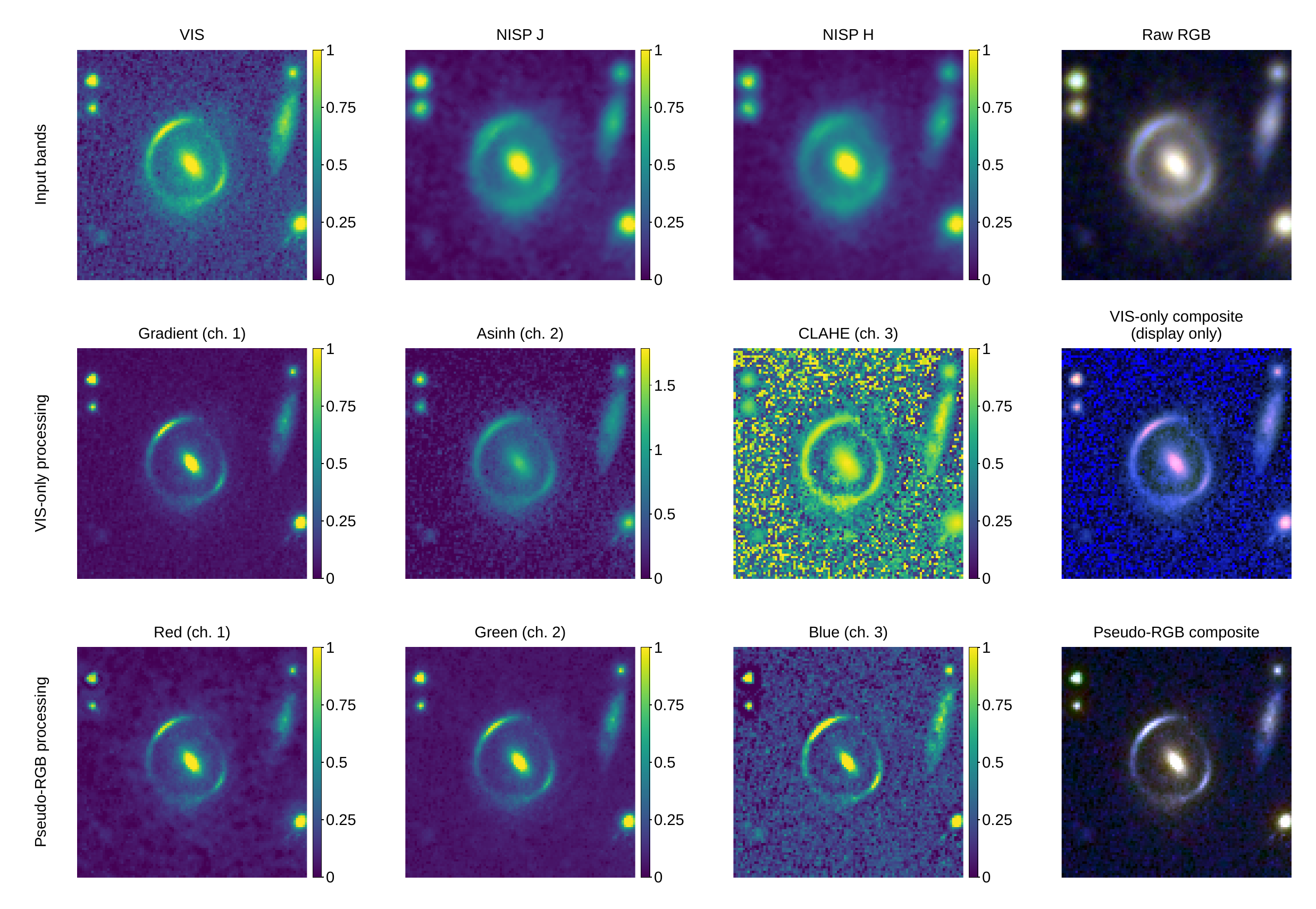}
\caption{Example of the input representations adopted in this work. The top row shows the original VIS \IE, NISP \JE, and \HE images, together with their RGB composite. The middle row presents the three VIS-only channels obtained through gradient enhancement, asinh stretching, and CLAHE, alongside their composite (shown for illustrative purposes only, as the individual channels rather than their combined RGB visualisation constitute the actual network input). The bottom row shows the individual red, green, and blue channels of the pseudo-RGB representation and the resulting colour composite. The original bands are displayed using an asinh stretch and percentile clipping, while the network channels fully reproduce the preprocessing described in Sect.~\ref{sc:trainingmethodology}. All panels show $10\arcsec \times 10\arcsec$ cutouts for system EUCL--DR1--L000783 in the ECLIPSE catalogue.}
\label{fig:appendix_input_representations}
\end{figure}

\noindent\begin{minipage}{0.48\textwidth}
Figure~\ref{fig:appendix_input_representations} illustrates the image preprocessing procedures described in Sect.~\ref{sc:trainingmethodology}, comparing the original \Euclid bands with the three channels used by the VIS-only and pseudo-RGB classifiers.
\end{minipage}

\label{LastPage}
\end{document}